\documentclass[a4paper,11pt]{article}
\pdfoutput=1 

\usepackage{wrapfig}
\usepackage{jcappub} 

\usepackage{caption}
\usepackage{changepage}
\usepackage{morefloats}

\usepackage{tablefootnote}
\usepackage[T1]{fontenc} 

\def\pmb#1{\setbox0=\hbox{#1}%
    \kern-.025em\copy0\kern-\wd0
    \kern.05em\copy0\kern-\wd0
    \kern-.025em\raise.0433em\box0}
\def\ltsima{$\; \buildrel < \over \sim \;$}
\def\gtsima{$\; \buildrel > \over \sim \;$}
\def\simlt{\lower.5ex\hbox{\ltsima}}
\def\simgt{\lower.5ex\hbox{\gtsima}}

\def\mask1{mask$\_$1}
\def\mask2{mask$\_$2}
\def\mask3{mask$\_$3}
\def\mask4{mask$\_$4}
\def\mask5{mask$\_$5}
\def\mask70{mask$\_$70}
\def\mask3pol{mask$\_$pol$\_$3}
\def\mask4pol{mask$\_$pol$\_$4}

\def\p2Y{\;_2Y}
\def\m2Y{\;_{-2}Y}

\numberwithin{equation}{section}
\numberwithin{figure}{section}

\providecommand{\Planck}{\textit{Planck}}

\providecommand{\text}[1]{\rm{#1}}

\newcommand{\Mpc}{\text{Mpc}}

\newcommand{\Hunit}{~\text{km}~\text{s}^{-1} \Mpc^{-1}}

\providecommand{\LCDM}{{$\rm{\Lambda CDM}$}}

\newcommand{\begm}{\begin{pmatrix}}
\newcommand{\enm}{\end{pmatrix}}

\newcommand\ba{\begin{eqnarray}}
\newcommand\ea{\end{eqnarray}}
\newcommand\bea{\begin{eqnarray}}
\newcommand\eea{\end{eqnarray}}

\newcommand\be{\begin{equation}}
\newcommand\ee{\end{equation}}

\def\pmb#1{\setbox0=\hbox{#1}%
    \kern-.025em\copy0\kern-\wd0
    \kern.05em\copy0\kern-\wd0
    \kern-.025em\raise.0433em\box0}
\def\ltsima{$\; \buildrel < \over \sim \;$}
\def\gtsima{$\; \buildrel > \over \sim \;$}
\def\simlt{\lower.5ex\hbox{\ltsima}}
\def\simgt{\lower.5ex\hbox{\gtsima}}

\def\p2Y{\;_2Y}
\def\m2Y{\;_{-2}Y}
\newcounter{parentequation1}

\title{A Lockdown Perspective on the Hubble Tension\\
{\vspace{-3mm}\large(with comments from the SH0ES team)}}

\author{G. Efstathiou}

\affiliation{Kavli Institute for Cosmology Cambridge and Institute of Astronomy \\
Madingley Road\\Cambridge CB3 OHA\\UK}

\emailAdd{gpe@ast.cam.ac.uk}

\abstract{This is a transcript of a talk that I gave in Cambridge
on 17th July 2020  on the  `Hubble tension', i.e. the
  discrepancy between traditional distance ladder measurements of the
  Hubble constant (which I will refer to as `late time measurements')
  and the value inferred from observations of the cosmic microwave
  background (CMB) and large-scale structure (`early time measurements'). 
  I review the SH0ES\footnote{SNe, $H_0$, for the Equation of State of
  dark energy} analyses by Riess
  and collaborators and point out some internal inconsistencies,
  including a discrepancy between the relative distances inferred from  Cepheids
 of two of the
  primary geometric distance anchors, the
  Large Magellanic Cloud (LMC) and NGC 4258. I then
 ask   `what would it take to make SH0ES compatible with
  early time measurements?'. The answer is a systematic bias of
  $\sim 0.1 - 0.15$ mag in the intercept of the Cepheid period-luminosity relations 
of  SH0ES galaxies.  Such a bias  resolves the Hubble tension,
the tension between the distance anchors, and the difference between SH0ES 
and the tip of the red giant branch (TRGB) distance ladder, as measured and calibrated by 
Freedman and collaborators.
I show that the difference between the  TRGB and SH0ES values of $H_0$ is caused mainly
  by a systematic calibration offset.   In the short term, observational efforts should
  be focussed on improving the calibrations of the distance anchors and nearby galaxies,
  rather than trying to measure distance moduli to more supernovae
  host galaxies. I argue that an independent distance estimate
  to NGC 4258 is particularly critical. With such observations, it should be possible, on a relatively short
timescale,  to
  establish definitively whether the Hubble tension really exists.  }

\begin{document}
\maketitle
\flushbottom

\vfill\eject

\section{Introduction}
\label{sec:introduction}

We are experiencing very strange times. One consequence of prolonged enforced isolation
 has been to fuel my obsession with `cosmic tensions'. Without the normal constraints of a `day job',
and the restraining influences of close colleagues,  this obsession has led to the work described here. 

By now, the `Hubble tension' has become well known to the astronomical community and beyond, and so needs
very little introduction. The latest analysis of the Cepheid-based distance scale measurement of $H_0$ from  the SH0ES collaboration \citep{Riess:2011, Riess:2016,
  Riess:2019} gives
\begin{equation}
H_0 = 74.03 \pm 1.42 \Hunit, \label{equ:H0_1}
\end{equation}
whereas the value inferred from \Planck\ assuming the standard six parameter  \LCDM\  model\footnote{I will refer to this model as base \LCDM.} is
\begin{equation}
H_0 = 67.44 \pm 0.58 \Hunit, \label{equ:H0_2}
\end{equation}
\cite{PCP18, Efstathiou:2019}. The difference of $6.9 \Hunit$
 between (\ref{equ:H0_1}) and (\ref{equ:H0_2}) is, apparently, a $4.3\sigma$ discrepancy. Another way of expressing the 
tension is to note that the difference between (\ref{equ:H0_1}) and (\ref{equ:H0_2}) is nearly $10\%$ and is much larger
than the $2\%$ error of the SH0ES estimate,  {\it which includes estimates of systematic errors}. This is, therefore, an
intriguing tension which has  stimulated a large (almost daily) literature on extensions to the
base \LCDM\ model, focussing mainly  on mechanisms to reduce the sound horizon \citep[for a review see][]{Knox:2020}.

Although the Hubble tension first became apparent following the
publication of the first cosmological results from
\Planck\ \cite{PCP13}, the low value of $H_0$ in the base
\LCDM\ cosmology can be inferred in many other ways. For example,
applications of the inverse distance ladder assuming either the
\Planck\ or WMAP \cite{Bennett:2013} values of the sound horizon,
$r_d$, or a CMB-free value of $r_d$ inferred from primordial
nucleosynthesis, consistently yield a low value of $H_0$
\citep[e.g.][]{Aubourg:2015, Verde:2017, Addison:2018, Abbott:2018,
  Macaulay:2019}. Furthermore, using BAO and SN data, it is possible
to reconstruct $H(z)$, independently of the late time behaviour of
dark matter and dark energy \cite{Bernal:2016, Lemos:2019}. The
resulting low value of $H_0$ strongly suggests that any departure from
the base \LCDM\ model must involve changes to the physics at early
times. This new physics must mimic the exquisite agreement of the base
\LCDM\ model with the \Planck\ temperature and polarization power
spectra and hopefully preserve the consistency of observed light element
abundances and primordial nucleosynthesis. Models invoking a transient
scalar field contributing to the energy density just prior to
recombination (early dark energy) have been suggested as a possible
solution to the Hubble tension \cite{Poulin:2019, Agrawal:2019,
  Smith:2020}, but these models fail to match the  shape of the
galaxy power spectrum \cite{Hill:2020, Ivanov:2020, D'Amico:2020}. The
shape of the galaxy power spectrum itself provides tight constraints
on the Hubble constant assuming the base \LCDM\ cosmology, consistent
with Eq. (\ref{equ:H0_2}) \cite{ D'Amico:2020b, Philcox:2020}. I think it
is fair to say that despite many papers, no compelling theoretical
solution to the the Hubble tension has yet emerged.

An alternative explanation\footnote{With all due respect to the SH0ES
  team.} is that the SH0ES result is biased by systematic errors that
are not included in the error estimate of Eq. (\ref{equ:H0_1}). As is
well known, distance ladder measurements of $H_0$ are extremely
challenging and have a long and chequered history. However, before
progressing further it is important to note that at the time of
writing the balance of evidence seems to be tilted in favour of the
SH0ES result. Gravitational lensing time delays \citep{Bonvin:2017,
  Wong:2019, Shajib:2020}, distant maser galaxies \citep{Pesce:2020}
and Mira variables \cite{Huang:2020} all yield high values of the
Hubble constant compared to the base \LCDM\ value of (\ref{equ:H0_2})
though with larger errors than the SH0ES estimate. However, the recent
TRGB measurements \citep{Freedman:2019, Freedman:2020} give
\begin{equation}
H_0 = 69.6 \pm 1.9 \Hunit, \label{equ:H0_3}
\end{equation}
apparently intermediate between the \Planck\ and SH0ES values (but see Sect. \ref{sec:TRGB}).

In this article, I address the following questions:

\smallskip

\noindent
(i)  Can we imagine systematics that would bias the  SH0ES analysis?

\smallskip

\noindent
(ii)  Is there any evidence for such  systematics?

\smallskip

\noindent
(iii) Are the TRGB results of \citep{Freedman:2019, Freedman:2020} compatible with SH0ES?

\smallskip

\noindent
(iv) What new observations are needed to resolve the discrepancies identified in this article?

\ \par
\begin{adjustwidth}{0.25in}{0.25in}
\noindent{\underline{\bf Response from SH0ES Team:} G.E. has kindly offered the SH0ES Team the opportunity 
to provide short responses to the points in his talk below.  We have kept these at the brief level
one might expect in the Q\&A after such a talk in times before ``Lockdowns'' when such exchanges were possible. 
We thank G.E. for providing a flavor of that interaction here.}
\ \par
\end{adjustwidth}

\section{Comments on the SH0ES data}
\label{sec:SH0ES_data}

I will refer to the sequence of SH0ES papers \cite{Riess:2011, Riess:2016, Riess:2019} as R11, R16, and R19 
 respectively and I  will begin with some representative numbers.
Suppose we wanted to explain the Hubble tension by invoking an additive error $\delta m$ 
in the magnitudes of SH0ES Cepheids. The shift in the value of $H_0$ would be
\begin{equation}
{\delta H_0 \over H_0} =  0.2\ln 10 \ \delta m. \label{equ:data1}
\end{equation}
The $6.9 \Hunit$ difference between (\ref{equ:H0_1}) and
(\ref{equ:H0_2}) requires $\delta m = 0.2 \ {\rm mag} $ (in the sense
that the SH0ES Cepheids would have to be too
bright)\footnote{Note that a shift of $\delta m = 0.1 \ {\rm mag.}$
  would reduce the Hubble tension to about $2\sigma$.}.  However, the
errors on individual distance moduli of SN host galaxies (e.g. from
Table 5 of R16) are typically quoted as $\delta \mu \sim 0.06 \ {\rm
  mag}$. Averaging over $19$ SN host galaxies, the error in the
Cepheid magnitude scale is of order $0.06 /\sqrt{19} \sim 0.014 \ {\rm
  mag.}$, {\it i.e. } about an order of magnitude smaller than the
required offset. At first sight it seems implausible that a magnitude
offset could have any bearing on the Hubble tension.

\begin{figure}
\centering
\includegraphics[width=150mm,angle=0]{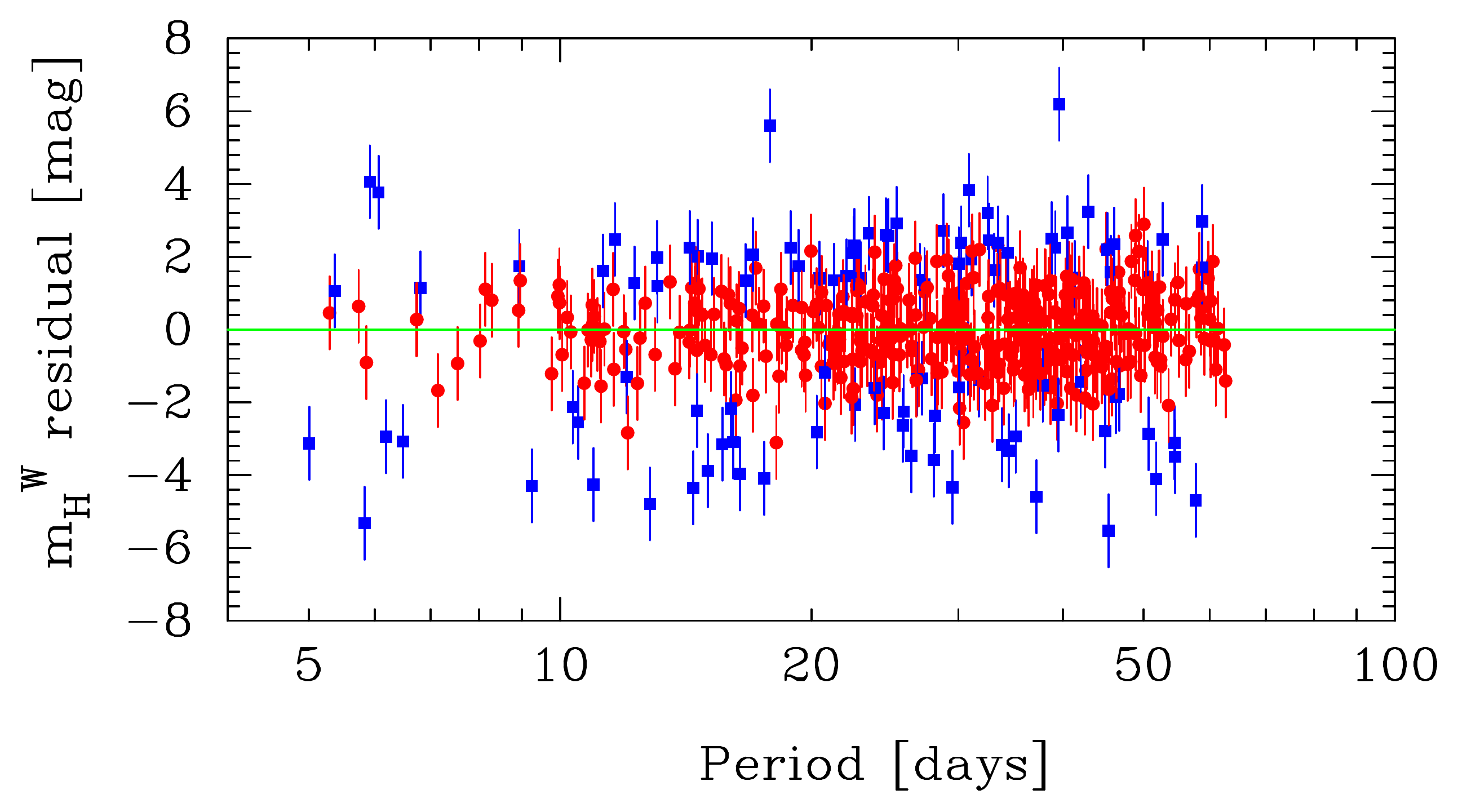} 
\caption {R11 PL magnitude residuals relative to the global fit 5 of Table 2 in E14.
  Red points show  residuals for Cepheids that are accepted by  R11 while 
  blue points are rejected by R11.}

\label{fig:HR11residual}

\end{figure}

\subsection{Outliers}
\label{subsec:outliers}

However, consider Fig. \ref{fig:HR11residual}. This is a combination of the two panels of Fig. 2 from \cite{Efstathiou:2014}
(hereafter E14). This shows residuals of the Wesenheit H-band ($m^W_H$) period luminosity (PL) relation from a global fit to the 9 galaxies in the R11 sample. As in R11, 
\begin{equation}
m_H^W = m_H - R(V-I), \label{equ:data2}
\end{equation}
where $H=F160W$, $V=F555W$ and $I=F814W$ in the HST photometric system. In Fig. 2, I used $R=0.41$ (consistent with R11), though for the rest of this article I use $R = 0.386$ to be consistent with R16 and later SH0ES papers. One can see from Fig. \ref{fig:HR11residual} that there are many `outliers', with residuals that differ by 3 mag or more from the best fit. R11 identified outliers by fitting the PL relation for each galaxy
 separately (rather than identifying outliers with respect to the global fit).

The R11 rejection algorithm works as follows:

\noindent
$\bullet$ The $m_H$ PL relations are fitted galaxy-by-galaxy, weighted by the magnitude errors in Table 2 of R11,
 to a power law with  slope fixed at $b=-3.1$ in the first iteration.
Cepheids with periods  $> 205$ days are excluded.

\noindent
$\bullet$  Cepheids are rejected if they deviate from the best-fit relation by $\ge 0.75 \ {\rm mag}$, or by more than 
2.5 times the magnitude error.

\noindent
$\bullet$ The fitting and rejection is repeated iteratively 6 times.

\smallskip

The Cepheids rejected by R11 are shown by the blue points in Fig. \ref{fig:HR11residual} and those accepted by R11
are shown by the red points. For the red points, the dispersion around the mean is $0.47$ mag  and it is  $0.60$ mag. for
all points. {\it To avoid a bias in the measurement of $H_0$, the mean must be determined to an accuracy much better
than $0.1$ mag (or a quarter of a tick mark on the y-axis in Fig. \ref{fig:HR11residual}}). The finite width of the
instability strip will produce an `intrinsic' scatter in the PL relation. In the H-band, the intrinsic scatter is
about $0.08$ mag and can be measured accurately from photometry of LMC and M31 Cepheids. The much higher dispersion 
apparent in Fig. \ref{fig:HR11residual} is a consequence of photometric errors and misidentification of Cepheids (type II
Cepheids misclassified as classical Cepheids) and is larger than the error estimates given in R11. In the H-band, 
Cepheid photometry of SN host galaxies is challenging even with HST because of crowding. Photometry
 requires deblending of the images, as illustrated in Fig. 5 of R16, and also crowding corrections to the local sky background. In addition, 
one can see from
Fig. \ref{fig:HR11residual} that the blue points  are distributed asymmetrically around the mean, and so the question that
has worried me is whether the PL relations are really  unbiased at the $0.1$ mag level, given that the underlying data 
are so noisy and contain so many outliers\footnote{E14 experimented with different ways of rejecting outliers relative to
{\it global} fits.}. The SH0ES methadology requires the absence of any systematic biases, relying on large numbers of Cepheids
to reduced the errors on distance moduli well below the $\sim 0.5$ dispersion seen in Fig. \ref{fig:HR11residual}.

\begin{figure}
\centering
\includegraphics[width=150mm,angle=0]{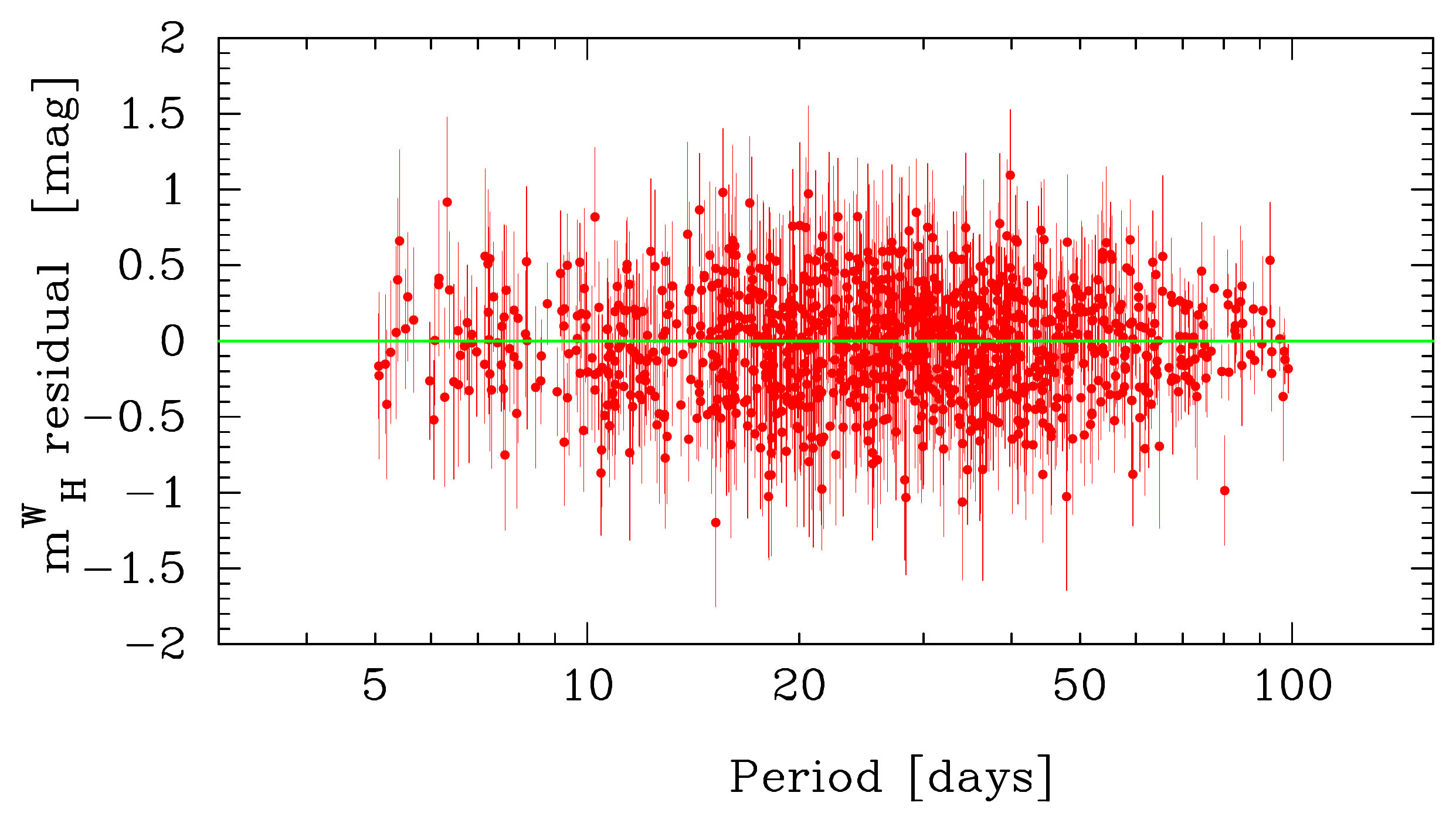} 
\caption {R16 magnitude residuals relative to a global fit of the PL relation for
the 19 SN host galaxies and NGC 4258.}

\label{fig:HR16residual}

\end{figure}

If we now fast forward to R16, the equivalent plot to
Fig. \ref{fig:HR11residual} is shown in Fig. \ref{fig:HR16residual}.
The number of SN host galaxies with Cepheid photometry was increased
from 8 in R11 to 19 in R16. The optical data (in the F350LP, F555W and
F814W filters) for the R16 sample is described in detail in
\cite{Hoffmann:2016}\footnote{In the optical,
the images are largely the same, but with WIFPC3 data for three galaxies (NGC 1309, 3021, 3370) supplementing earlier ACS and WFPC2 images.}. For the 9  galaxies in common between R11
and R16, the F160W images are identical in the two analyses, 
but were reprocessed for R16. There
are now no obvious outliers in Fig. \ref{fig:HR16residual}. and the dispersion around the mean is
$0.35 \ {\rm mag}$ with $\chi^2=1114$ for 1037 Cepheids. What happened to the outliers?
The outliers in R16 were removed in two stages: first for each
individual galaxy, Cepheids were rejected if their F814W-F160W colours
lay outside a $1.2 \ {\rm mag}$ colour range centred on the median
colour. An additional $\approx 5\%$ of the Cepheids were rejected if
their magnitudes differed by more than $2.7\sigma$ from a global
fit. Since colour selection removes a large fraction of the outliers,
R16 argue that outlier rejection is not a serious issue. However, R16
did not publish magnitudes for the rejected Cepheids and so it is not
possible to check the impact of colour selection and outlier rejection.
It is not obvious to this author that applying
symmetric rejection criteria to exclude outliers will produce unbiased
results. The R16 photometric data has been investigated by
\cite{Follin:2018} and (extensively) by myself and appears to be
self consistent.

\vfill\pagebreak\newpage
\begin{adjustwidth}{0.25in}{0.25in}
\noindent{\narrower{\underline{\bf Response from SH0ES Team:} The discovery of fundamental-mode Cepheid variables from time-series imaging requires a number
of winnowing steps from non-variables which outnumber them more than 1000-fold and other variable types which rival their frequency.  Colors are useful to distinguish Cepheids (which have spectral types F-K, a modest color range) from other variables like Miras and Semi-regular variables which are redder (spectral type M) or RR Lyraes which are bluer (spectral types A-F). Colors may also remove strong blends of a Cepheid with a red or blue Giant.  R11 did not use $I-H$ colors in this way and subsequently rejected many ``outliers'' off of the period-luminosity relation though these were published for further analyses, e.g., by G.E.  R16 added $I-H$ color selection and at the same rejection threshold of the PL relation found very few PL relation outliers, about 2\% of the sample and their inclusion or exclusion has little impact on $H_0$.   So most of the R11 outliers were too red or blue to be Cepheids and are not outliers in R16 because they never make it to the PL relation.  (We consider outliers only as objects that pass selection criteria.)
We think the R16 selection is a clear and defined improvement in method and supersedes the R11 results which are no longer relevant.}}
\ \par
\ \par
\end{adjustwidth}

 Nevertheless, we can test the R16 analysis by invoking additional data. Here I describe four tests:

\noindent
(i) Consistency of the slope of the SH0ES PL relation with that of nearby  galaxies.

\noindent
(ii) Consistency of the R11 and R16 photometry.

\noindent
(iii) Consistency of the distance anchors.

\noindent
(iv) Consistency of Cepheid and TRGB distance moduli.

\begin{figure}
\centering
\includegraphics[width=130mm,angle=0]{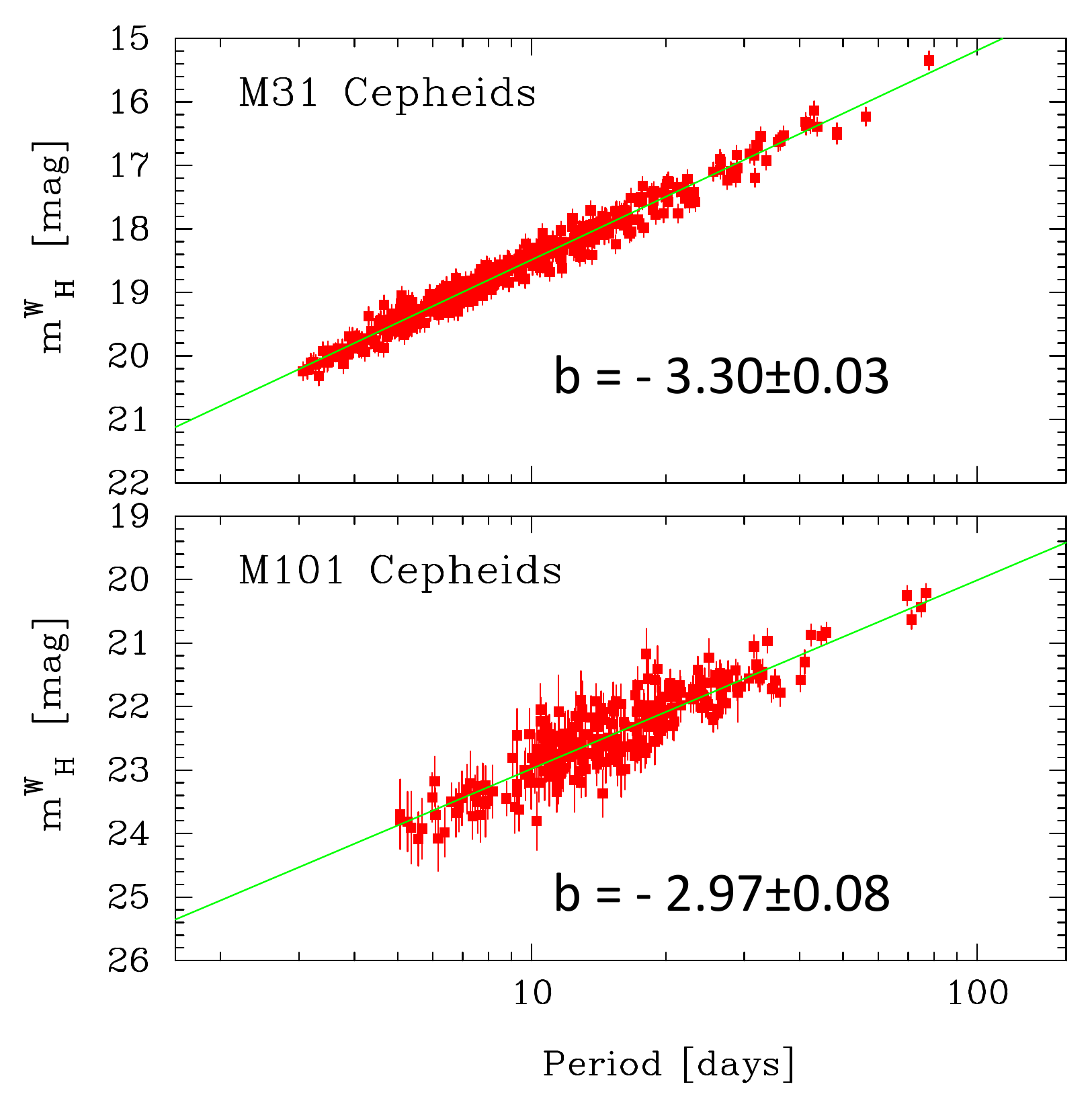} 
\caption {PL relations for M31 (upper plot) and M101 (lower plot). The best fit slope and its $1\sigma$ error are
listed in each plot. }

\label{fig:slopes}

\end{figure}

\subsection{The slope of the PL relation}
\label{subsec:slopes}

 The upper panel
of Fig. \ref{fig:slopes} shows the PL relation for M31\footnote{The M31 photometry is 
listed  in R16 and is based on the following sources \cite{Dalcanton:2012, Riess:2012, Kodric:2015, Wagner-Kaiser:2015}.}. The solid line shows
a fit
\begin{equation}
m_H^W = a + b \log_{10} \left [ { P \over 10 \ {\rm days}} \right ],  \label{equ:dataP}
\end{equation}
where $P$ is the Cepheid period. For M31, the slope is\footnote{There has been some discussion in the literature, e.g.
\cite{Kodric:2015} of a break in the slope of the PL relation at $P \sim \ 10 \ {\rm days}$. There is no evidence of such a break
at H-band. Fitting the M31 PL relation to Cepheids with $P \ge 10 \ {\rm days}$ gives $b = -3.30 \pm 0.06$. Note that for the LMC, using the 70 Cepheids with HST photometry listed in R19, I find $b=-3.32 \pm 0.04$.} 
\begin{equation}
  b = -3.30 \pm 0.03.       \label{equ:data4}
\end{equation}
The lower panel in Fig. \ref{fig:slopes} shows the PL relation for
M101 using the photometry from R16. The slope for M101 is much shallower
than that for M31, differing by $3.9 \sigma$, strongly suggesting a
bias. One finds similar results for other R19 galaxies, including NGC
4258.  Global fits to the PL relation using either the R11 or R16 data
therefore give a slope that is shallower than (\ref{equ:data4}). The
reason for this is clear from Fig. 15 of \cite{Hoffmann:2016} which shows that
the  Cepheid sample is incomplete at short periods. The
short period Cepheids are faint and close the photometric limits of the R16
observations. As a consequence, the optical selection is biased
towards brighter Cepheids. \cite{Hoffmann:2016} impose lower limits on the periods
for the final Cepheid sample, but it is clear from their Fig. 15 that these limits
are not sufficiently conservative. If we assume that the Cepheid PL relation is universal\footnote{Clearly a necessary assumption if we are to use Cepheids in distance ladder measurement.} then we can assess the impact of this incompleteness bias
by comparing values of $H_0$ with and without imposing a prior on the slope  (see E14). We will show below
that imposing a strong slope prior of $b=-3.300 \pm 0.002$\footnote{The width of this prior is not a typo. To force $b=-3.3$, the prior must be tight enough 
to counteract the shallow slow of the R16 PL relation.}  {\it lowers} $H_0$ by  $1.7 \Hunit$.
 This is a systematic shift, which is larger than the $1\sigma$ error
quoted in Eq. (\ref{equ:H0_1}).

\begin{adjustwidth}{0.25in}{0.25in}
\ \par
\noindent{\underline{\bf Response from SH0ES Team:} We find the statistical uncertainty in individual slopes is often an underestimate because individual slopes are strongly influenced by the presence or absence of rare very long period Cepheids ($P>70$ days).   
For example, in the LMC we measured the H-band slope from 43 Cepheids with $P>10$ days (Riess et al. 2019) to be $-3.38 \pm 0.07$ mag while Persson et al. (2004) measured it to be $-3.16 \pm 0.1$ from 39 $P>10$ day Cepheids, a highly significant difference considering this is the same galaxy and mostly same objects.  The difference comes from the inclusion of 2 rare ones at $P>50 $ days, $P=99$ and $P=134$ days, which were too bright for us to observe with HST.  At long periods Cepheids are too rare to well populate 
the instability strip and may even depart somewhat in slope.  We also note that the two G.E. compared, M101 and M31 lie at opposite extremes among the sample of 20 plus hosts.  A fair test of the uniformity of slopes should be made in a narrower period range where all hosts are well-populated.}
\ \par
\end{adjustwidth}

\subsection{Comparison of the R11 and R16 photometry}
\label{subsec:photometry}

R11 and R16 contain 9 galaxies in common. As a consistency check, it is interesting to compare the R11 and R16 photometry on an object-by-object basis. It is not possible to do this for all Cepheids within a particular galaxy,
because the R16 data have been pre-clipped: data for Cepheids rejected as outliers are not presented. Also, there are 
Cepheids listed in R16 that are not listed in R11. The overlap between the two samples
is high enough, however,  that one can draw reliable conclusions. The details of this comparison are given in Appendix \ref{sec:appendix}. I summarize the results in this subsection.

\begin{figure}
\centering
\includegraphics[width=74mm,angle=0]{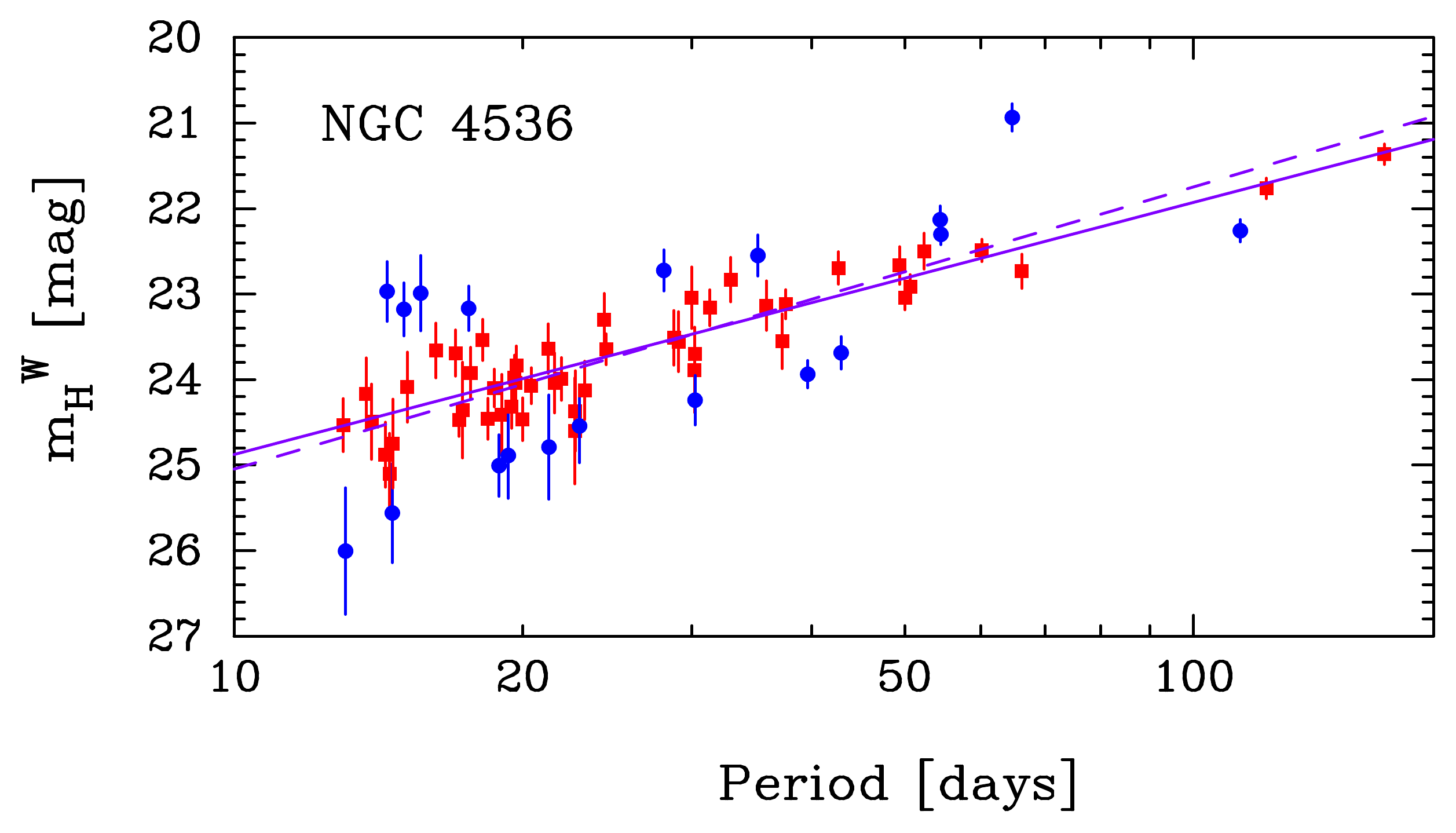}  \includegraphics[width=74mm,angle=0]{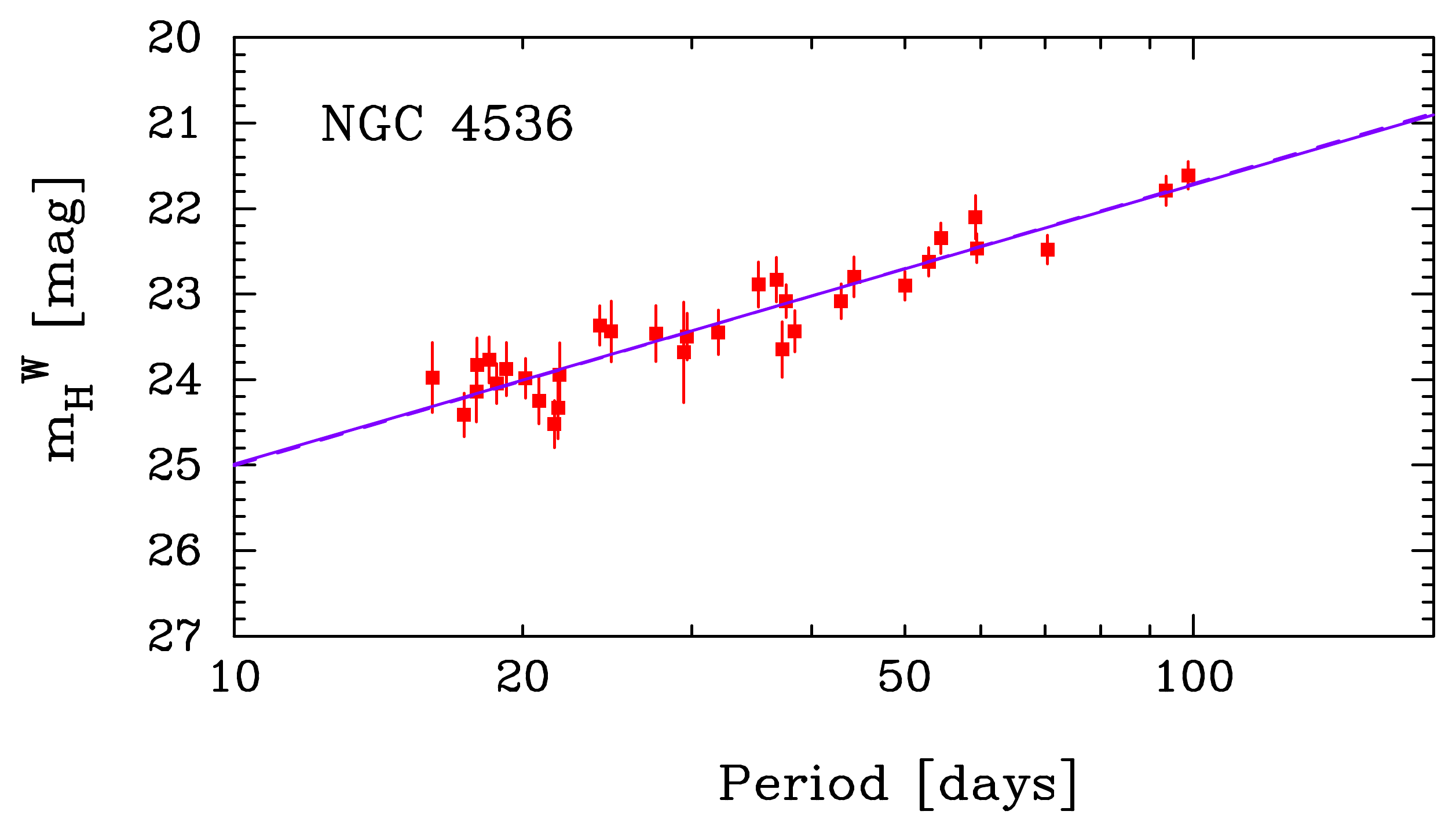} 
\caption {Left plot shows the PL relation for R11 Cepheids in the SN host galaxy NGC 4536. Blue points show Cepheids rejected by R11 (IFLAG =1);
red points show Cepheids accepted by R11 (IFLAG = 0). The solid line shows the best fit linear
relation fitted to the red points. The dashed line shows the best fit with the slope constrained to $b=-3.3$.
Right plot shows the PL relation for R16 Cepheids. The solid line shows the best fit linear
relation and the dashed line shows the best fit with the slope constrained to $b=-3.3$.}
\label{fig:R4536a}

\end{figure}

\begin{table}[h]
\begin{center}

\begin{tabular}{lllllll} \hline
             &          \multicolumn{3}{c}{R11} & \multicolumn{3}{c}{R16} \\
   galaxy    & \qquad $b$ & \qquad  $a$ & \qquad  $a_{-3.3}$ & \qquad $b$ & \qquad $a$ & \qquad $a_{-3.3}$ \\ \hline
   N4536  &    $-2.95 \pm 0.18$ &  $24.88 \pm 0.10$  & $25.05 \pm 0.05$ &   $-3.27 \pm 0.11$ & $24.99 \pm 0.11$ & $25.01 \pm 0.04$ \\ 
   N4639  &    $-2.68 \pm 0.50$ &  $25.41 \pm 0.30$  & $25.78 \pm 0.08$ &   $-2.33 \pm 0.47$ & $25.03 \pm 0.29$ & $25.61 \pm 0.06$ \\ 
   N3370  &    $-3.17 \pm 0.25$ &  $26.18 \pm 0.17$  & $26.28 \pm 0.04$ &   $-3.34 \pm 0.21$ & $26.21 \pm 0.14$ & $26.18 \pm 0.04$ \\ 
   N3982  &    $-3.77 \pm 0.56$ &  $26.18 \pm 0.32$  & $25.91 \pm 0.08$ &   $-2.54 \pm 0.21$ & $25.40 \pm 0.29$ & $25.85 \pm 0.06$ \\ 
   N3021  &    $-2.86 \pm 0.49$ &  $26.08 \pm 0.28$  & $26.31 \pm 0.10$ &   $-2.29 \pm 0.60$ & $26.04 \pm 0.34$ & $26.58 \pm 0.09$ \\ 
   N1309  &    $-2.35 \pm 0.60$ &  $26.08 \pm 0.45$  & $26.78 \pm 0.07$ &   $-3.09 \pm 0.38$ & $26.47 \pm 0.29$ & $26.63 \pm 0.04$ \\ 
   N5584  &    $-2.87 \pm 0.23$ &  $25.57 \pm 0.16$  & $25.85 \pm 0.04$ &   $-3.07 \pm 0.18$ & $25.72 \pm 0.12$ & $25.88 \pm 0.03$ \\ 
   N4038  &    $-2.84 \pm 0.29$ &  $25.45 \pm 0.26$  & $25.84 \pm 0.08$ &   $-3.81 \pm 0.75$ & $25.79 \pm 0.63$ & $25.37 \pm 0.11$ \\ 
   N4258  &    $-3.22 \pm 0.15$ &  $23.39 \pm 0.06$  & $23.43 \pm 0.04$ &   $-3.15 \pm 0.10$ & $23.35 \pm 0.04$ & $23.40 \pm 0.03$ 

\cr \hline
\end{tabular}

\caption{Fits to the PL relation for data given in R11 and R16. }
\label{table:PL}
\end{center}

\end{table}

To give an idea of the differences between R11 and R16, Fig. \ref{fig:R4536a}\footnote{
This figure is reproduced from Appendix \ref{sec:appendix}, which includes 
equivalent plots for all 9 galaxies in common between R11 and R16.}
shows the PL relations for NGC 4536. 
The solid lines show fits to Eq. (\ref{equ:data1}) (fitted only to the IFLAG = 0 Cepheids for the  R11 data).
The dashed lines show fits with slope constrained to $b=-3.3$;
the intercept of these fits is denoted as  $a_{-3.3}$. The parameters for these fits are listed in Table 1.
The error weighted averages of the slopes are:
\begin{subequations}
\begin{eqnarray}
\langle b \rangle & = & -2.95 \pm 0.10, \quad {\rm R11 \ excluding \ NGC \ 4258}, \\
\langle b \rangle & = & -3.04 \pm 0.08, \quad {\rm R11 \ including \ NGC \ 4258}. \\
\langle b \rangle & = & -3.09 \pm 0.09, \quad {\rm R16 \ excluding \ NGC \ 4258}, \\
\langle b \rangle & = & -3.11 \pm 0.07, \quad {\rm R16 \ including \ NGC \ 4258}.
\end{eqnarray}
\end{subequations}
consistent with the shallow slopes determined from global fits.

\begin{wrapfigure}[]{l}{3.0in}
\vspace{-0.00in}
\includegraphics[width=0.4\textwidth, angle=0]{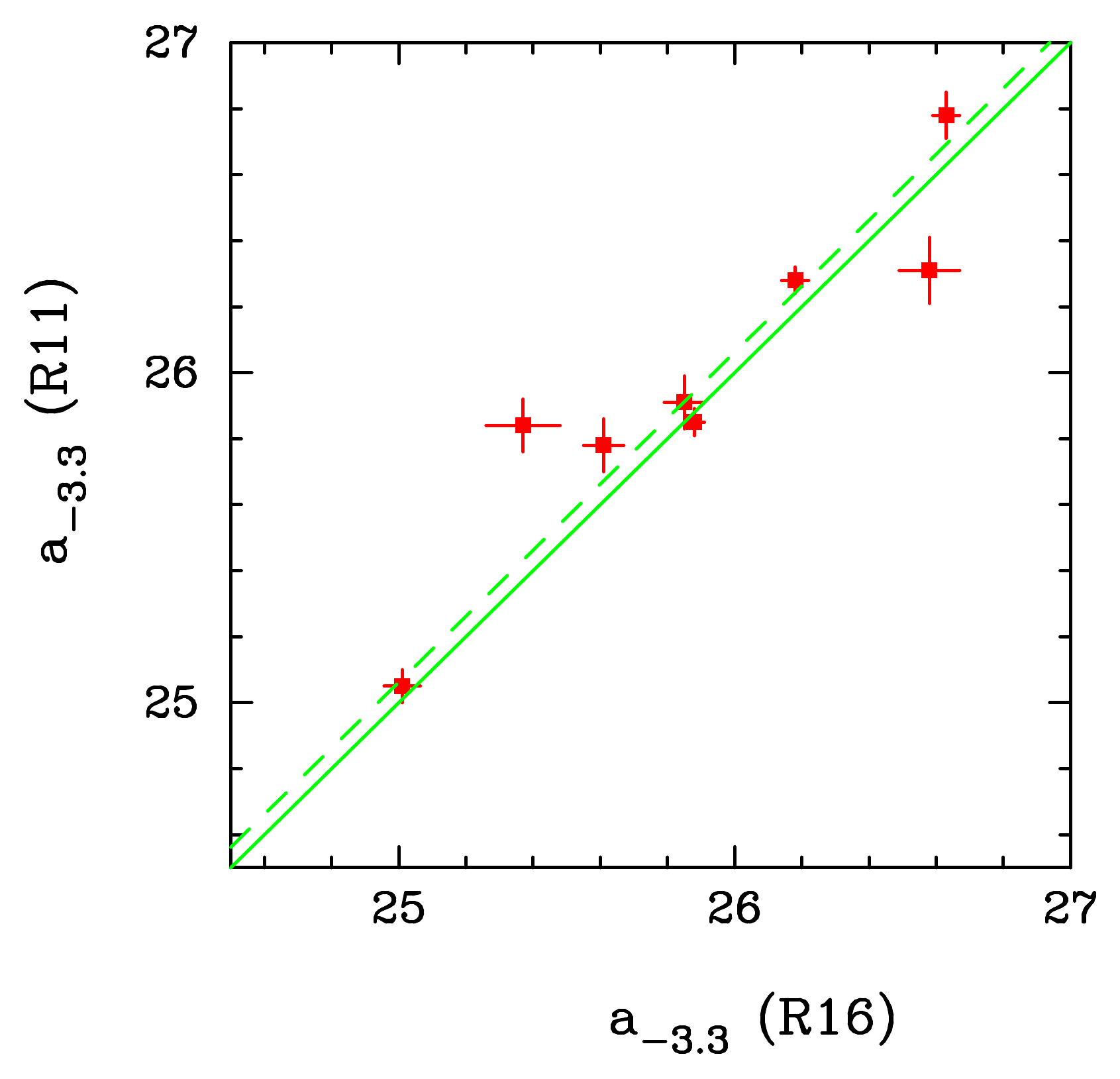}
\caption
{The intercept $(a_{-3.3})_{R11}$ plotted against $(a_{-3.3})_{R16}$.  The dashed line shows the best fit relation with a slope of unity. Solid
line shows $(a_{-3.3})_{R11} = (a_{-3.3})_{R16}$. }
\label{fig:intercept}
\vspace{0.09in}
\end{wrapfigure}
The uncertainties in the intercepts are very large if the slopes are allowed to vary and so we focus on the constrained intercepts $a_{-3.3}$. These are plotted in Fig. \ref{fig:intercept} for the 8 SN host galaxies. The dashed line shows the `best fit'
relation with  a slope of unity and an offset,
\begin{subequations}
\begin{equation}
(a_{-3.3})_{R16} =  (a_{-3.3})_{R11} + \Delta a, \label{equ:data3a}
\end{equation}
where
\begin{equation}
\Delta a = -0.062 \pm 0.027.  \label{equ:data5}
\end{equation}
\end{subequations}
The fit  Eq. (\ref{equ:data5}) assumes that the errors on $(a_{-3.3})_{R11}$ and $(a_{-3.3})_{R16}$ 
are independent, which is clearly not  true since the imaging data is largely common to both samples. In reality the
offset is much more significant than the $2.3 \sigma$ suggested by Eq. (\ref{equ:data5}).
It is also clear from Fig. \ref{fig:intercept} that the errors are underestimated\footnote{Assuming independent errors, the
dashed line gives $\chi^2= 21.27$ for 8 degrees of freedom, which is high by $\sim 3.3\sigma$.}. The offset of Eq. (\ref{equ:data5}) translates to a systematic shift in $H_0$ of about $2 \Hunit$, in the sense that the R16 data gives a higher $H_0$, for all
distance anchors\footnote{This can easily be verified by repeating the $H_0$ analysis for the 9 galaxies in R11 using the R16 photometry.}. Again, this systematic shift is larger than the error given in Eq. (\ref{equ:H0_1}).

The origin of this offset is surprising and is discussed in detail in Appendix \ref{sec:appendix}. The object-by-object
comparison of the photometry between R11 and R16 yields mean offsets (after removal of a few outliers) of 
\begin{subequations}
\begin{eqnarray}
(m_H^W)_{\rm R16} &= & (m_H^W)_{\rm R11} + \langle \Delta m_H^W \rangle, \qquad  \langle \Delta m_H^W \rangle = -0.051 \pm 0.025, \label{equ:data6a}\\
(V-I)_{\rm R16} &= & (V-I)_{\rm R11} + \langle \Delta C \rangle, \qquad  \langle \Delta C \rangle = 0.14 \pm 0.03. \label{equ:data6b}
\end{eqnarray} 
\end{subequations}
The offset in Eq. (\ref{equ:data5}) comes almost entirely from the difference in colours. As discussed in Appendix \ref{sec:appendix}, the colour offset is very significant: the errors in photometric scene reconstruction are correlated between different
passbands (and, in any case, are smaller in $V$ and $I$ than in $H$ band),  so the errors in $V-I$ colours are much smaller than the errors for individual passbands. This large systematic difference in colours  is not discussed by R16.

\begin{adjustwidth}{0.25in}{0.25in}
\ \par
\noindent{\underline{\bf Response from SH0ES Team:} There have been many updates to the official STScI pipeline between R11 in 2011 (produced shortly after WFC3 was installed) and R16 in 2016 (when WFC3 was mature) including changes in zeropoints, geometric distortion files, flat fields, CTE correction procedures, dark frames, hot pixel maps, bias frames, count linearity calibration, and count-rate non-linearity corrections which change photometry.  We believe the sum of these improved accuracy and the comparison to the early knowledge of WFC3 (or earlier methods of SH0ES) is not informative on present accuracy (or else we are all doomed by the past!).  We note with levity better communicated as a verbal touche that the Planck calibration of the overall power spectrum, indicated by the parameter $A_s e^{-2 \tau}$ changed by 3.5 $\sigma$ between the 2013 and 2015 Planck release as documented in Table 1 of the 2015 paper.  We celebrate the improved accuracy of Planck and promise we do not see this as any indication that improved accuracy cannot be established (stated with tongue in cheek).  We also note an important improvement to the SH0ES data since 2016 (e.g., Riess et al. 2018,2019) is to have now measured Cepheids in the LMC and the Milky Way on the same WFC3 system as extragalactic Cepheids which makes $H_0$ insensitive to future changes in the calibration of WFC3 since $H_0$ is only sensitive to differences in Cepheid measurements.}
\ \par
\end{adjustwidth}
 
\section{Tension between the distance anchors}
\label{sec:anchors}

R11 and R16 use the LMC, NGC 4258  and Milky Way (MW) Cepheid parallaxes as the primary  distance anchors.
In  Sect. \ref{subsec:LMCanchor}, we  discuss the LMC and NGC 4258 anchors and show that their geometrical distances are in 
tension with the R16 photometry. The MW distance anchor  is discussed in Sect. \ref{subsec:MWanchor}. 

\subsection{The LMC and NGC 4258 anchors}
\label{subsec:LMCanchor}

   Recently, there have been substantial improvements in the accuracies of the distance anchors.
Using 20 detached eclipsing binary (DEB) systems, \cite{Pietrzynski:2019} determine a distance modulus for 
the Large Magellanic Cloud (LMC) of 
\begin{equation}
     \mu_{\rm LMC} = 18.477 \pm 0.026 \ {\rm mag}.   \label{equ:mu1}
\end{equation}
A  reanalysis of  VLBI observations \cite{Humphreys:2013} of water masers in NGC 4258 by \cite{Reid:2019}
gives a distance modulus of
\begin{equation}
    \mu_{\rm N4258} = 29.397 \pm 0.033 \ {\rm mag}, \label{equ:mu2}
\end{equation}
substantially reducing the systematic error in the geometric distance to this galaxy compared to the distance
estimate used in R16.
In addition, R19 have published HST photometry for
LMC Cepheids using the same photometric system as that used for the
cosmological analysis of $H_0$ in R16. With
these observations, calibration errors associated with ground based photometry of LMC Cepheids
are effectively eliminated as a significant source of systematic error.

\begin{figure}
	\centering
	\includegraphics[width=150mm, angle=0]{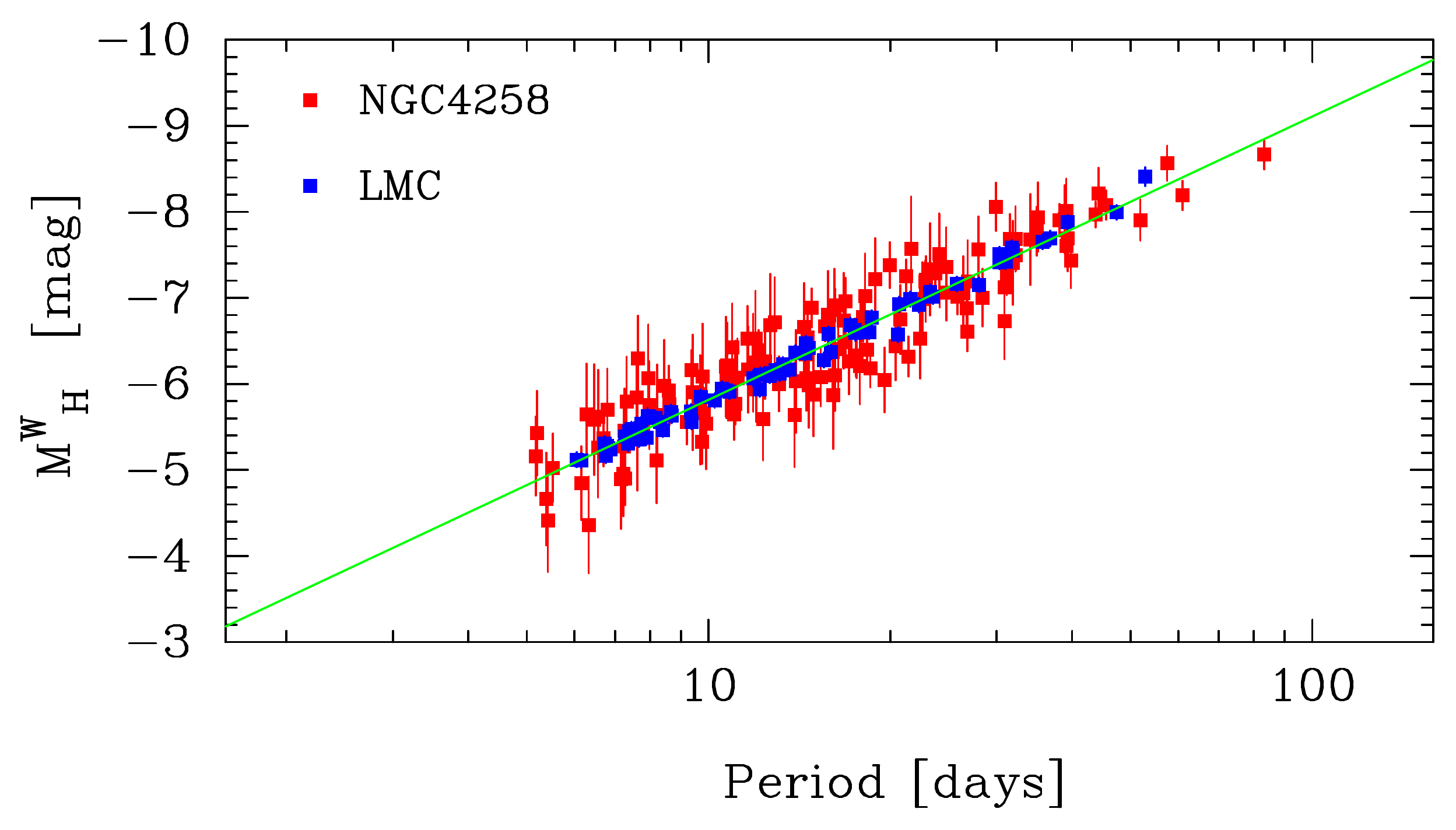}	
	\caption{The joint PL relation for the LMC and NGC 4258 Cepheids.
 The line shows the best fit from
Eqs. (\ref{equ:M4a}) -- (\ref{equ:M4c}).}

	\label{fig:PLanchor}

\end{figure}

I use the data for the 70 LMC Celpheids listed in R19 
and the data for the 139 Cepheids in NGC 4258 from R16 to perform a joint $\chi^2$ fit:
\begin{subequations}
\begin{eqnarray}
(m^W_H)_{\rm LMC} & = &   \mu_{\rm LMC} + c + b \ {\rm log}_{10} \left ({ P \over 10 \ {\rm days} } \right),  \label{equ:M3a} \\
(m^W_H)_{\rm N4258} & = &   \mu^P_{\rm N4258} + c + b \  {\rm log}_{10} \left ({ P \over 10 \ {\rm days} } \right),  \label{equ:M3b}
\end{eqnarray}
\end{subequations}
where  $\mu_{\rm LMC}$ is fixed at the value of Eq. (\ref{equ:mu1}) and $\mu^P_{\rm N4258}$, $c$, $b$ 
are parameters to be determined from the fit\footnote{For the LMC Cepheids an  intrinsic scatter
of $0.08$ mag is added to the error estimates given in R19 which is necessary to produce a reduced $\chi^2$ of unity.}. The results are as follows:
\begin{subequations}
\begin{eqnarray}
\mu^P_{\rm N4258} & = & 29.220 \pm 0.029, \label{equ:M4a} \\
c    & =  & -5.816 \pm 0.011, \label{equ:M4b}\\ 
b    & = & -3.29 \pm 0.04.  \label{equ:M4c}
\end{eqnarray}
\end{subequations}
The difference between (\ref{equ:M4a}) and  (\ref{equ:mu2}) is
\begin{equation}
 \Delta \mu_{\rm N4258} = (\mu_{\rm N4258} - \mu^P_{\rm N4258}) = 0.177 \pm 0.051, \label{equ:M4}
\end{equation}
and  differs from zero by nearly $3.5 \sigma$. In other
words, the DEB LMC distance together with SH0ES Cepheids is placing NGC 4258 at a distance of
$6.98 \ {\rm Mpc}$ if metallicity effects are ignored, whereas the maser distance is $7.58 \ {\rm Mpc}$.
The best fit PL relation to the combined LMC and NGC 4258 data is shown in Fig. \ref{fig:PLanchor}.

There are a number of possible explanations of this result:

\smallskip

\noindent
(i) There may be unidentified systematic errors in the geometric distance estimates of Eqs.  (\ref{equ:mu1})
and (\ref{equ:mu2}).

\smallskip

\noindent
(ii) Part of the discrepancy may be attributable to a metallicity dependence
of the PL relation.

\smallskip

\noindent
(iii) There may be a photometric offset between the R16 and LMC Cepheid magnitudes.

\smallskip

Since this discrepancy involves two of the three primary distance anchors used by the
SH0ES team, it needs to be considered extremely seriously.

Point (i) can be tested by using independent distance indicators. For example,  a recent
paper \cite{Huang:2018} used the near-infrared PL relation of Mira variables in the LMC and NGC 4258 to 
determine a relative distance modulus of 
\begin{equation}
     \mu_{\rm N4258} - \mu_{\rm LMC} = 10.95 \pm 0.06, \qquad {\rm Miras}, \qquad \qquad \label{equ:mu3}
\end{equation}
(where the error is dominated by systematic errors). This estimate is in good agreement with
the geometric estimates of Eqs. \ref{equ:mu1} and \ref{equ:mu2}:
\begin{equation}
       \mu_{\rm N4258} - \mu_{\rm LMC} = 10.92 \pm 0.04, \qquad {\rm DEB+maser}. \label{equ:mu4}
\end{equation}

\subsection{Milky Way parallaxes and NGC 4258}
\label{subsec:MWanchor}

\begin{figure}
	\centering
	\includegraphics[width=150mm, angle=0]{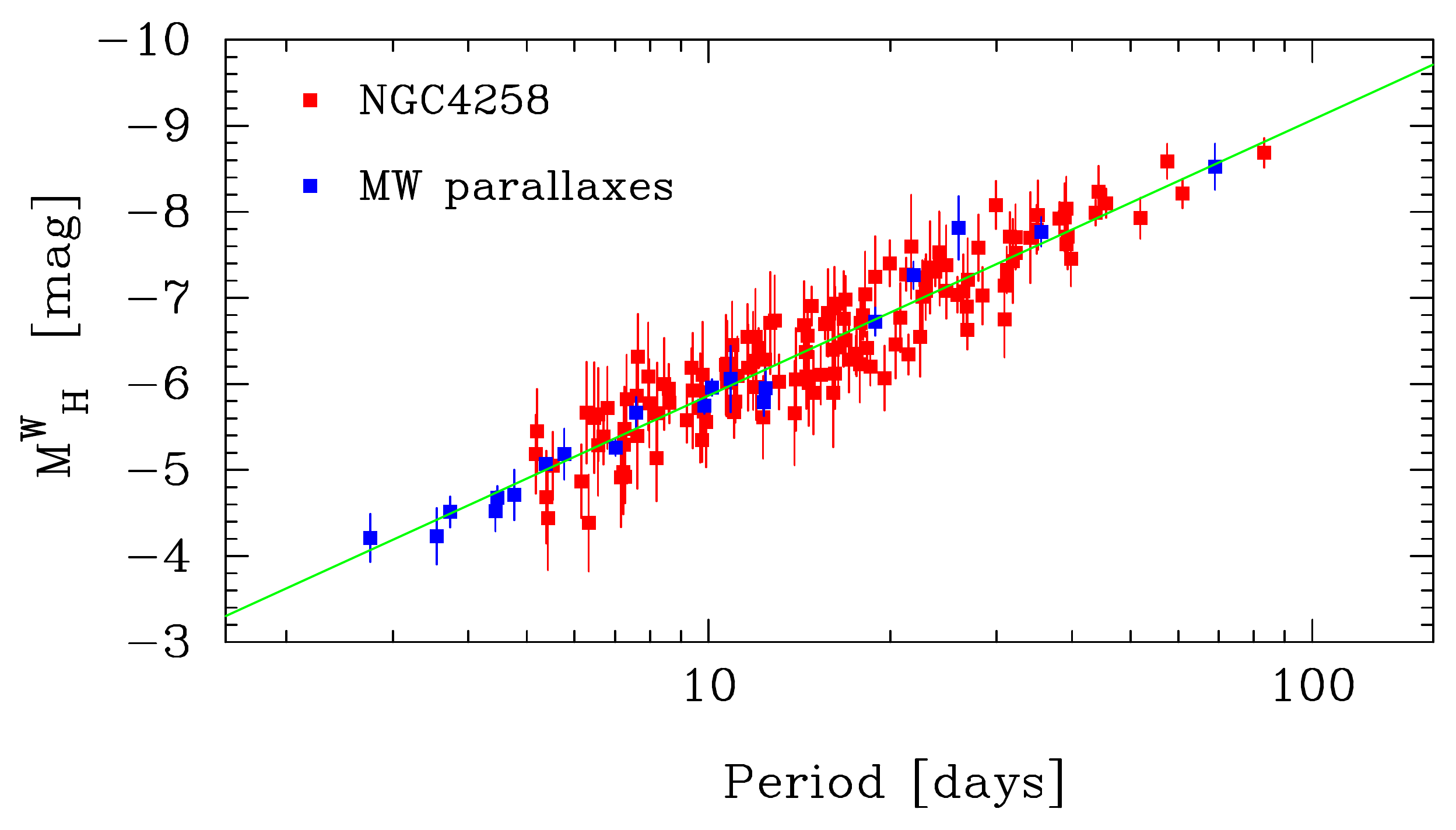}	
	\caption{The PL relations for the MW Cepheids with parallaxes and NGC 4258 Cepheids.
The solid line shows the best fit to the joint data.}

	\label{fig:MWPL}

\end{figure}

Parallaxes of Milky Way (MW) Cepheids have been used by the SH0ES team
as a third distance anchor.  Following the publication of R16,
\cite{Riess:2018} have measured parallaxes for an additional 7 MW
Cepheids with periods greater than 10 days, supplementing the parallax
measurements of 13 MW Cepheids from \cite{Benedict:2007,
  vanLeeuwen:2007}. Figure \ref{fig:MWPL} shows the equivalent of
Fig. \ref{fig:PLanchor} but now using the 20 MW Cepheids as an anchor
in place of the LMC\footnote{Note that I repeated the GAIA DR2 analysis
  of \cite{Riess:2018b} finding identical results for the DR2 parallax
  zero point offset. I therefore do not consider GAIA parallaxes any
  further.}. The best fit gives
\begin{subequations}
\begin{eqnarray}
\mu^P_{\rm N4258} & = & 29.242 \pm 0.052, \label{equ:M5a} \\
c    & =  & -5.865 \pm 0.033, \label{equ:M5b}\\ 
b    & = & -3.21 \pm 0.08.  \label{equ:M5c}
\end{eqnarray}
\end{subequations}
As with the LMC, this comparison disfavours the maser distance, though because the error bar is larger this is significant 
at only $2.2\sigma$.
However, since the metallicities of the MW and NGC 4258 Cepheids are very similar, this comparison
suggests that  metallicity differences of the PL relation cannot explain the shift reported in Eq. (\ref{equ:M4}).

\begin{adjustwidth}{0.25in}{0.25in}
\ \par
\noindent{\underline{\bf Response from SH0ES team:} The full analysis of the SH0ES Cepheid data includes an empirically determined correction
for Cepheid metallicity as shown in equation 4.1 below.  The difference stated here of 3.5$\sigma$ between the LMC 
and NGC 4258 appears so only when ignoring the metallicity term.  Including it brings the LMC and NGC 4258 closer
together by $\sim$ 0.06 mag and reduces the significance of the difference in these anchors to slightly over 2$\sigma$. 
 Because the metallicity term is measured internal to the data and included as part of the $H_0$ analysis, we think the 2$\sigma$ number is the fair statement of the significance.  
 We note there are now 5 other geometric anchors from different methods of Milky Way parallax (from HST FGS, HST spatial scan, Gaia DR2 Cepheids, Gaia DR2 Cepheid companions, and Gaia DR2 cluster hosts, which yield $H_0$=76.2 (2.2\%), 75.7 (3.3\%), 73.7 (3.3\%), 72.7 (3.8\%) and 73.3 (3.5\%) (see Breuval et al. 2020) which makes it appear that the difference between 72.0 (NGC 4258, 1.5\%) and 74.2 (LMC,1.3\%) is not significant.  However, this point which G.E. shared with us motivated us to acquire new Cepheid observations of the outer regions of NGC 4258 with HST to measure the anchor distance at the same (i.e., low) metallicity so we can revisit this issue with an improved characterization of metallicity.  We appreciate G.E.'s suggestions and we expect an update on this in the next SH0ES paper.}  
\ \par
\end{adjustwidth}

\section{The SH0ES degeneracy}
\label{sec:degeneracy}

One interpretation of the results of the previous section is that
\cite{Reid:2019} have overestimated the maser distance to NGC 4258
and/or underestimated the error. The maser analysis is considerably
more complicated than the DEB distance estimates, and so
it is extremely important that the maser analysis is revisited and, if
possible, checked against independent techniques (such as TRGB
\cite{Mager:2008, Jang:2017} and Mira \cite{Huang:2018} distance
measurements). In this Section, I want to discuss another possibility, which I will call
the `SH0ES degeneracy'.

\subsection{Global fits and values of $H_0$}
\label{subsec:global_fits}

I will begin by analysing global fits to determine $H_0$ in (almost) the same way as in the SH0ES papers.
A metallicity dependence of the PL relation is included by adding an extra term to Eq. (\ref{equ:dataP}) so that the
magnitude of Cepheid $j$ in host galaxy $i$ is
\begin{equation}
(m_H^W)_{i, j}  = a_i + b \log_{10} \left [ { P_{i,j} \over 10 {\rm days}} \right ] + Z_w \Delta \log_{10} (O/H)_{i, j},  
\end{equation}
where $Z = 12 + \log_{10} (O/H)_{i,j}$ is the metallicity listed in Table 4 of R16, $\Delta \log_{10} (O/H)_{i,j}$ is the difference
relative to Solar metallicity, for which I adopt $Z_\odot = 8.9$. For the LMC I adopt a uniform metallicity of $Z_{\rm LMC} = 8.65$.

First I list the results\footnote{The fits discussed here were computed using the {\tt MULTINEST} sampling algorithm \cite{Feroz:2009, Feroz:2011}.  }
 using NGC 4258 as a distance anchor, adopting the distance modulus of Eq. (\ref{equ:mu2}). 
In these fits, I use all Cepheids with periods in the
range $10-200 \ {\rm days}$. In the first solution,  labelled  `no priors', no slope or metallicity priors are imposed. 
In this solution,  the best fit slope is
much shallower than the slope of the M31 PL relation (as discussed in Sect. \ref{subsec:photometry}). For the solution
labelled `slope prior', I impose a tight prior on the slope of $b = -3.300 \pm 0.002$ to counteract the HST photometry
and force the slope to match the M31 and LMC slopes. 
\begin{subequations}
\begin{eqnarray}
{\rm NGC} \ {\rm 4258}\ {\rm  anchor},   \ {\rm no} \ {\rm priors:}  & &   H_0 = 72.0 \pm 1.9 \ {\rm km} {\rm s}^{-1}  {\rm Mpc}^{-1}, \qquad \qquad  \label{equ:deg2a}\\
                          & &   b = -3.06 \pm 0.05,  \nonumber \\
                          & &   Z_w = -0.17 \pm 0.08,    \nonumber \\
{\rm NGC} \ {\rm 4258}\ {\rm  anchor},   \ {\rm slope}  \ {\rm prior:}   & &   H_0 = 70.3 \pm 1.8 \ {\rm km} {\rm s}^{-1}  {\rm Mpc}^{-1},  \label{equ:deg2b}\\
                          & &   b = -3.30 \pm 0.002, \nonumber \\
                          & &   Z_w = -0.15 \pm 0.08.    \nonumber 
\end{eqnarray}
\end{subequations}
These solutions are not strongly discrepant with the \Planck\ value of $H_0$; the `no prior' solution for $H_0$ is high by $2.3\sigma$ and the `slope prior' solution is high by $1.5\sigma$ (see also E14).

Using the LMC as a distance anchor, the LMC photometry from R19, and adopting the distance modulus of Eq. (\ref{equ:mu1}) I find
\begin{subequations}
\begin{eqnarray}
{\rm LMC} \ {\rm  anchor},    \ {\rm no} \ {\rm priors:}  & &   H_0 = 76.5 \pm 1.7 \ {\rm km} {\rm s}^{-1}  {\rm Mpc}^{-1}, \qquad \qquad \label{equ:deg3a}\\
                          & &   b = -3.17 \pm 0.04,  \nonumber \\
                          & &   Z_w = -0.18 \pm 0.08,    \nonumber \\
{\rm LMC} \ {\rm  anchor},    \ {\rm slope}  \ {\rm prior:}   & &   H_0 = 74.8 \pm 1.6 \ {\rm km} {\rm s}^{-1}  {\rm Mpc}^{-1},  \label{equ:deg3b}\\
                          & &   b = -3.30 \pm 0.002, \nonumber \\
                          & &   Z_w = -0.18 \pm 0.08,    \nonumber 
\end{eqnarray}
\end{subequations}
and using both anchors:
\begin{subequations}
\begin{eqnarray}
{\rm NGC}\ 4258 + {\rm LMC} \ {\rm  anchors},    \ {\rm no} \ {\rm priors:}  & &   H_0 = 74.6 \pm 1.5 \ {\rm km} {\rm s}^{-1}  {\rm Mpc}^{-1}, \qquad \qquad \label{equ:deg4a}\\
                          & &   b = -3.18 \pm 0.04,  \nonumber \\
                          & &   Z_w = -0.22 \pm 0.07,    \nonumber \\
{\rm  NGC} \ 4258 + {\rm LMC} \ {\rm  anchors},    \ {\rm slope}  \ {\rm prior:}   & &   H_0 = 73.5 \pm 1.3 \ {\rm km} {\rm s}^{-1}  {\rm Mpc}^{-1},  \label{equ:deg4b}\\
                          & &   b = -3.30 \pm 0.002, \nonumber \\
                          & &   Z_w = -0.21 \pm 0.07.    \nonumber 
\end{eqnarray}
\end{subequations}

These fits differ slightly from those given in R16 and R19 because the
SH0ES team include M31 Cepheids in the fits. The only effect of adding
the M31 data is to pull the best fit slope closer to $b=-3.3$; as a
consequence their best fits are intermediate between the `no prior'
and `slope prior' results
\footnote{There are other minor differences, for example in R19, the
  SH0ES team include the ground based LMC data to better constrain the
  slope of the PL relation; they also include the NGC 4258 photometry
  when using the LMC or MW parallaxes as distance anchors,  which only
  affects the slope of the PL relation. These differences are
  unimportant for our purposes.}.
The joint solution of Eq. (\ref{equ:deg4a}) is actually quite close to the R19 solution of Eq. ({\ref{equ:H0_1}) and is
higher than the \Planck\ value by $4.5 \sigma$.

\begin{figure}
	\centering
 	\includegraphics[width=75mm, angle=0]{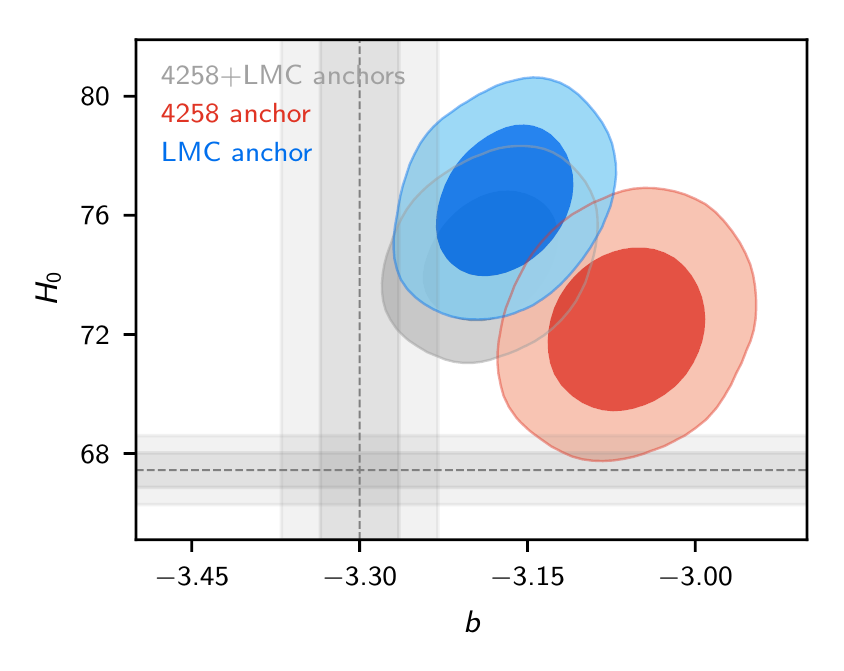}		\includegraphics[width=75mm, angle=0]{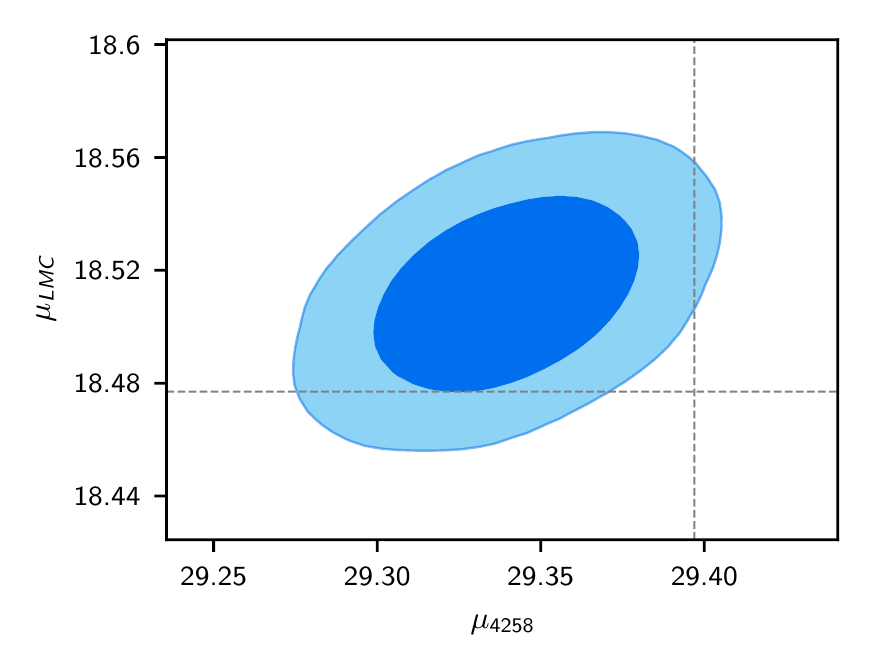}
	\caption{The panel to the left shows 68 and 95\% contours in the $H_0-b$ plane for the solutions of Eqs. (\ref{equ:deg2a}) (red contours)
 and (\ref{equ:deg3a}) (blue contours). The joint solution for the NGC 4258 and LMC anchors ( Eq. (\ref{equ:deg4a})) is
shown by the grey contours. The horizontal and vertical bands shows $1\sigma$ and $2\sigma$ ranges of the Planck value of $H_0$ (Eq. 
(\ref{equ:H0_2}) ) and M31 PL slope (Eq. (\ref{equ:data4})). The plot to the right shows the 68\% and 95\% constraints on the NGC 4258 and LMC distance moduli from the joint fit. The geometric distance moduli of Eqs. (\ref{equ:mu1}) and
(\ref{equ:mu2}) are shown by the dotted lines.  }

	\label{fig:H0_contour}

\end{figure}

The `no prior' solutions are shown by the left hand panel of Fig. \ref{fig:H0_contour}.  One might think that the
blue and red contours are consistent with each other, but in fact, {\it all} of SN data and {\it almost all} of 
the Cepheid data are common to both analyses. The difference between these two  solutions reflects the tension between 
the LMC and NGC 4258 anchor distances discussed in Sect. \ref{subsec:LMCanchor}. The joint fit (grey contours) tries
to share this tension, but lies closer to  the LMC fit because the LMC anchor carries more weight than NGC 4258.
The joint fit then leads to a value of $H_0$ that is strongly in tension with \Planck. However, there is a statistical inconsistency in the joint fit. This is illustrated by the right hand plot in Fig. \ref{fig:H0_contour} which shows constraints on the
LMC and NGC 4258 distance moduli from the joint fit. These parameters are, of course, highly correlated, but one can see
that the geometrical best fit values (shown by the intersection of the dotted lines) sits well outside the 95\% contours. This is the statistically
more rigorous way of quantifying the discrepancy discussed in Sect. \ref{subsec:LMCanchor}, including metallicity effects.

\subsection{The SH0ES degeneracy and values of $H_0$}
\label{subsec:SH0ES degeneracy}

At this point, one might follow R16 and argue that the safest way to proceed is to average over distance anchors.
However, this is extremely dangerous if one or more of the distance anchors is affected by systematic errors,  or
if there are systematics in the photometry that affect some distance anchors but not others.

\begin{figure}
	\centering
 	\includegraphics[width=75mm, angle=0]{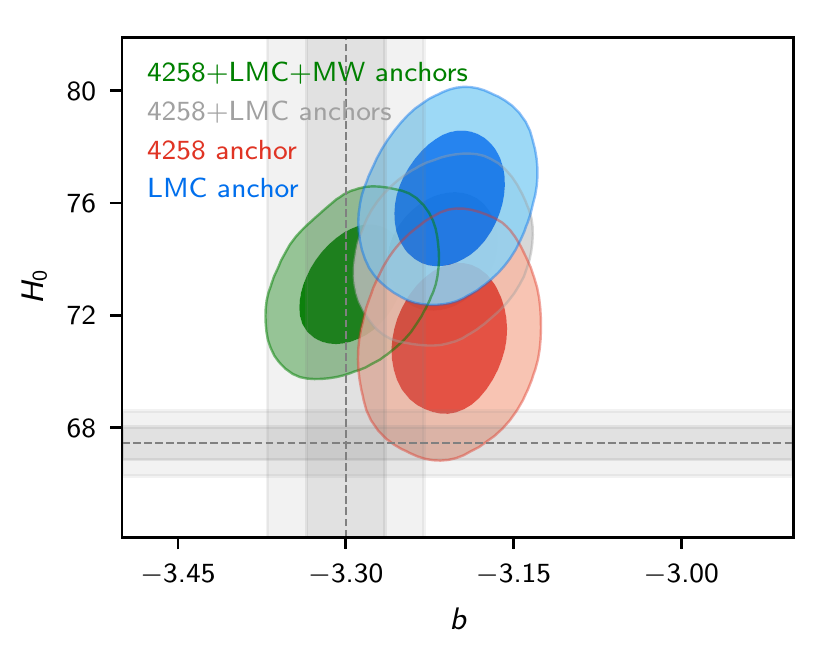}		
\includegraphics[width=75mm, angle=0]{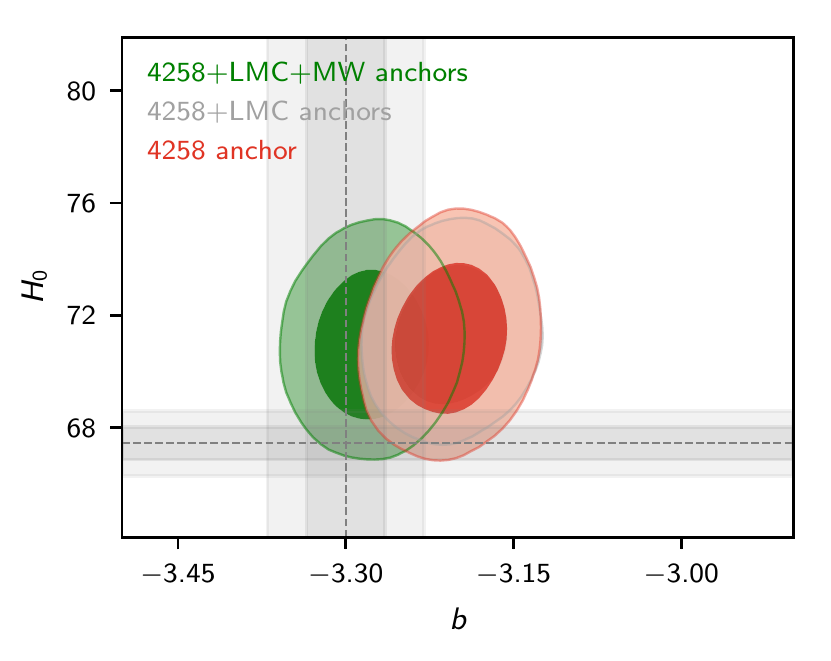}
	\caption{The plot to the left shows 68 and 95\% contours in the $H_0-b$ plane for different combinations of distance
anchors. In this analysis, I have added M31 Cepheids which shifts the contours towards $b=-3.3$ and slightly lower 
values of $H_0$.  The plot to the right shows the effect of subtracting a constant offset $\delta a$ from the distant Cepheid
magnitudes, illustrating the SH0ES degeneracy.}
	\label{fig:H0_contour1}

\end{figure}

The SH0ES degeneracy is an example of the latter type of systematic. Imagine that there is a constant
offset between the SH0ES  PL intercepts, and the intercepts  of the
MW and LMC  distance anchors. Write the true intercept of the PL relation  as 
\begin{equation}
   a^T =  a_{\rm SH0ES} - \delta a. \label{equ:deg5}
\end{equation}
 There are a
number of possible sources of error that might produce such a shift;
for example, systematic errors in crowding corrections (though see
\cite{Riess:2020}), 
asymmetries in the distribution of outliers, asymmetries in the
magnitude errors, scene reconstruction errors,  or selection biases such as the short period incompleteness
discussed in Sect. \ref{subsec:slopes}.  
The assumption here is that the shift
$\delta a$ is the same for all R16 galaxies, whereas for well resolved nearby bright Cepheids
such as those in the LMC and MW, there is no such shift. This model should not be taken literally because the data 
quality in the SH0ES sample varies (for example, although NGC 4258 is nearer than most of the SN host galaxies,
the Cepheids are more crowded in this galaxy).  We will approximate the model of Eq. (\ref{equ:deg5}) by subtracting a
constant offset from all SH0ES H-band magnitudes
\begin{equation}
   m_H =  m_{H,{\rm SH0ES}} - \delta a. \label{equ:deg6}
\end{equation}
Since $\delta a$ is assumed to be a constant, it will cancel if we use NGC 4258 as a distance anchor. However, a
constant offset will lead to
a bias in the value of $H_0$ if the LMC, MW or M31 are used as distance anchors. This is the SH0ES degeneracy\footnote{The
possible cancellation of systematic errors if NGC 4258 is used as a distance anchor was recognised in the early SH0ES papers. 
\cite{Riess:2009, Riess:2011}.}.

\begin{wrapfigure}[]{l}{3.0in}
\vspace{-0.14in}
\includegraphics[width=0.4\textwidth, angle=0]{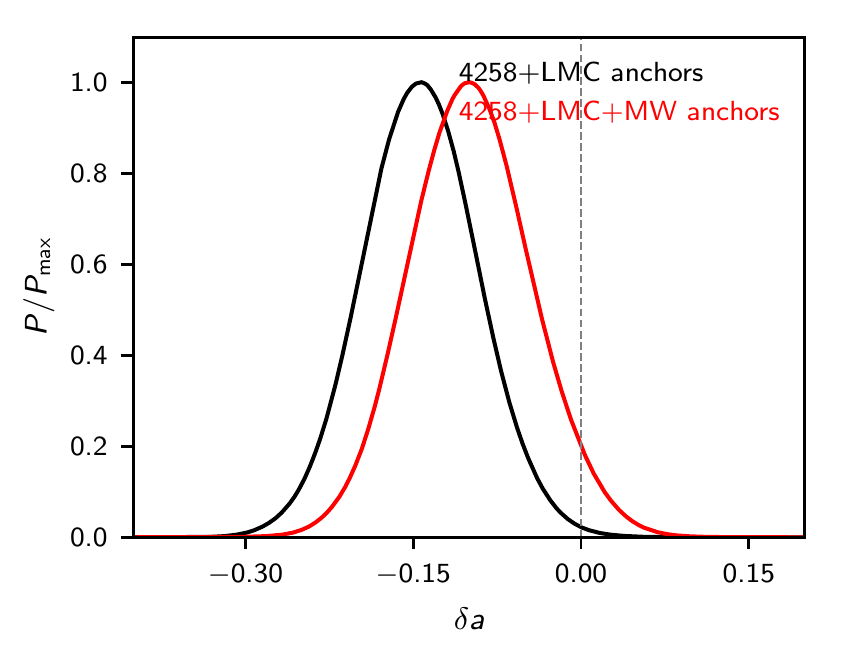}
\caption
{Posterior distributions for the offsets $\delta a$ for the 4258+LMC and 4258+LMC+MW solutions
show in the right hand panel of Fig. \ref{fig:H0_contour1}.}
\label{fig:offset}

\end{wrapfigure}
The effect of this degeneracy is illustrated in Fig. \ref{fig:H0_contour1}. The panel to the left shows the progressive combination of NGC 4258, LMC and MW anchors in the usual way. Note that for this comparison, 
 I have added Cepheids from M31, which pushes the contours closer to $b=-3.3$. The panel to the right shows what happens if we
add a constant offset $\delta a$ as a free parameter. Clearly, this has no impact on the results using NGC 4258 as an anchor. However, if
we combine the 4258 anchor with any combination of   LMC, MW anchors, all that happens is that $\delta a$ adjusts to bring the
combined constraints close to those from the NGC 4258 anchor.  All of these contours have substantial overlap with the 
\Planck\ value of $H_0$.

The posterior distributions of $\delta a$ for these solutions are shown in Fig. \ref{fig:offset}. The offsets are as follows:
\begin{subequations}
\begin{eqnarray}
{\rm NGC}\ 4258 + {\rm LMC} \ {\rm  anchors},  & &  \delta a = -0.14 \pm 0.05, \\
{\rm NGC}\ 4258 + {\rm LMC} + {\rm MW}\ {\rm  anchors},  & & \delta a = -0.10 \pm 0.05.
\end{eqnarray}
\end{subequations}
The first of these constraints is another way of expressing the nearly $3 \sigma$ tension between the LMC and NGC 4258 anchors. 
As expected from the discussion in Sect. \ref{sec:SH0ES_data}, we see that an offset of $\approx 0.1 \ {\rm mag}$ (in the sense that the R16 photometry is too bright)  can largely resolve the Hubble tension. {\it Increasing the precision of the MW and LMC distance anchors
(as has been done in R19 and R20)  does not strengthen the case for the Hubble tension, unless one can rule out 
the SH0ES degeneracy convincingly}. This can be done: (a) by independently determining the distance modulus of NGC 4258, and/or
(b) comparing Cepheid distance moduli for SN hosts with distance moduli determined from other techniques, to which we turn next.

\begin{adjustwidth}{0.25in}{0.25in}
\ \par
\noindent{\underline{\bf Response from SH0ES team: } We agree that use of NGC 4258 as the only anchor, excluding the LMC and the 5 sets of Milky Way parallaxes reduces the tension.  Further fixing the slope to be steeper than these fits provide by what is more than a few $\sigma$ as G.E. does in this example reduces the tension further.  We don't think this approach is reasonable (not to mention the danger of CMB-confirmation bias) but we have made all the photometry for these exercises available to the Community for their own analyses and the Community has reanalyzed the data, consistently concluding that the data are not easily moved to a place of diminished tension.  See Follin and Knox (2018), Cardona et al. (2017) or Burns et al. (2018) for examples.

On the second point G.E. hypothesizes a ``common mode'' error where measurements of nearby, bright Cepheids such as those in the LMC or MW are measured differently than other, extragalactic Cepheids in SN hosts or NGC 4258.  This is a concern that keeps us up at night!  The full test of this as G.E. notes is comparing $H_0$ using only the anchor NGC 4258 where we get $H_0=72.0$ and as discussed above only 2 $\sigma$ different than the LMC, hence no real evidence of an issue.  We expect this test to improve with the new NGC 4258 Cepheid data now in hand.  The most likely way such a common mode error would arise is by ``crowding''.   We just published a new paper, Riess et al. (2020) that used the amplitudes of Cepheid light curves to measure such a unrecognized, crowding-induced, common mode error and constrained it to 0.029 $\pm$ 0.037 mag, ruling out crowding as the source of such an error.  Count-rate non-linearity is another potential source of common-mode error but was recently calibrated to 0.3\% precision in $H_0$ making it an unsuitable source.  We welcome any specific hypotheses for another mechanism for such an error so we can test for it.  We also note that we neglect the possibility that Cepheids in nearby hosts (Milky Way, LMC, NGC 4258 and M31) are actually different than those in SN hosts as it would seem to violate the principle that we should not live in a special region of space where Cepheids are fainter.}
\ \par
\end{adjustwidth}

\section{Comparison of Cepheid and TRGB distance moduli}
\label{sec:TRGB}

As discussed in the Sect. \ref{sec:introduction}, recently \cite{Freedman:2019} (herafter F19) have determined a value for
$H_0$ using  TRGB as a standard candle. There are $10$ SN host galaxies in common between R16 and F19,  which are listed in
Table \ref{table:distance_moduli}. Five of these have TRGB distance measurements as part of the Carnegie-Chicago Hubble Program (CCHP); the remaining
five galaxies labelled `JL' (which are more distant) have TRGB distances from data analysed by \cite{Jang:2017a} but reanalysed by F19.
These distance moduli are listed in Table \ref{table:distance_moduli} together with the distance moduli for the solution
of Eq. (\ref{equ:deg2a}) using NGC 4258 as a distance anchor and the solution of  Eq. (\ref{equ:deg3a}) using the LMC as an anchor.

\begin{figure}
\centering
\includegraphics[width=120mm,angle=0]{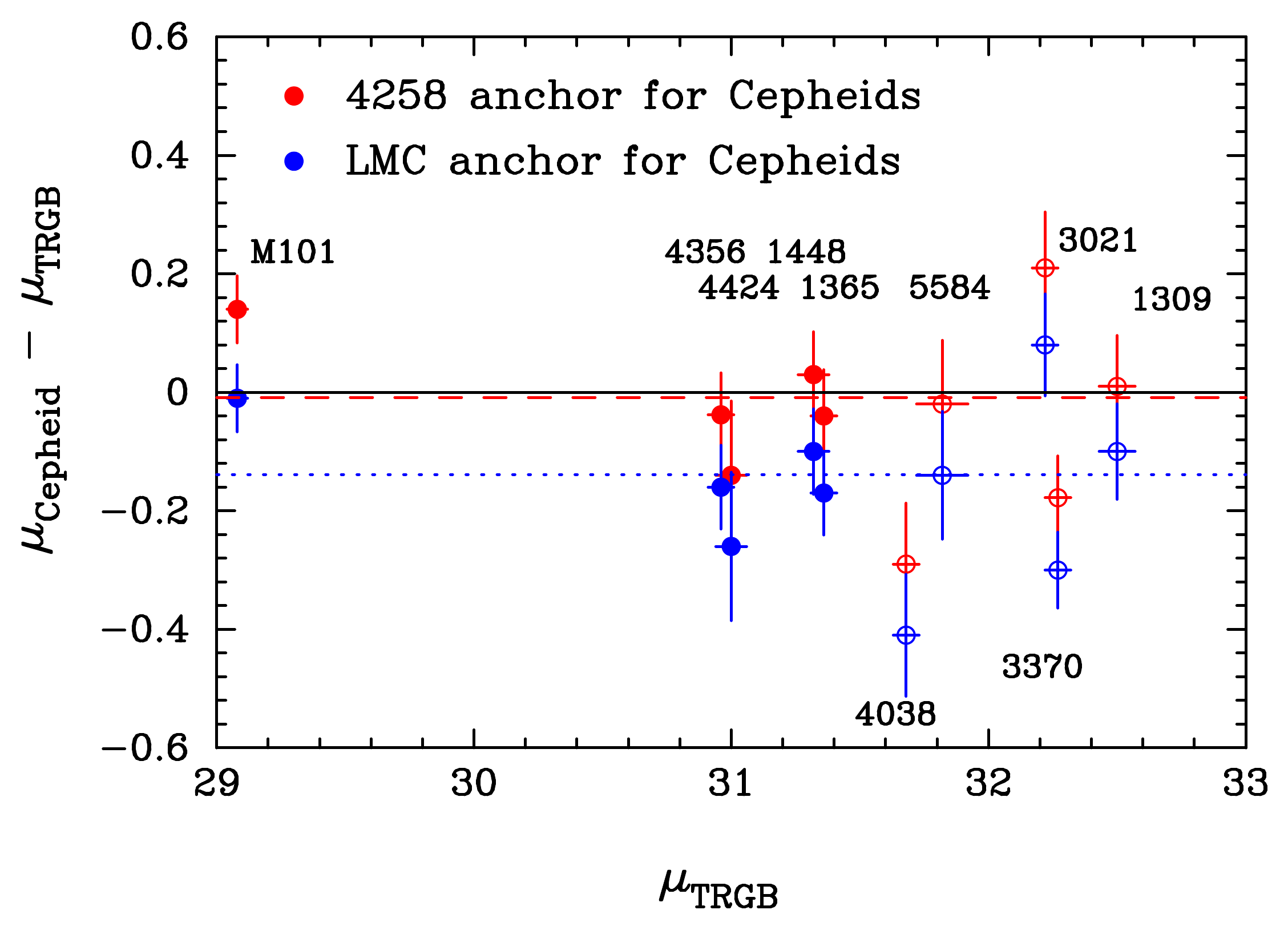} 
\caption {Differences in the TRGB and Cepheid distance moduli from Table \ref{table:distance_moduli} plotted
against $\mu_{\rm TRGB}$. Filled and open symbols are for the galaxies labelled `CCHP' and `JL' respectively in Table
\ref{table:distance_moduli}. The red and blue dotted lines shows the best-fit offsets to the red points and blue points
respectively.}

\label{fig:distance_modulus}

\end{figure}

\begin{table}[h]
\begin{center}

\begin{tabular}{lllll} \hline
             &            &  {\rm LMC} \ {\rm anchor} &   {\rm N4258} \ {\rm anchor} &  {\rm LMC} \ {\rm anchor} \\
   galaxy    & {\rm TRGB} &  $\mu_{\rm TRGB}$ & $\mu_{\rm Cepheid}$  & $\mu_{\rm Cepheid}$   \\ \hline
   N4536  &    CCHP & $30.96 \pm 0.05$ &  $30.92  \pm 0.05$  & $30.80 \pm 0.05$ \\
   N3370  &   JL &  $32.27 \pm 0.05$ &  $32.09 \pm 0.05$  & $31.97 \pm 0.04$ \\
   N3021  &   JL & $32.22 \pm 0.05$ &  $32.43  \pm 0.08$  & $32.30 \pm 0.07$ \\
   N1448  &  CCHP &   $31.32 \pm 0.06$ &  $31.35 \pm 0.04$  & $31.22 \pm 0.04$ \\
   N1309  &   JL &   $32.50 \pm 0.07$ &  $32.51 \pm 0.05$  & $32.40 \pm 0.05$ \\
   N5584  &   JL &  $31.82 \pm 0.10$ &  $31.80 \pm 0.04$  & $31.68 \pm 0.04$ \\
   N4038  &   JL &  $31.68 \pm 0.05$ &  $31.39 \pm 0.09$  & $31.27 \pm 0.09$ \\
   M101  &    CCHP &$29.08 \pm 0.04$ &  $29.22 \pm 0.04$  & $29.07 \pm 0.04$ \\
   N4424  &    CCHP &  $31.00 \pm 0.06$ &  $30.86 \pm 0.11$  & $30.73 \pm 0.11$ \\
   N1365  &    CCHP &  $31.36 \pm 0.05$ &  $31.32 \pm 0.06$  & $31.19 \pm 0.05$

\cr \hline
\end{tabular}
\caption{Galaxies with F19 TRGB and SH0ES Cepheid distance moduli. The second column denotes
the source of the TRGB data (see text). Third column lists the TRGB distance modulus and error from
Table 3 of F19,  calibrated with the LMC DEB distance. The fourth and fifth columns list the
distance moduli for the solutions of Eqs. (\ref{equ:deg2a}) and (\ref{equ:deg3a}). The errors on these
estimates reflect errors arising from the PL fits only. They do not include errors in the anchor distances, 
which would shift all distance moduli up or down by the same number.}
\label{table:distance_moduli}
\end{center}

\end{table}

The dotted lines in Fig. \ref{fig:distance_modulus} show least squares fits of a constant offset. For the red points,
the offset is close to zero. However, for the blue points, there is an offset
\begin{equation}
\mu_{\rm Cepheid} - \mu_{\rm TRGB} = -0.139 \pm 0.024 \ {\rm mag} ,   \label{equ:dist}
\end{equation}
and since both sets of distance moduli are based on the geometric
distance to the LMC, the error in the DEB distance cancels.  If the
calibration of the TRGB is correct (a topic of some controversy
\cite{Yuan:2019, Freedman:2019}), this comparison reveals a
statistically significant ($\approx 6 \sigma$) offset compared to the
LMC calibration of the Cepheid distances. The TRGB distance scale is,
however, compatible with the NGC 4258 calibration of the Cepheid
distances\footnote{Interestingly, \cite{Reid:2019} reached a similar
  conclusion.}. The tension between these two calibrations leads to
the offset of Eq. (\ref{equ:dist}) which, of course, is almost
identical to the offset $\delta a$ found in Sect. \ref{subsec:SH0ES
  degeneracy}. As a consequence of these offsets, the TRGB value of
$H_0$ must be close the value inferred from the NGC  4258 calibration of
the Cepheid distance scale and should strongly disfavour the value derived
from the LMC calibration.

\begin{figure}
\centering
\includegraphics[width=48mm,angle=0]{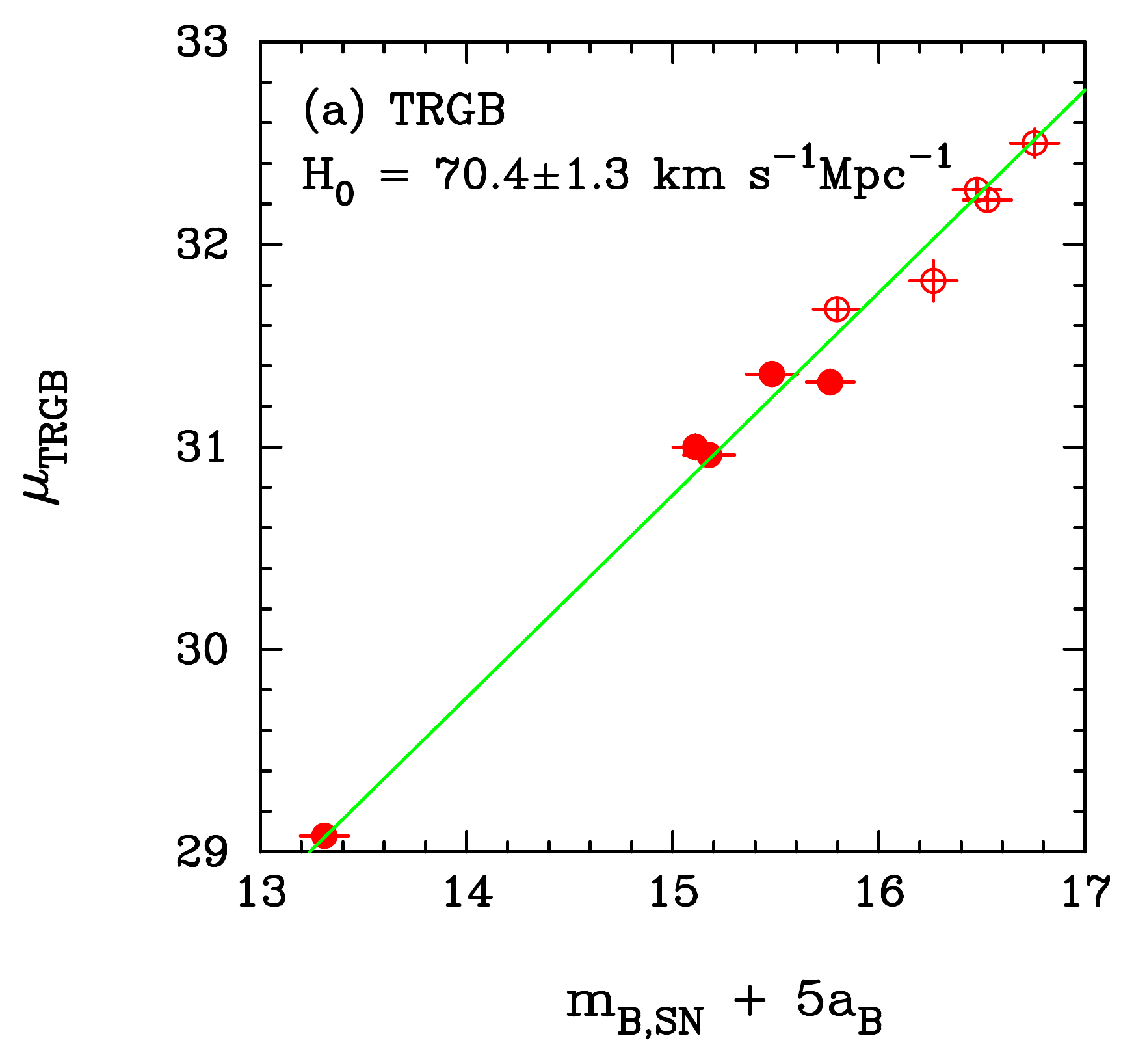} \includegraphics[width=48mm,angle=0]{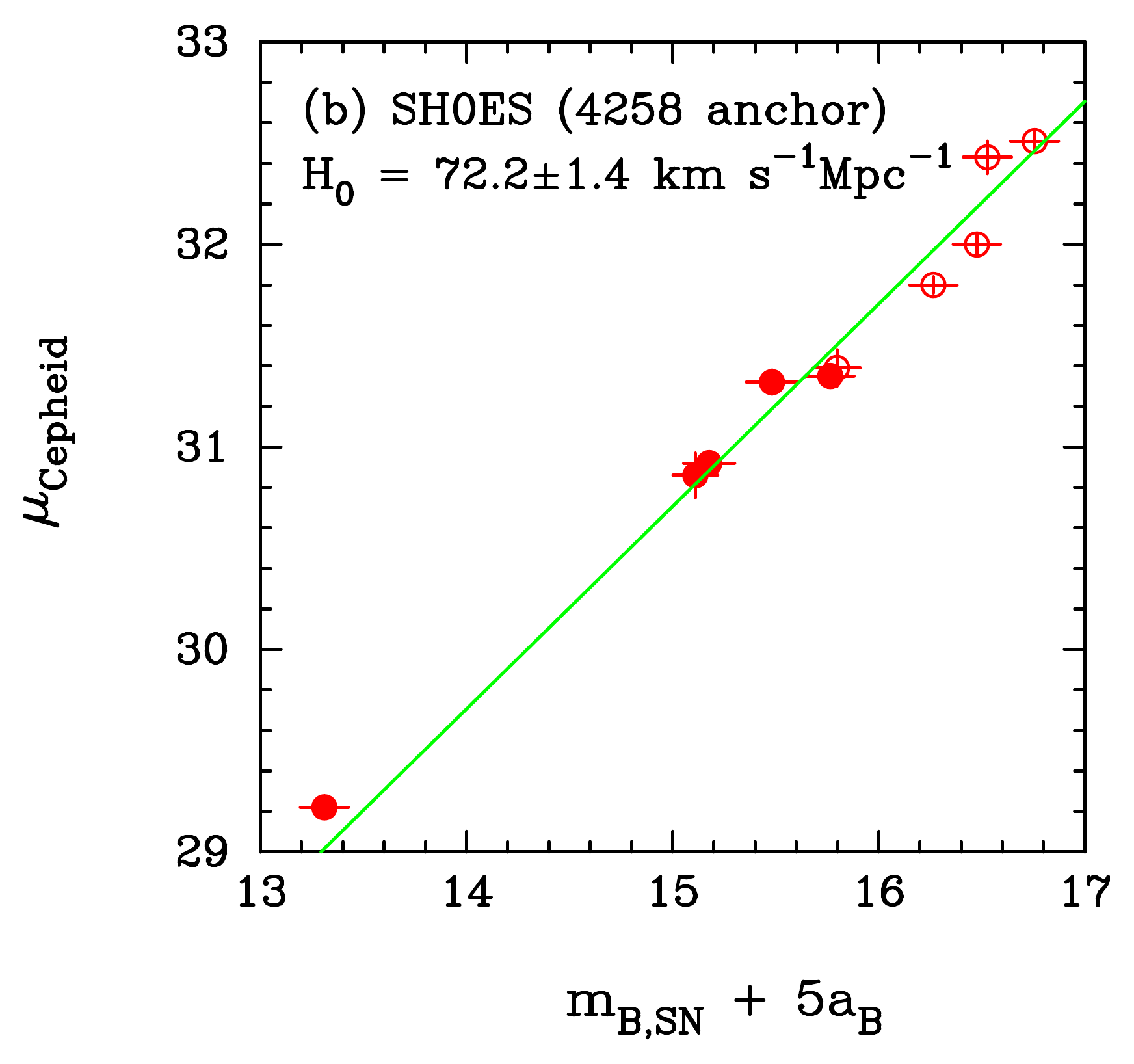} \includegraphics[width=48mm,angle=0]{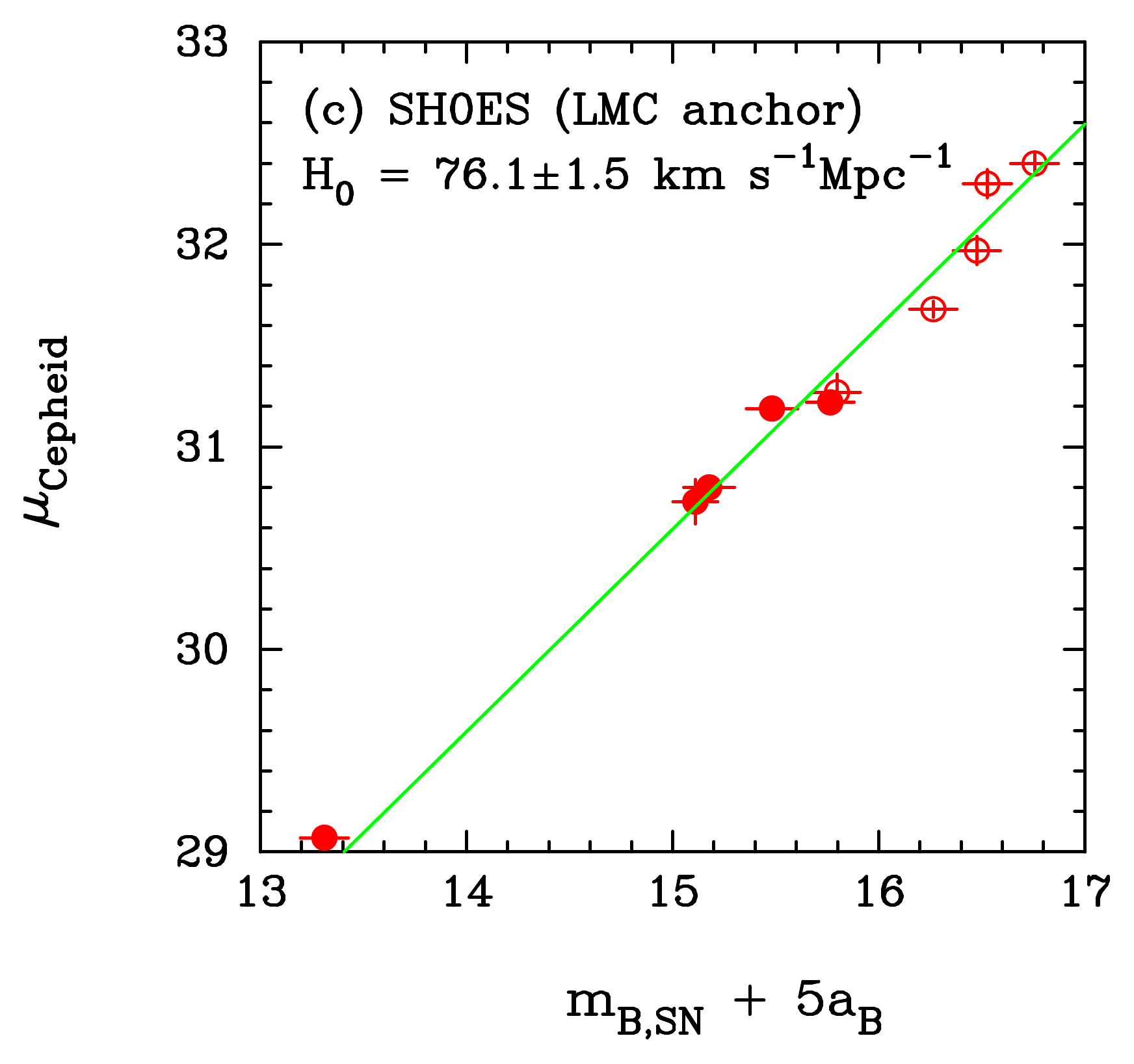} 
\caption {Distance moduli from Table \ref{table:distance_moduli} plotted against {\tt Supercal} Type Ia SN B-band peak magnitudes from  Table 5 of R16. The solid lines show the best fit linear relation of Eq. (\ref{equ:H0a}). The best fit values for $H_0$
from Eq. (\ref{equ:H0b}) for each relation are given in each panel. Note that the quoted errors on $H_0$ includes only the
photometric errors on $\mu$ and $m_{\rm B, SN} +5a_{\rm B}$.}

\label{fig:H0}

\end{figure}

This is illustrated in  Fig. \ref{fig:H0}. Here values and errors on $m_{\rm B, SN} + 5a_{\rm B}$ are from Table 5 
of R16, where $m_{\rm B, SN}$ is the {\tt Supercal} B-band peak magnitude and $a_{\rm B}$ is the intercept of the SN Ia
magnitude-redshift relation. These are plotted against the SN host distance moduli from Table 
\ref{table:distance_moduli}. We perform a least squares fit to determine the offset $\alpha$,
\begin{equation}
 \mu = m_{\rm B, SN} + 5a_{\rm B} - \alpha,  \label{equ:H0a}
\end{equation}
which gives a value for $H_0$, 
\begin{equation}
 H_0 = 10^{(0.2\alpha + 5)}, \quad \delta H_0 = 0.2 H_0 \log10 \delta\alpha.  \label{equ:H0b}
\end{equation}
Only the errors in $m_{\rm B, SN} +5a_{\rm B}$ and the errors in the distance moduli listed in Table \ref{table:distance_moduli}
are included in the fits. The  best fit values of $H_0$ and $1\sigma$ error are listed in each panel of Fig. \ref{fig:H0}.
Note that these error estimates do not include  the  error in the anchor distance. The resulting values of $H_0$ can be easily understood.
The value of $H_0$ in panel (a) for the TRGB distance moduli is consistent with the F19 value of Eq. (\ref{equ:H0_3}), showing that the subsample
of galaxies and SN data used in Fig. \ref{fig:H0} gives similar results to the full sample analysed by F19. Likewise, the
fit to panel (b) agrees well with the solution Eq. (\ref{equ:deg2a}) and the fit to panel (c) agrees with 
Eq. (\ref{equ:deg3a}). Since these results use the same SN data, the low value of $H_0$ derived from  the TRGB distance moduli is
caused almost entirely by a {\it calibration} difference. The TRGB calibration 
strongly disfavours the LMC calibration of the R16 Cepheids,
as quantified by Eq. (\ref{equ:dist}).

\begin{adjustwidth}{0.25in}{0.25in}
\ \par
\noindent{\underline{\bf Response from SH0ES team:} In our own analyses we find that the {\it relative} distances measured by Cepheids and TRGB agree well, matching G.E. here, thus the difference in derived $H_0$ (about 1.5 $\sigma$ as stated in Freedman et al. 2020) must occur from a disparity in their absolute calibration.  We think the best way to compare their calibrations is by using the same geometric distance calibration for each.  Doing so using NGC 4258 we see no difference as Cepheids give $H_0$=72.0 and TRGB gives 71-72 (see Jang and Lee 2017 and Reid et al. 2019) and if we use the same SN Ia measurements we find that they are spot on.  The disparity then arises only for the LMC where TRGB with the F20 calibration gives $H_0$=70 and Cepheids give 74.    We think it likely that this difference may be traced to the extinction of TRGB in the LMC that is used in the F20 analysis.  F20 finds $A_I=0.16 \pm 0.02$ mag and $H_0=70$ while Yuan et al. (2019) finds $A_I=0.10$, as did Jang and Lee (2017) and $H_0=72.7$.  (Cepheids use NIR Wesenheit magnitudes so determination of absolute extinction is not an issue for Cepheids.) The determination of extinction of TRGB in the LMC is challenging and its resolution is required to conclude whether TRGB and Cepheid agree when using the LMC as an anchor.  The comparison using NGC 4258 is cleaner for TRGB since the dust is negligible, the measurement is thus more similar to TRGB in SN hosts, and suggests excellent agreement with Cepheids.  It is also worth noting that Freedman et al. (2012) recalibrated the HST KP Cepheid distance ladder with the revised LMC distance, fortuitously the same distance value used today, and found $H_0=74.2 \pm 2.2$ but using different Cepheids, a different HST instrument, optical data, software, etc indicating Cepheids scales have been consistent.}
\ \par
\end{adjustwidth}

\section{Conclusions}
\label{sec:conclusions}

In the abstract, I asked the question `what would it take to make
SH0ES compatible with early time measurements?'.  The right hand panel
of Fig. \ref{fig:H0_contour1} provides an answer.  A bias in the
intercept of the PL relation of the crowded field SH0ES photometry
common to all SH0ES galaxies of about 0.1 - 0.14 mag, which I have
termed the SH0ES degeneracy, resolves the tension between late and
early time measurements of $H_0$ without the need for new
physics. Such a bias also resolves the tension between the geometric
distance of the LMC, MW parallaxes and the maser distance to NGC 4258,
and resolves the tension between the TRGB distance scale of F19 and
the Cepheid scale of the SH0ES team (which is based on combining the LMC, MW and NGC 4258
anchors). To my knowledge, there is no convincing way at present of
ruling out such a common mode systematic in the SH0ES data. {\it
  Excluding the SH0ES degeneracy as a possibility should be a priority
  for future research.}

Fortunately, this can be done by concentrating on {\it calibrations},
since the Hubble tension is a near 10\% effect.  There is really no
need to focus on acquiring data for more SN host galaxies (which may
cause $\sim 2$\% changes to the value of $H_0$, see
Fig. \ref{fig:H0}).  Here are some possible ways forward:

\smallskip

\noindent
(i) The CCHP and SH0ES teams disagree on the calibration of  the TRGB \cite{Yuan:2019, Freedman:2020}. The main issue
concerns corrections for rededening and extinction to the LMC. As discussed by \cite{Freedman:2020}
Gaia parallaxes of MW globular clusters should improve the Galactic calibration of the TRGB and JWST should 
allow precision measurements of additional calibrators. Calibrating the TRGB ladder using NGC 4258 would provide
an important consistency check of the LMC calibration.

\smallskip

\noindent
(ii) The discrepancy between the LMC and NGC 4258 anchors identified
in Sect. \ref{subsec:LMCanchor} needs to be resolved. Even if this turns
out to be a consequence of a rare statistical fluctuation (which I doubt), one needs to
establish which of these anchors is closer to the truth. In my view,
the DEB LMC distance appears to be reliable and so it would be worth rechecking the
NGC 4258 maser VLBI analysis \cite{Reid:2019}. Independent checks of
the maser distance, for example, refining the accuracy of the TRGB
distance to NGC 4258 \cite{Mager:2008}, would be particularly
valuable. Another interesting test would be to obtain HST data on
Cepheids in the outskirts of NGC 4258, reducing the impact of crowding
corrections and metallicity difference with the LMC (Riess, private
communication). If the distance to NGC 4258 can be shown to be lower
than the $7.58 \ {\rm Mpc}$ found by \cite{Reid:2019}, this would
strengthen the case for the Hubble tension.

\smallskip

\noindent
(iii) I have included the Appendix on the R11 and R16 photometric
comparison because I am far from convinced that the crowded field
photometry is unbiased and returns realistic errors. It should be
possible to apply MCMC techniques for scene reconstruction and to
calculate posterior distributions on the magnitude errors (which will
almost certainly be asymmetric). It would also be helpful if researchers
published as much data as possible, including data on outliers,  to allow
independent checks.

\smallskip

\noindent
(iv) In the longer term, other techniques such as strong lensing
time delays, distant masers, and gravitational wave sirens will hopefully become
competitive with the traditional distance ladder approaches\footnote{As this article
was nearing completion \cite{Birrer:2020} presented a new analysis of strong gravitational 
lensing time delays analysing the mass-sheet degeneracy and sensitivity to galaxy mass profiles.
These authors find  $H_0=67.4^{+4.1}_{-3.2} \Hunit$, lower than the value derived by \cite{Shajib:2020} but with a larger error. Also,
the Atacama Cosmology Telescope collaboration released the DR4 maps and cosmological parameters \cite{Naess:2020, Aiola:2020}.
Their results are in very good agreement with the results from \Planck. Athough this was not a surprise to me, it
surely lays to rest any remaining concerns of even the most hardened of skeptics that the \Planck\ results are affected by systematic errors.}.

\begin{adjustwidth}{0.25in}{0.25in}
\ \par
\noindent{\underline{\bf Response from SH0ES team: } 
The case that Vanilla $\Lambda$CDM calibrated in the Early Universe predicts a value of $H_0$ 
of 67-68 appears very sound.  It is a mystery to us why the distance ladder methods consistently find a 
higher value, but one we feel worth paying attention to.
  It is worth highlighting a recent result that is independent of all of G.E.'s considerations
in this talk and all other rungs, Cepheids, TRGB, SNe Ia, etc 
and which has not received much attention.  That is the final result from the Maser Cosmology 
Project, Pesce et al. 2020, which measures geometric distances to 6 masers in the Hubble flow and finds $73.9 \pm 3.0$,
which corroborates prior indications that the local value of $H_0$ exceeds the predicted value with $\sim$ 98\% confidence
if G.E. sees no problems with it (tongue back in cheek).  
}
\ \par
\end{adjustwidth}

\medskip

\medskip

\vspace{ -0.2 truein}

\acknowledgments Parts of this article have been circulated to various
people over the last two years. Since the `Hubble tension' is spawning
an ever increasing literature (and media attention) I felt that it
would be useful to put this work in the public domain. Researchers
should not simply accept a value of $H_0$ quoted in the abstract of a
paper, but should be encouraged to `look under the hood' and
understand the data that they are trying to explain. I hope that this
article will add to that understanding. I should add that I have
tremendous respect for the SH0ES and CCHP teams, and their respective
PIs. I would especially like to thank Wendy Freedman and Adam Riess
for patiently answering my questions over the last few years. Some of the
plots in this article were made with Antony Lewis's {\tt GetDist} software
({\tt https://getdist.readthedocs.io/en/latest/} ).

\appendix
\section{Object-by-object comparison of the R11 and R16 Cepheid photometry}
\label{sec:appendix}

\begin{table}[h]
\begin{center}

\begin{tabular}{llllllll} \hline
             &   & &       &  \multicolumn{2}{c}{all Cepheids} & \multicolumn{2}{c}{outliers removed} \\
   galaxy     & $N_{\rm R11}$ &  $N_{\rm R16}$ &  $N_{\rm match}$ &   $\qquad \langle \Delta m \rangle$&  $\langle \Delta C \rangle$ & $ \qquad \langle \Delta m \rangle$ & $\langle \Delta C \rangle$ \\ \hline
   N4536  &  69 & 33 &  28 &   $-0.114 \pm 0.057$ &  $0.153$  & $-0.069 \pm 0.062$ &   $0.153$\\ 
   N4639  &  32 & 25 &  17 &   $-0.071 \pm 0.100$ &  $0.091$  & $-0.071 \pm 0.100$  &  $0.091$ \\    
   N3370  &  79 & 63 & 51 &   $-0.105 \pm 0.055$ &  $0.146$  & $-0.090 \pm 0.055$ &   $0.145$\\ 
   N3982  &  29  & 16 & 12 &   $-0.178 \pm 0.090$ &  $0.092$  & $-0.081 \pm 0.094$ &   $0.092$\\ 
   N3021  & 29 &  18 & 13 &   $+0.120 \pm 0.146$ &  $0.196$  & $+0.120 \pm 0.146$ &   $0.196$ \\
   N1309  & 36 &  44  & 16 &   $-0.087 \pm 0.091$ &  $0.330$  & $-0.087 \pm 0.091$ &   $0.330$ \\
   N5584  & 95 &  83 & 65 &   $-0.028 \pm 0.049$ &  $0.039$  & $+0.001 \pm 0.051$  &   $0.038$ \\
   N4038  & 39 & 13 &  11 &   $-0.239 \pm 0.153$ &  $0.109$  & $-0.239 \pm 0.153$  &  $0.109$ \\
   N4258  & 165  & 139 & 73 &   $-0.217 \pm 0.055$ &  $0.145$  & $-0.020 \pm 0.062$ &   $0.143$
\cr \hline
\end{tabular}

\caption{Offsets to the magnitudes and colours. $N_{\rm R11}$ is the number of Cepheids in Table 2 of R11. $N_{\rm R16}$ is the number of Cepheids in Table 4 of R16. $N_{\rm match}$ is the number of Cepheids common to both tables.}
\label{table:fits}
\end{center}

\end{table}

This analysis is based on matching Cepheids from Table 2  of R11 and Table 4 of R16.
Note the following:

\smallskip

\noindent
(i) The R11 table contain Cepheids with a rejection flag. Cepheids with IFLAG = 0 were accepted
by R11 and those with IFLAG = 1 were rejected.

\smallskip

\noindent
(ii) The data in the R16 table has been `pre-clipped' by the authors and  does not list
data for  Cepheids that were rejected by R16. The R16 table contains Cepheids
that do not appear in the R11 table.

\smallskip

\noindent
(iii) The R16 magnitudes have been corrected for bias and blending errors from scene reconstruction. 
Each Wesenheit F160W magnitude  has an error estimate:
\begin{equation}
 \sigma_{\rm tot} = (\sigma^2_{\rm sky} + \sigma^2_{\rm ct} + \sigma^2_{\rm int} + (f_{\rm ph} \sigma_{\rm ph}^2)) ^{1/2},
\end{equation}
where $\sigma_{\rm sky}$ is the dominant error and comes from contamination of the sky background by blended images,
$\sigma^2_{ct}$ is the error in the colour term $R(V-I)$, which is small and of order $0.07$ {\rm mag};
 $\sigma_{\rm int}$ is the internal scatter from the width of the instability strip, which is known
from the LMC and M31 to be small ($\approx 0.08$ mag); $f_{\rm ph} \sigma_{\rm ph}$ is the error in the phase correction
of the Cepheid light curves. 

\smallskip

\noindent
(iv) The positions in R11 are not listed to high enough precision to uniquely identify a Cepheid in the R16 table.
There are ID numbers listed by R11 and R16, but for three galaxies (NGCs 3370, 3021, 1309) these numbers do not match.
Where possible, we have matched Cepheids using their ID numbers. For the remaining three galaxies, we 
have used a combination of positional coincidence and agreement in periods to match Cepheids. (This gives
perfect agreement for  the six galaxies with matching ID numbers.)

Outliers can have a significant effect on fits to the magnitude and colour differences. We fit:
\begin{subequations}
\begin{eqnarray}
(m_H^W)_{\rm R16} &= & (m_H^W)_{\rm R11} + \langle \Delta m_H^W \rangle, \\
(V-I)_{\rm R16} &= & (V-I)_{\rm R11} + \langle \Delta C \rangle,
\end{eqnarray} 
\end{subequations}
with and without outliers, where outliers are defined as having
\begin{equation}
\left  \vert {((m_H^W)_{\rm R11} - (m_H^W)_{\rm R16}) \over (\sigma_{\rm tot})_{\rm R16}} \right \vert > 2.5.
\end{equation}
The results are given in Table \ref{table:fits}. The rest of this appendix shows the equivalent plots for each of the nine galaxies.

\begin{figure}

\includegraphics[width=150mm, angle=0]{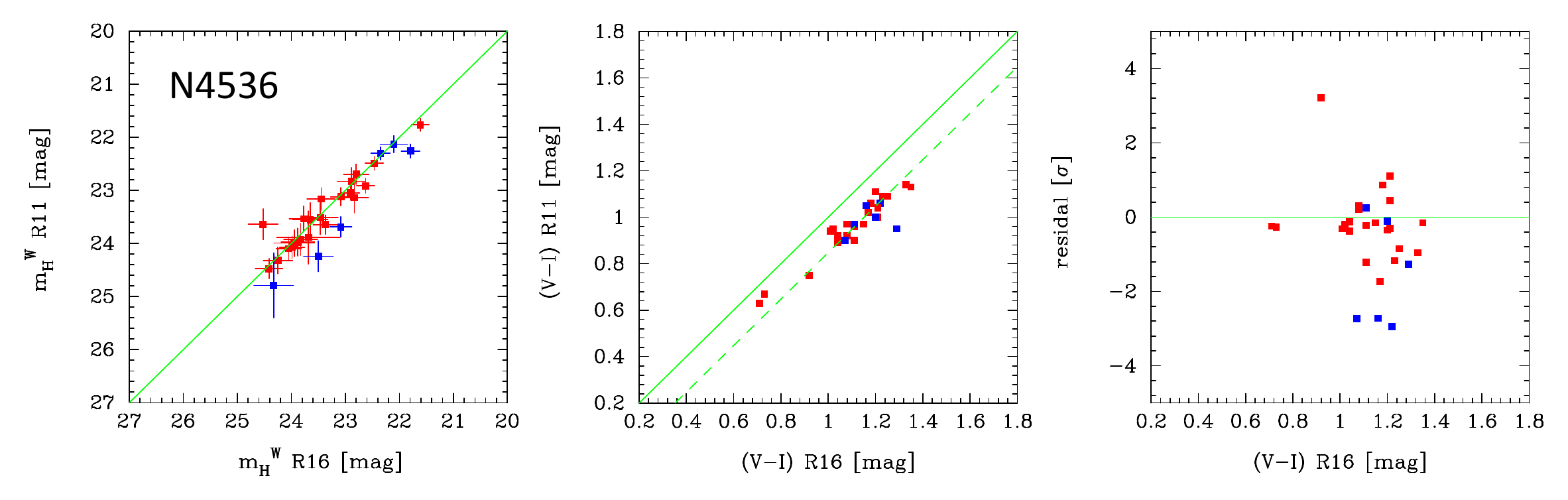}
\caption
{The plot to the left shows R11 Wesenheit $H$ magnitudes plotted against R16 Wesenheit H magnitudes. Central plot shows
R11 (V-I) colours plotted against R16 (V-I) colours. The dashed line in the central
panel shows the best fit offset. Plot to the right shows the difference in $H$-band magnitudes
in units of the R16 error, $((m_H^W)_{\rm R11} - (m_H^W)_{\rm R16})/(\sigma_{\rm tot})_{\rm R16}$, plotted against R16 $(V-I)$ colour.
 Blue points show Cepheids  with IFLAG=1 in R11 (i.e. these
were rejected by R11 but accepted by R16). Red points show Cepheids with IFLAG=0 in R11. }

\label{fig:4536}
\end{figure}

\begin{figure}

\includegraphics[width=75mm, angle=0]{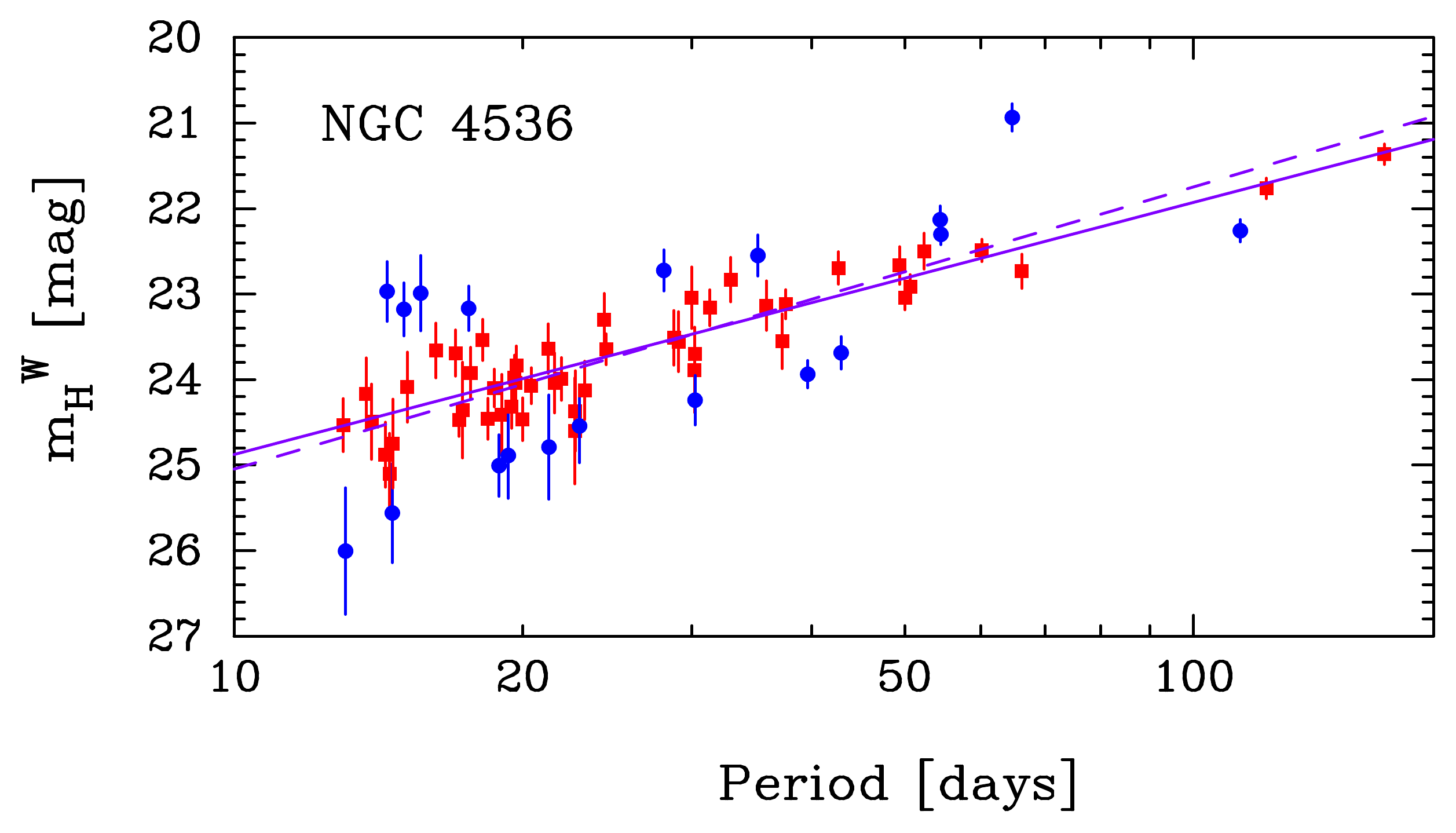} \includegraphics[width=75mm, angle=0]{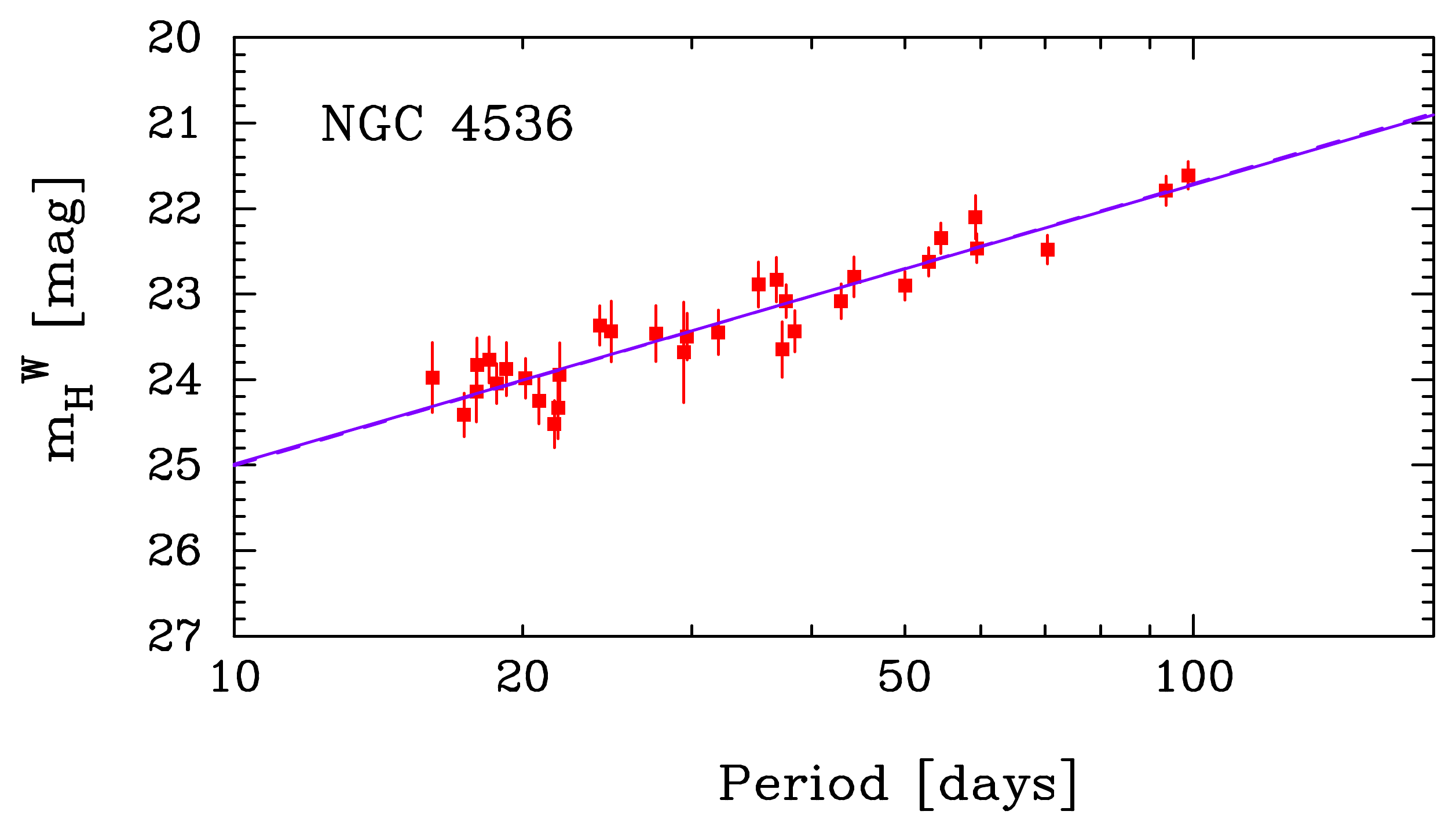}
\caption
{Left plot shows the PL relation for R11 Cepheids. Blue points show Cepheids rejected by R11 (IFLAG =1);
red points show Cepheids accepted by R11 (IFLAG = 0). The solid line shows the best fit linear
relation fitted to the red points. The dashed line shows thebest fit with the slope constrained to $b=-3.3$.
Right plot shows the PL relation for R16 Cepheids. The solid line shows the best fit linear
relation fitted to the red points. The dashed line shows the best fit with the slope constrained to $b=-3.3$. The parameters
of these fits are given in Table \ref{table:PL}.}
\label{fig:R4536}
\end{figure}

\begin{figure}

\includegraphics[width=150mm, angle=0]{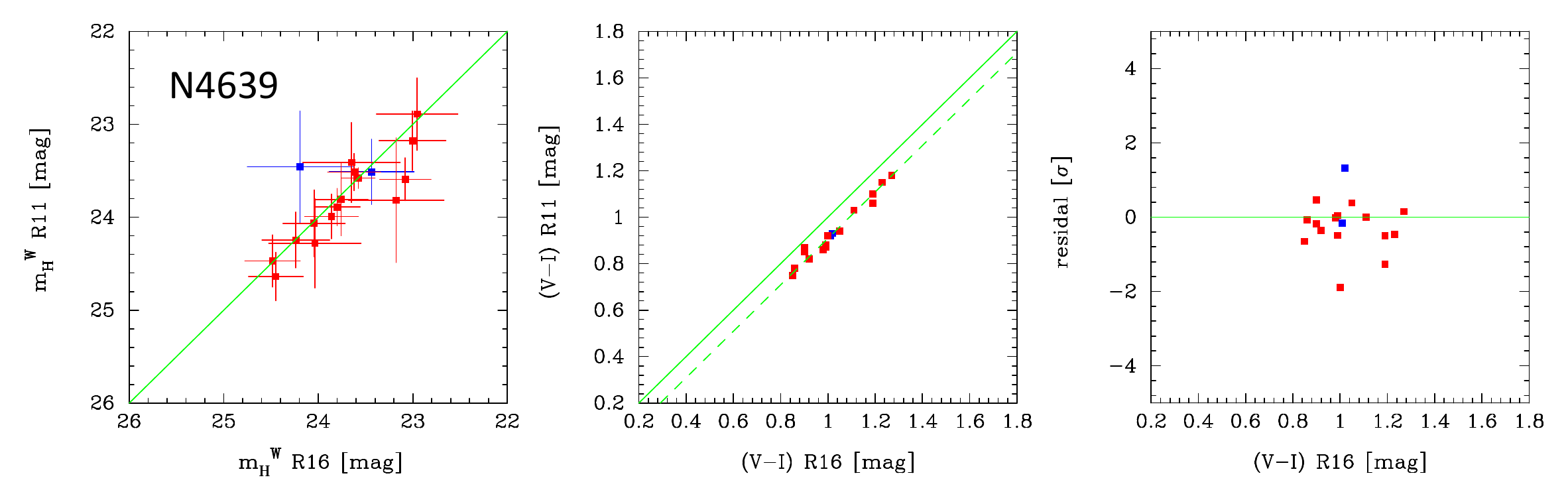}
\caption
{As Fig. \ref{fig:4536}.}
\label{fig:4639}
\end{figure}

\begin{figure}

\includegraphics[width=75mm, angle=0]{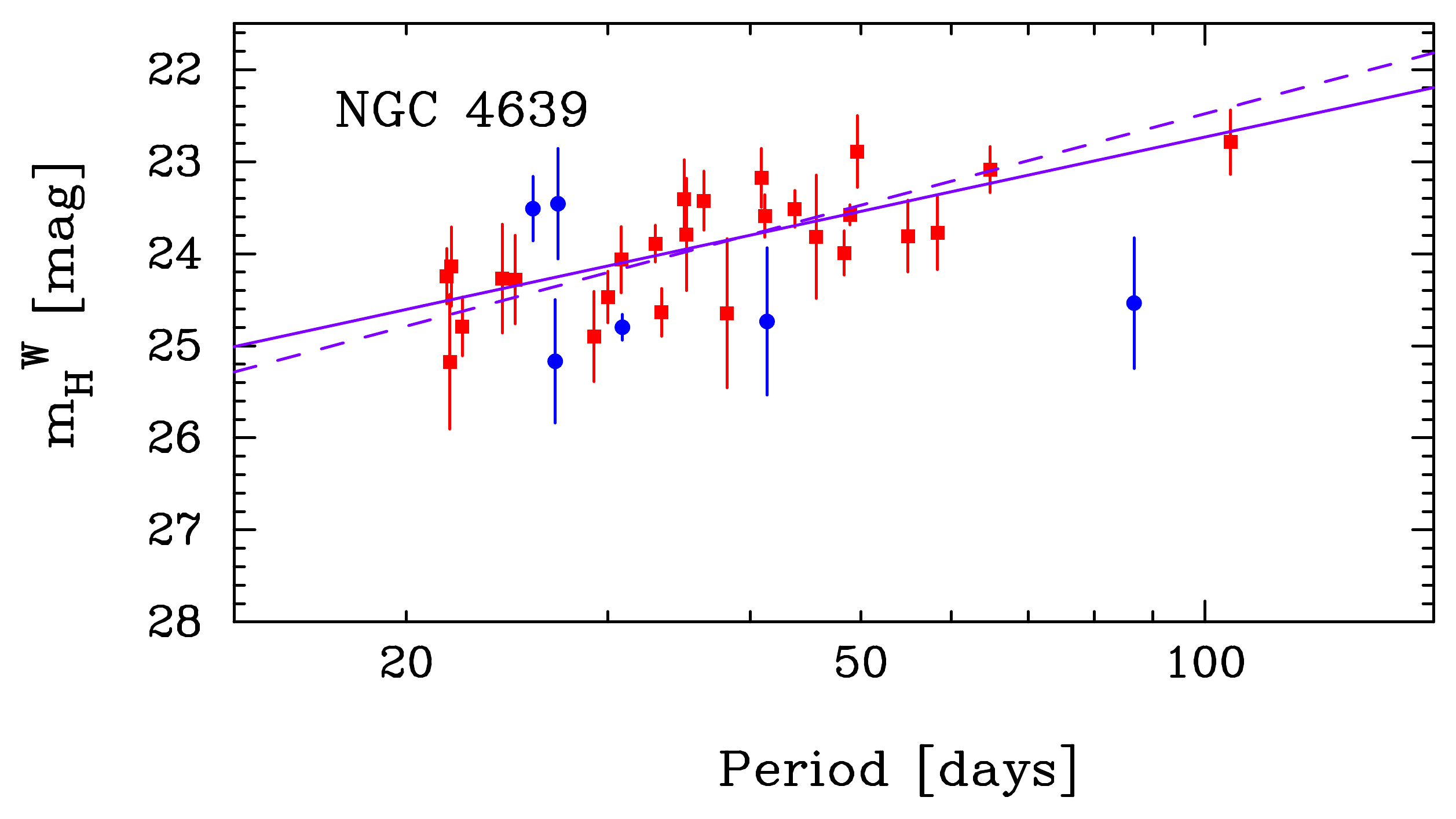} \includegraphics[width=75mm, angle=0]{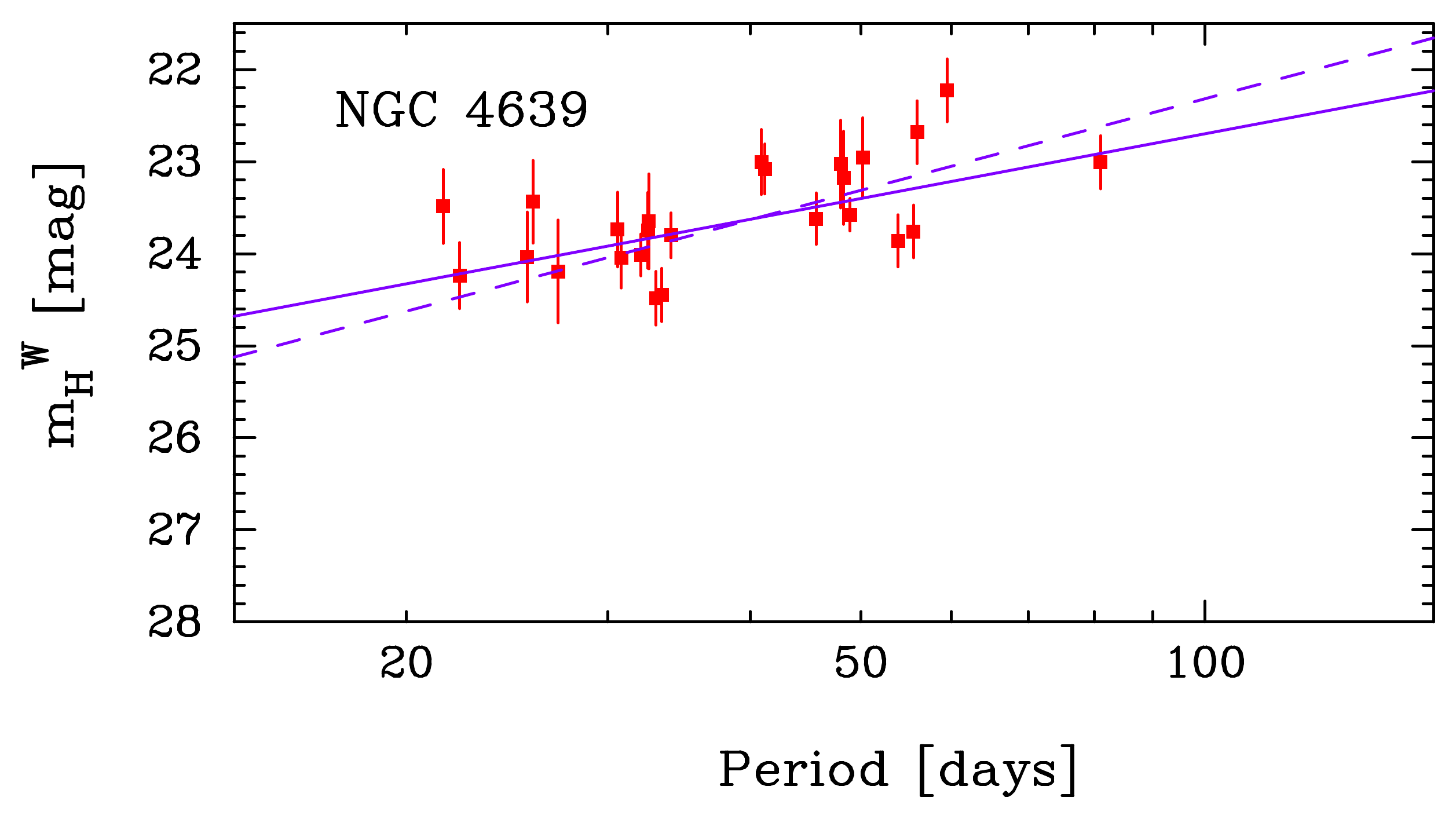}
\caption
{As Fig. \ref{fig:R4536}}

\label{fig:R4639}
\end{figure}

\begin{figure}

\includegraphics[width=150mm, angle=0]{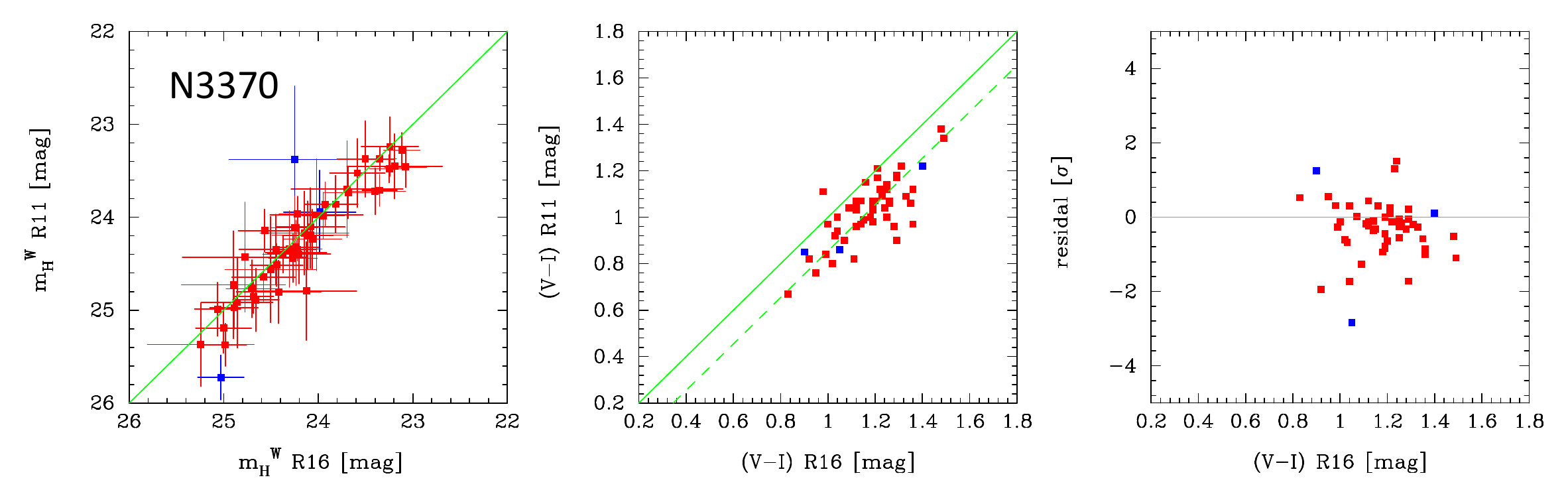}
\caption
{As Fig. \ref{fig:4536}.} 
\label{fig:3370}
\end{figure}

\begin{figure}

\includegraphics[width=75mm, angle=0]{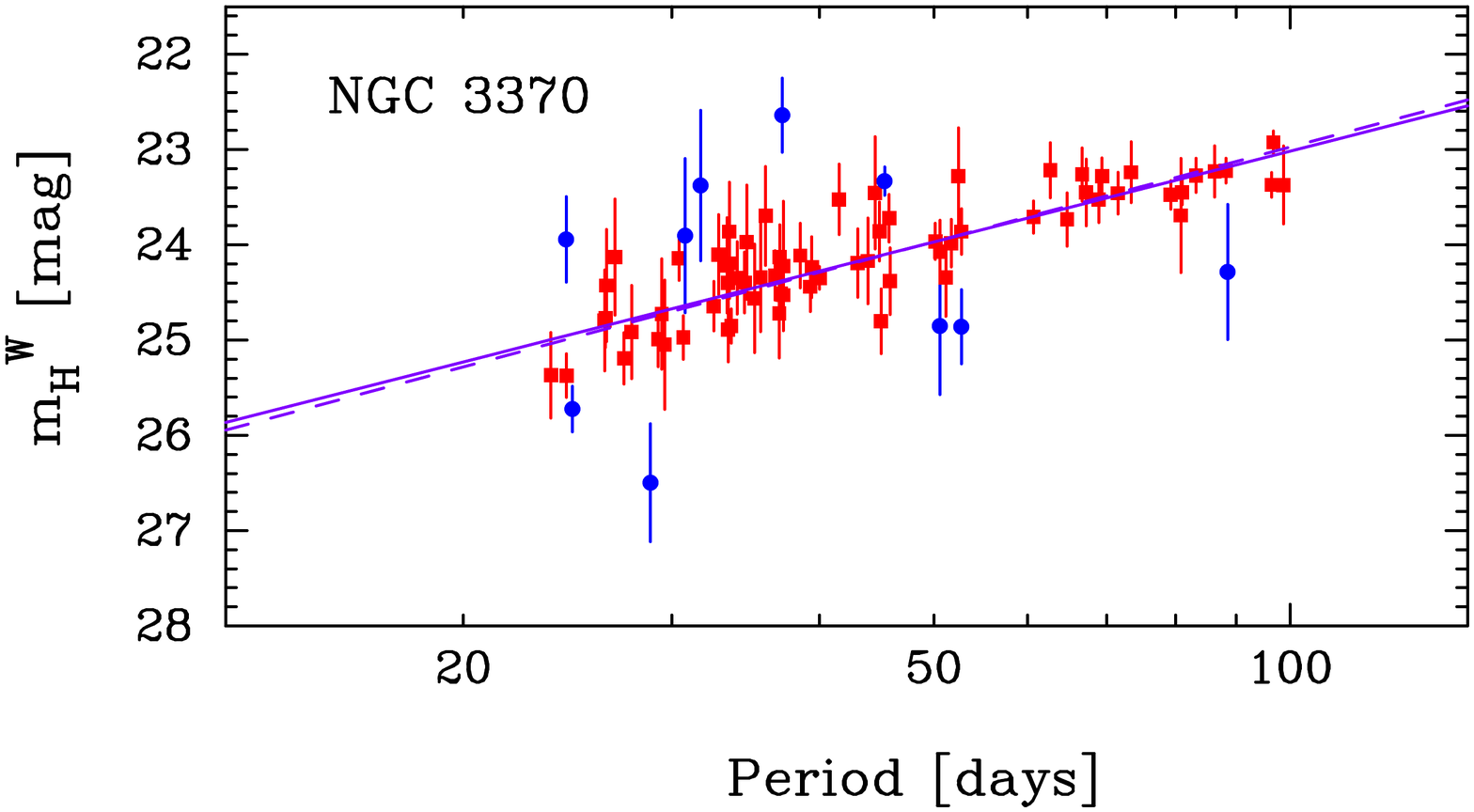} \includegraphics[width=75mm, angle=0]{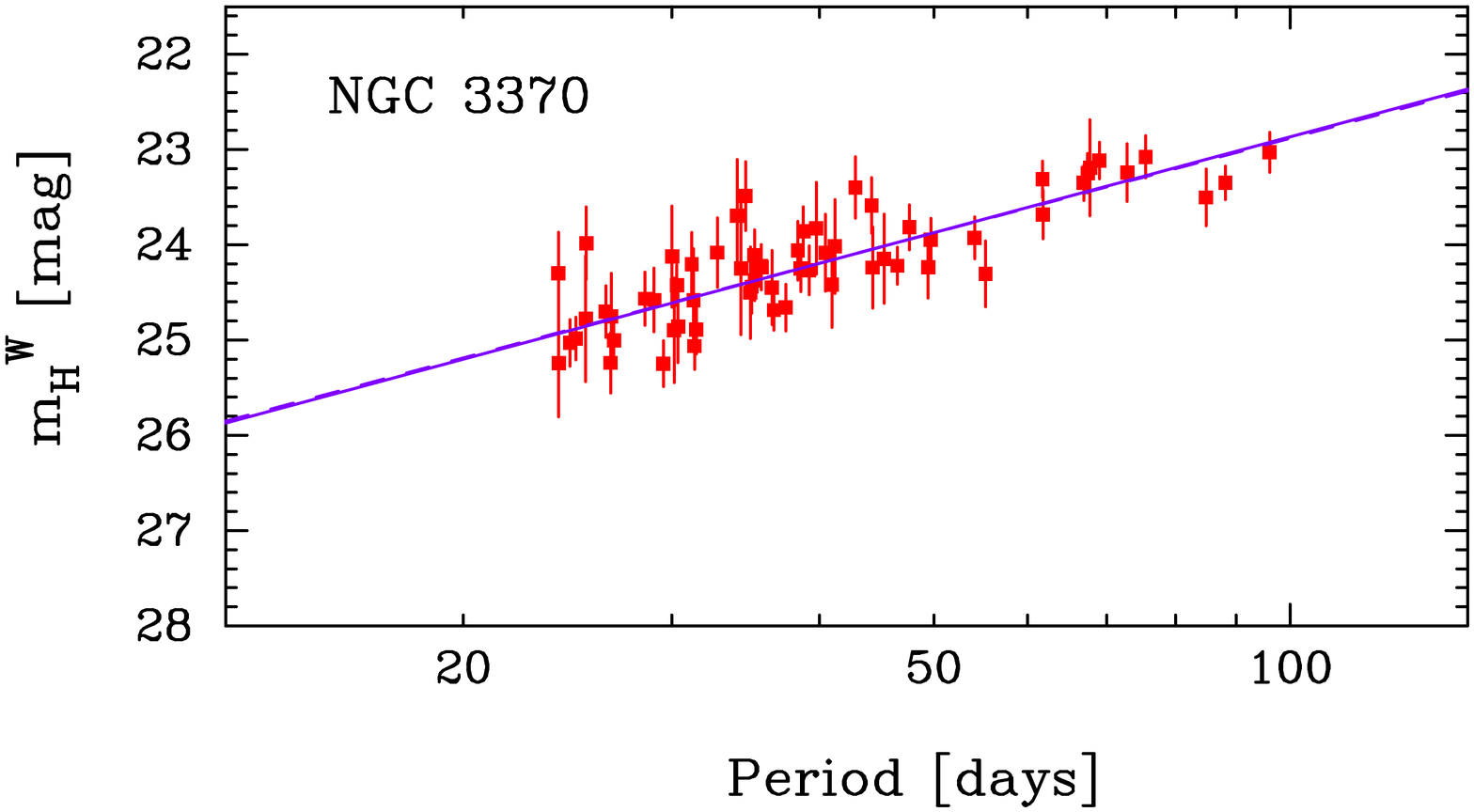}
\caption
{As Fig. \ref{fig:R4536}}

\label{fig:R3370}
\end{figure}

\begin{figure}

\includegraphics[width=150mm, angle=0]{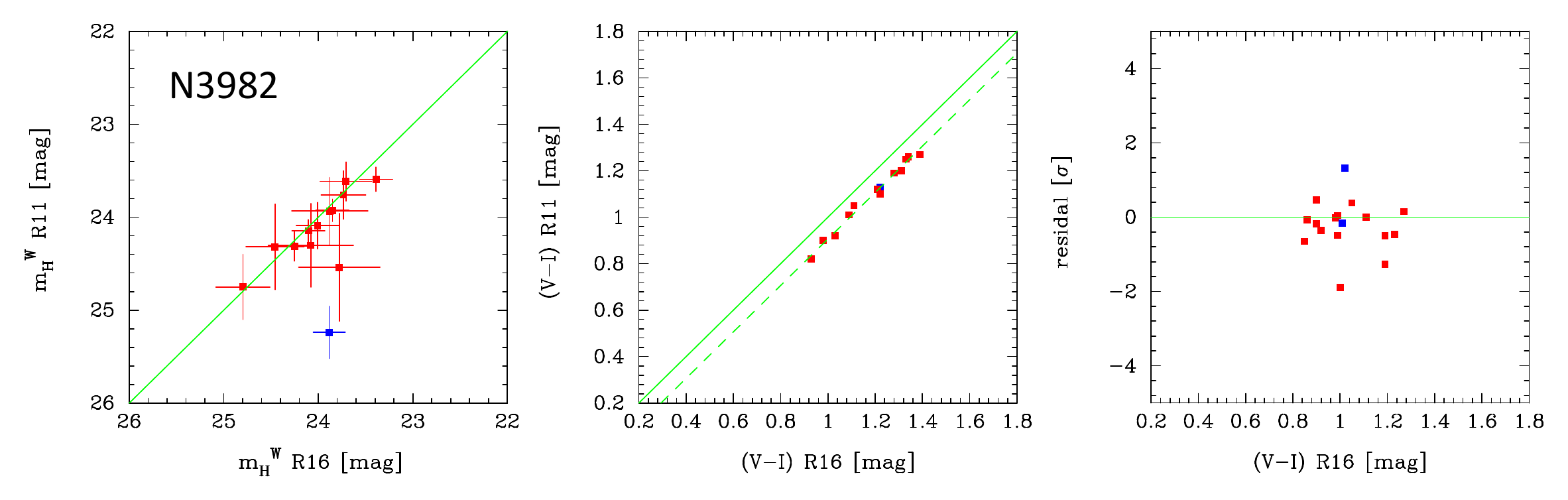}
\caption
{As Fig. \ref{fig:4536}.}
\label{fig:3982}
\end{figure}

\begin{figure}

\includegraphics[width=75mm, angle=0]{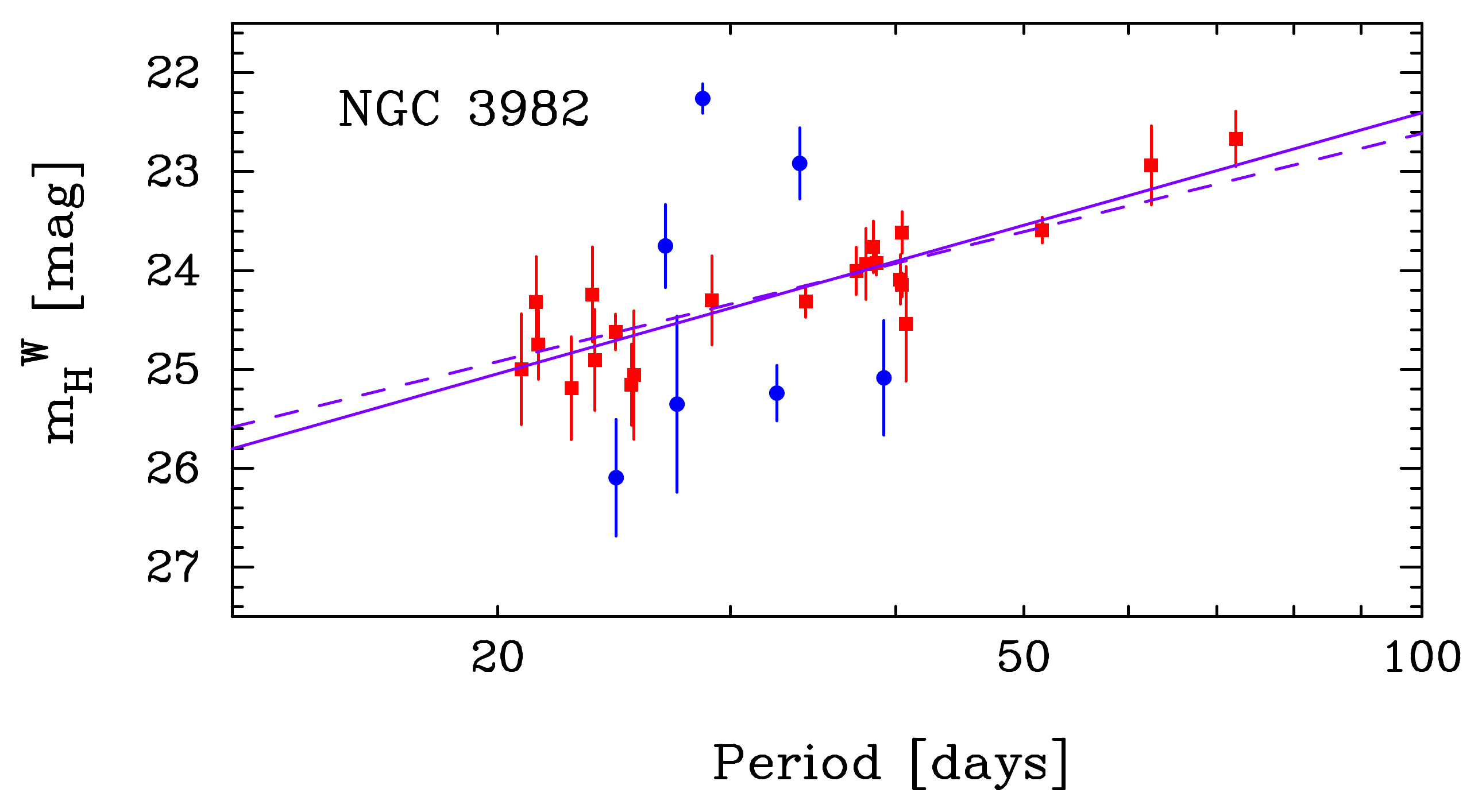} \includegraphics[width=75mm, angle=0]{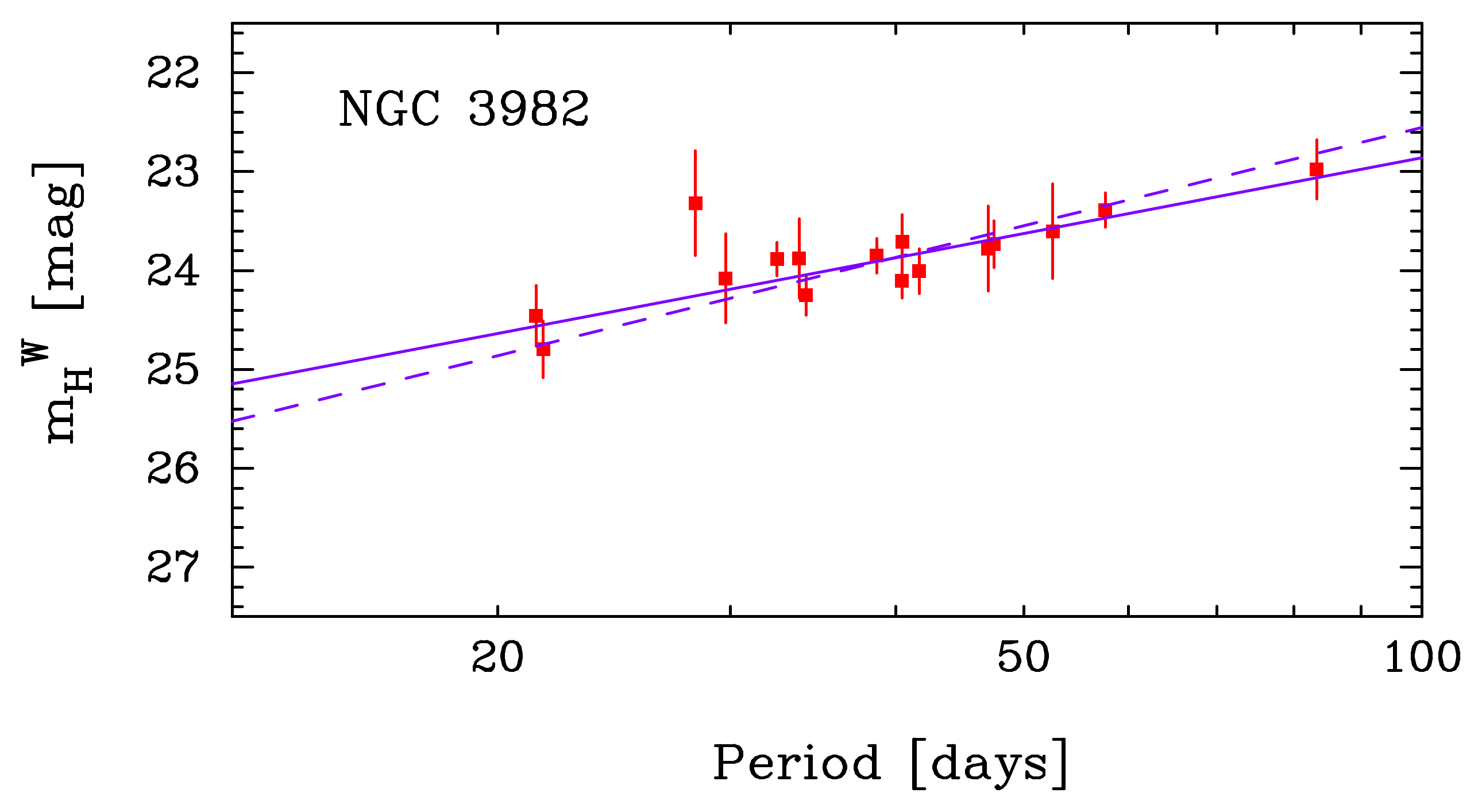}
\caption
{As Fig. \ref{fig:R4536}}

\label{fig:R3982}
\end{figure}

\begin{figure}

\includegraphics[width=150mm, angle=0]{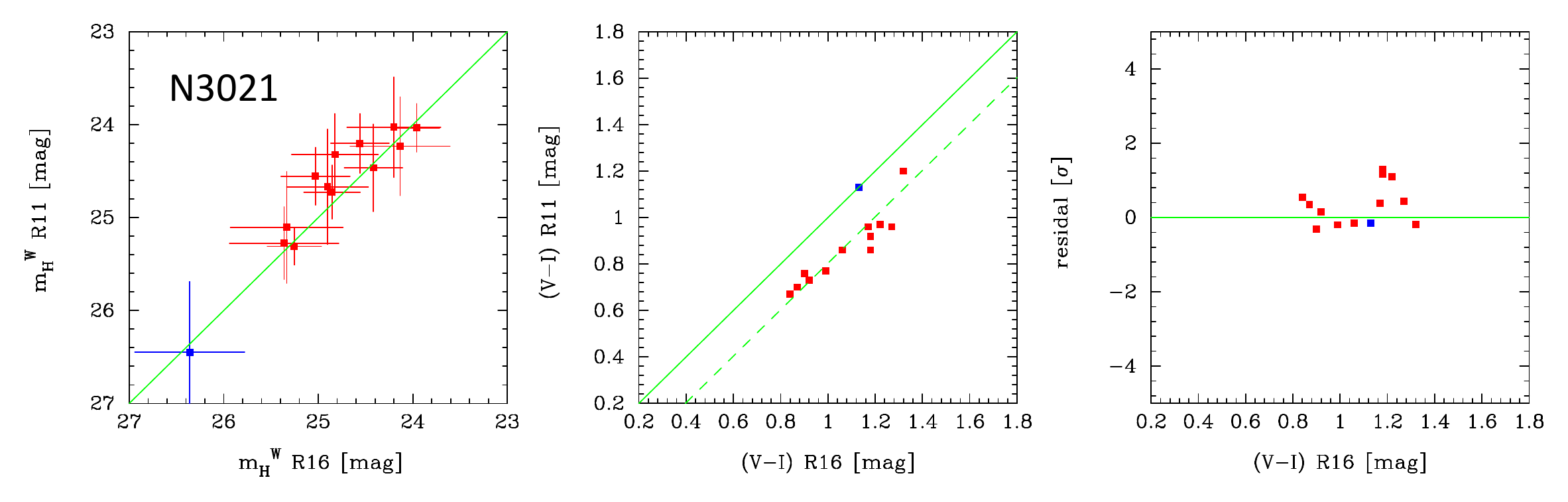}
\caption
{As Fig. \ref{fig:4536}.}
\label{fig:3021}
\end{figure}

\begin{figure}

\includegraphics[width=75mm, angle=0]{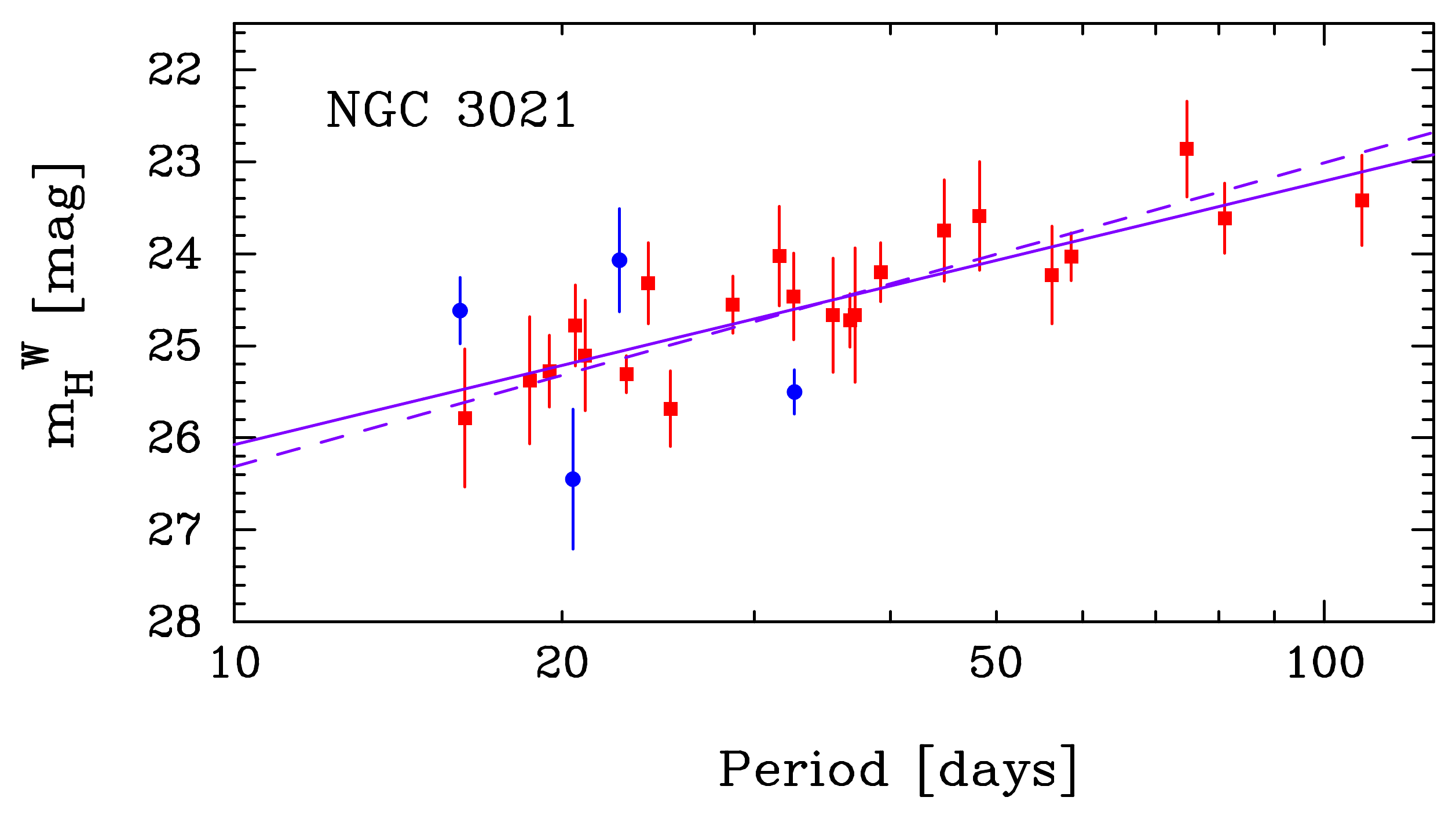} \includegraphics[width=75mm, angle=0]{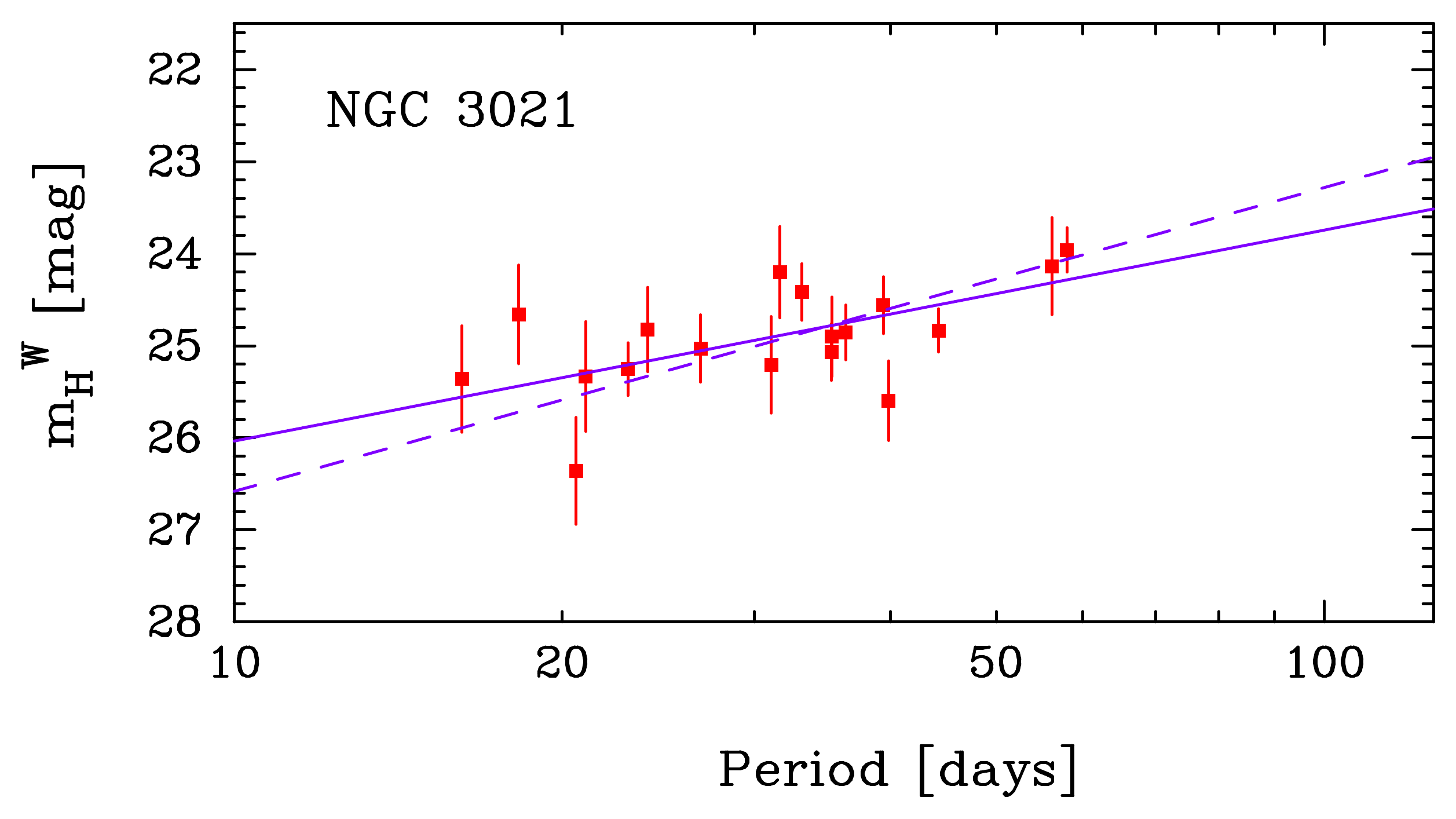}
\caption
{As Fig. \ref{fig:R4536}}

\label{fig:R3021}
\end{figure}

\begin{figure}

\includegraphics[width=150mm, angle=0]{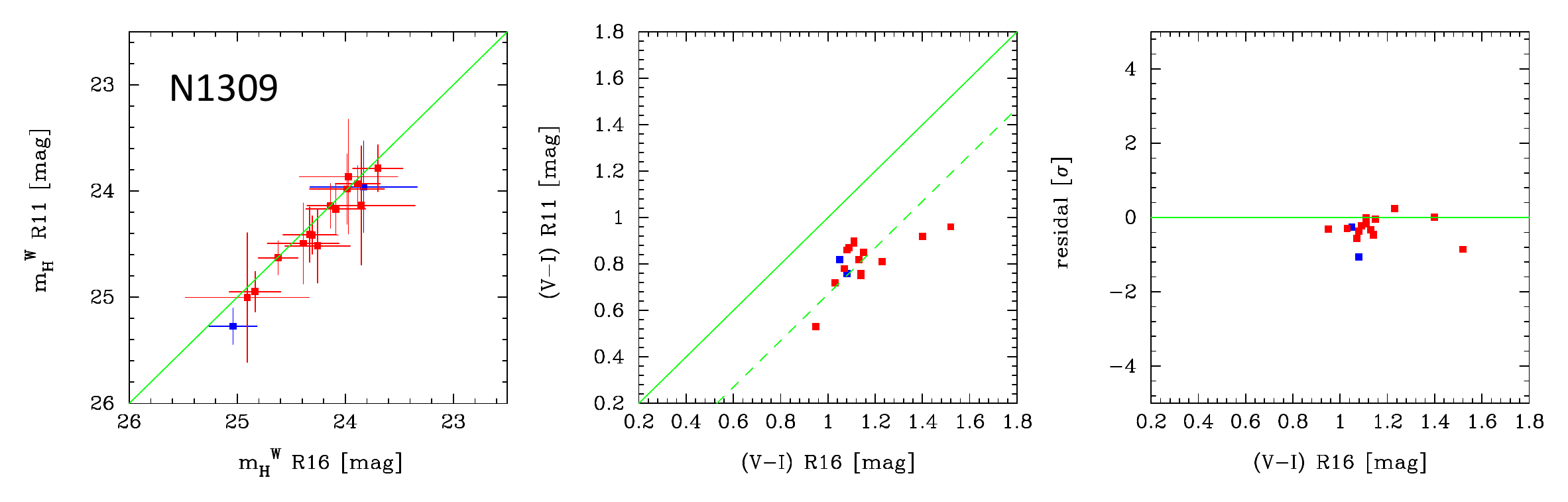}
\caption
{As Fig. \ref{fig:4536}.}
\label{fig:1309}
\end{figure}

\begin{figure}

\includegraphics[width=75mm, angle=0]{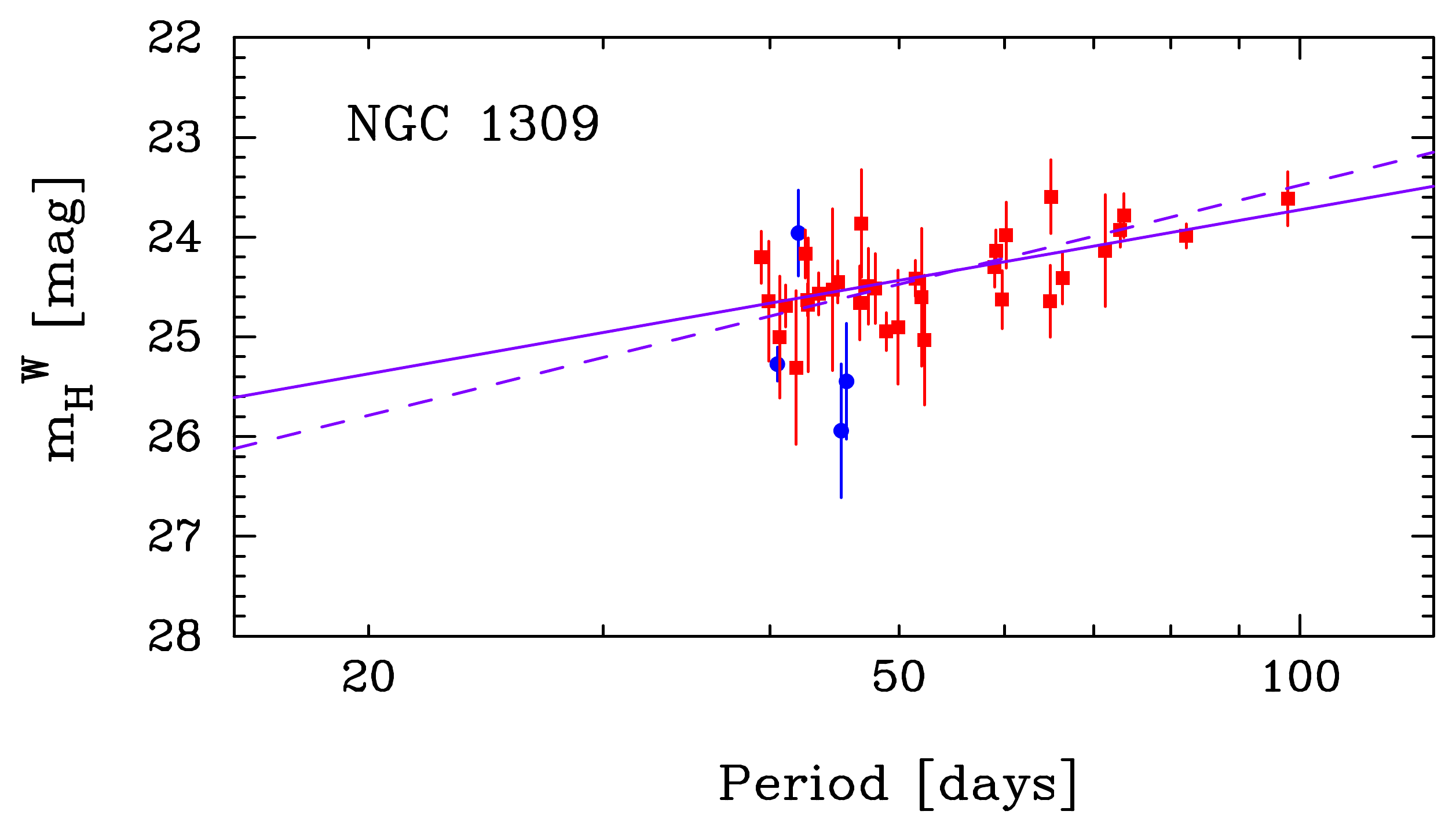} \includegraphics[width=75mm, angle=0]{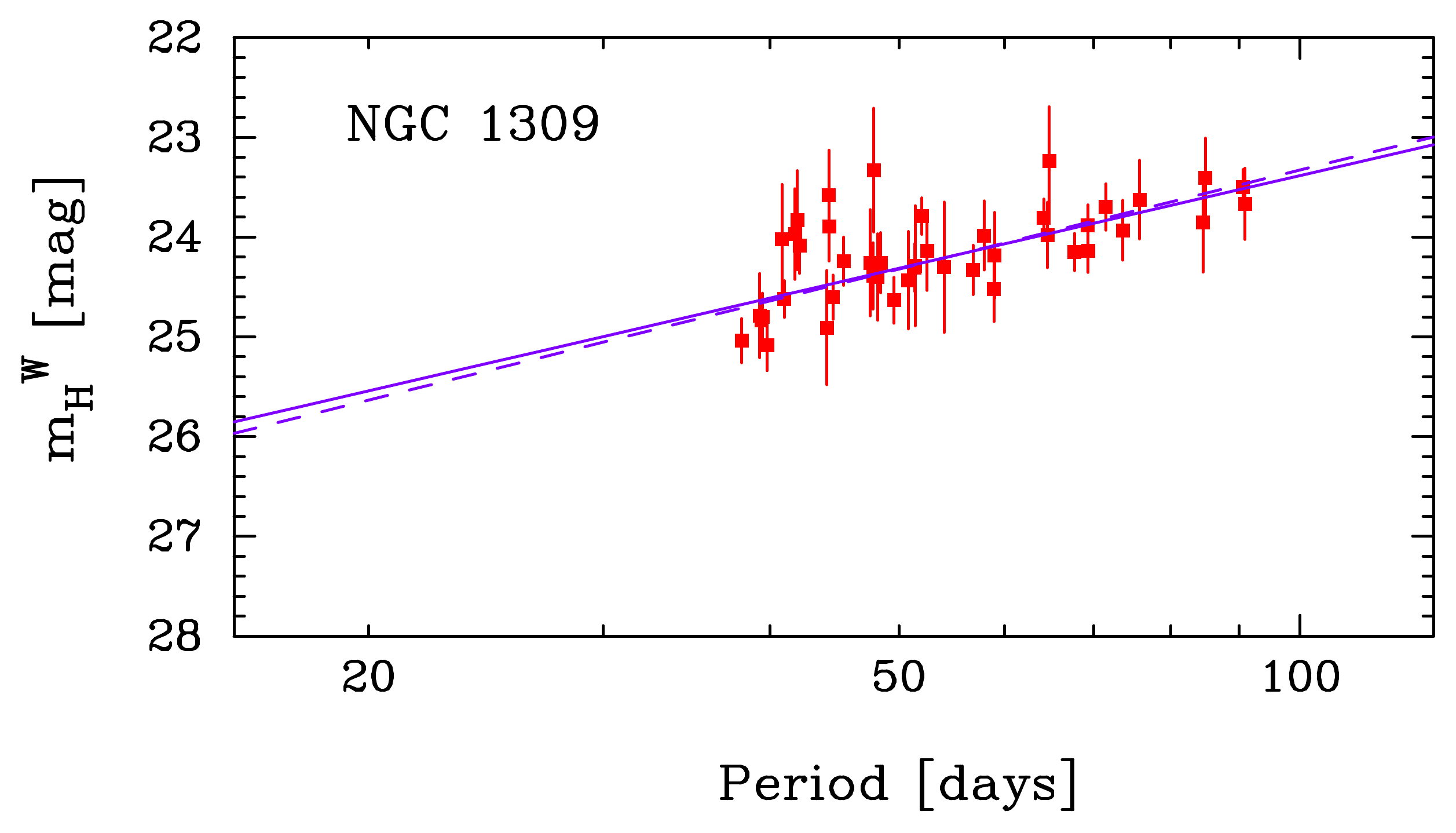}
\caption
{As Fig. \ref{fig:R4536}}

\label{fig:R1309}
\end{figure}

\begin{figure}

\includegraphics[width=150mm, angle=0]{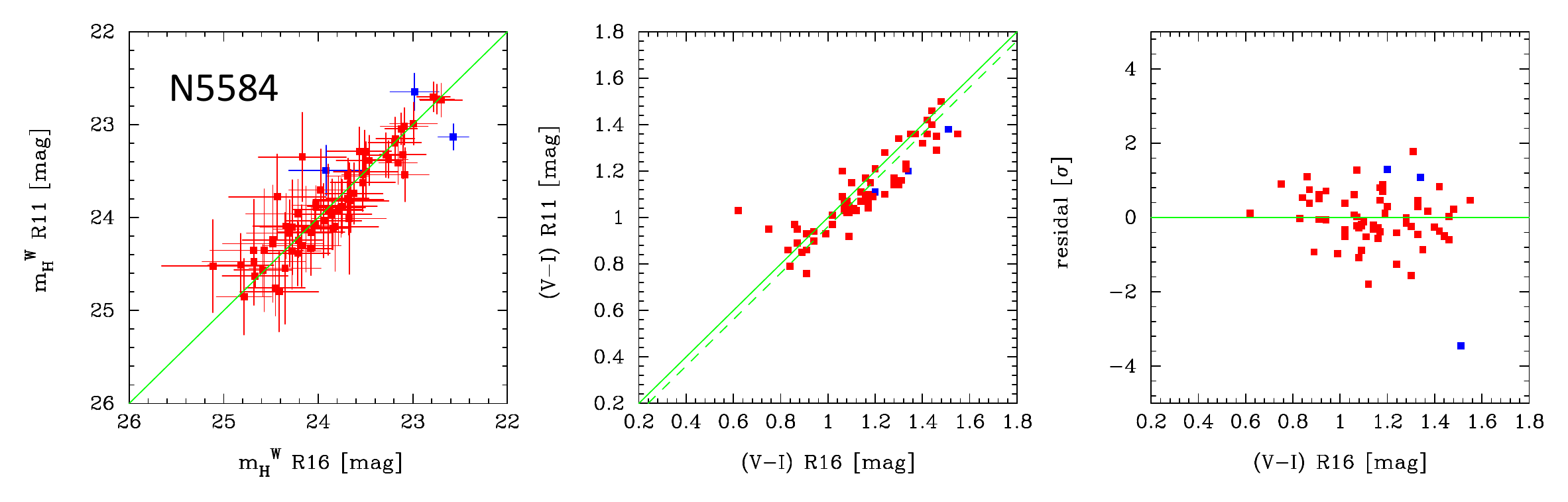}
\caption
{As Fig. \ref{fig:4536}.}
\label{fig:5584}
\end{figure}

\begin{figure}

\includegraphics[width=75mm, angle=0]{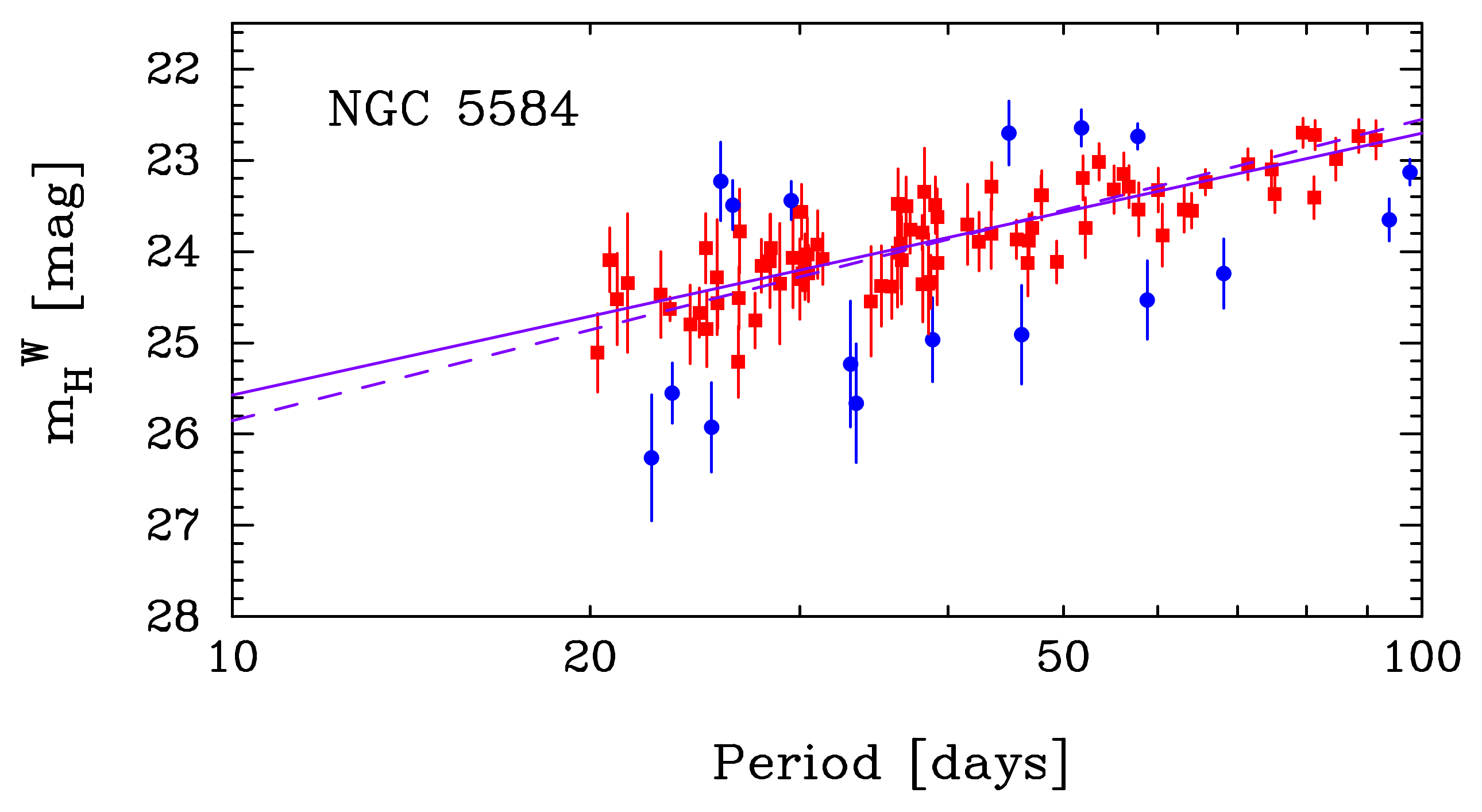} \includegraphics[width=75mm, angle=0]{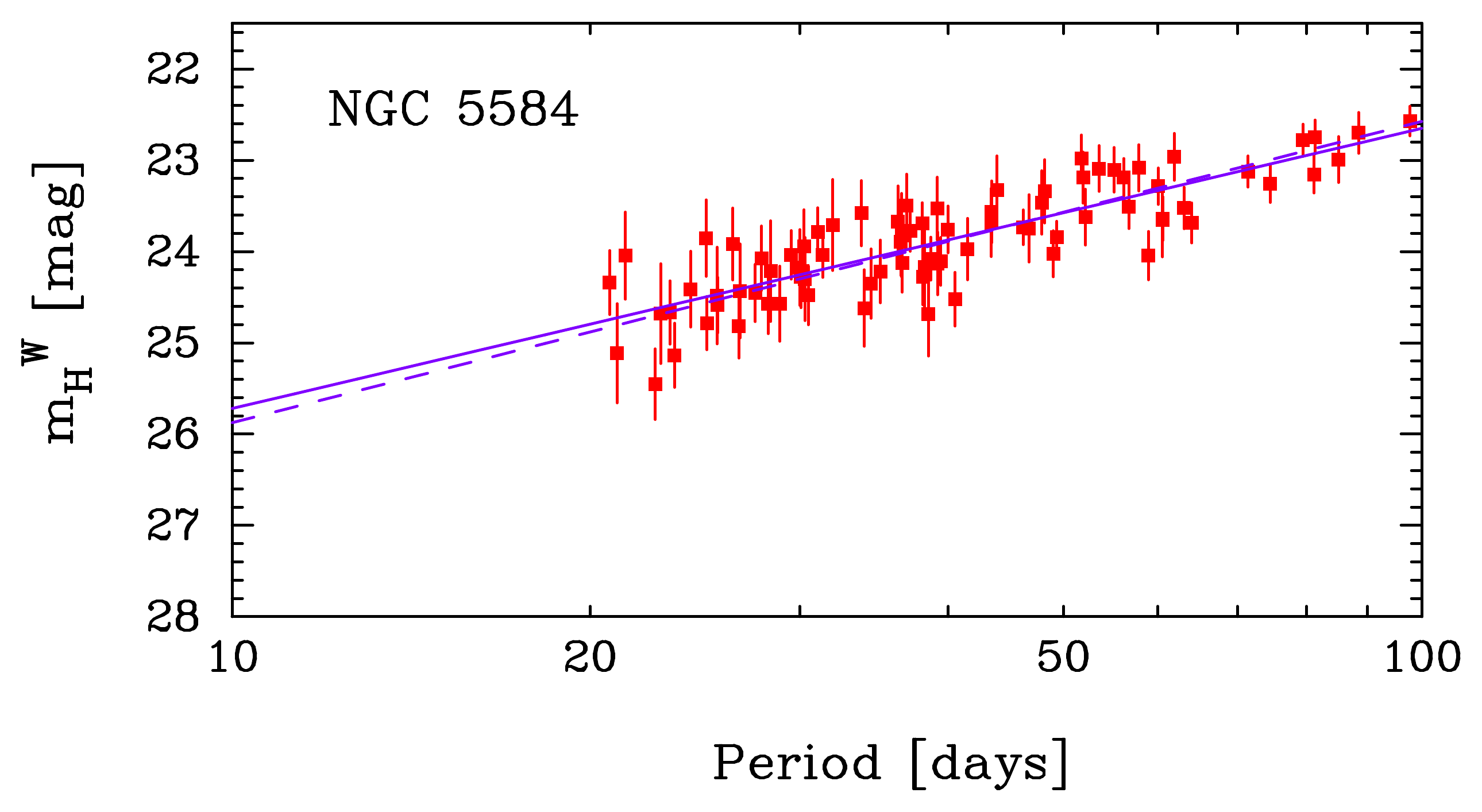}
\caption
{As Fig. \ref{fig:R4536}}

\label{fig:R5584}
\end{figure}

\begin{figure}

\includegraphics[width=150mm, angle=0]{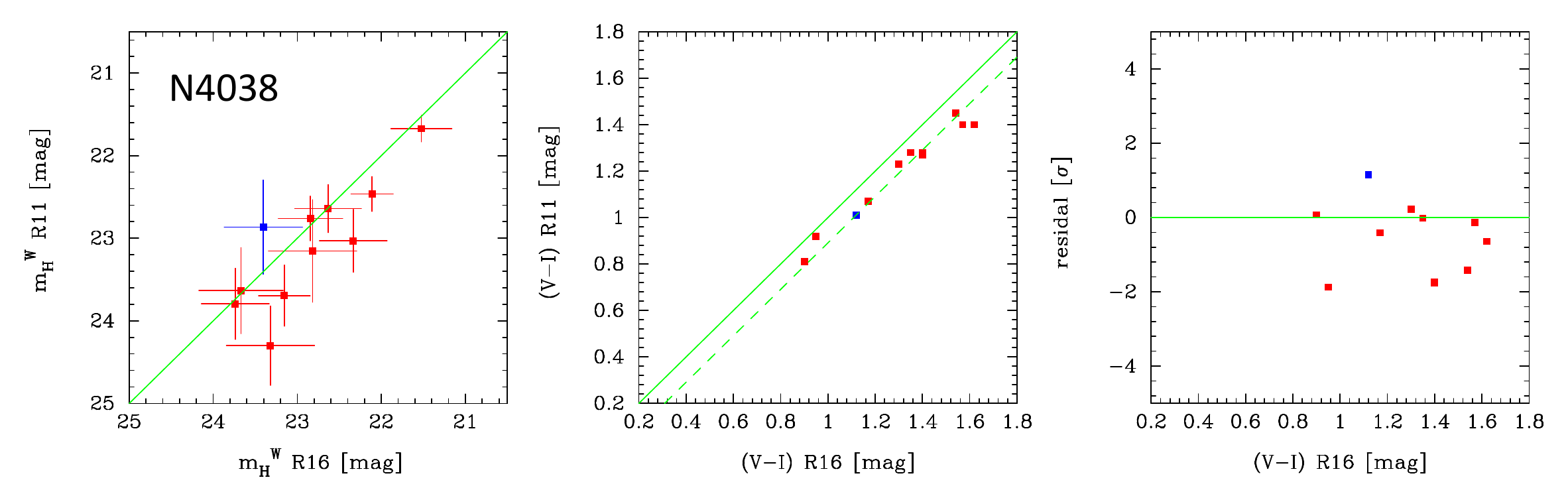}
\caption
{As Fig. \ref{fig:4536}. There are 39 Cepheids listed in R11 and 13 in R16 with 11 matches.}
\label{fig:5584}
\end{figure}

\begin{figure}

\includegraphics[width=75mm, angle=0]{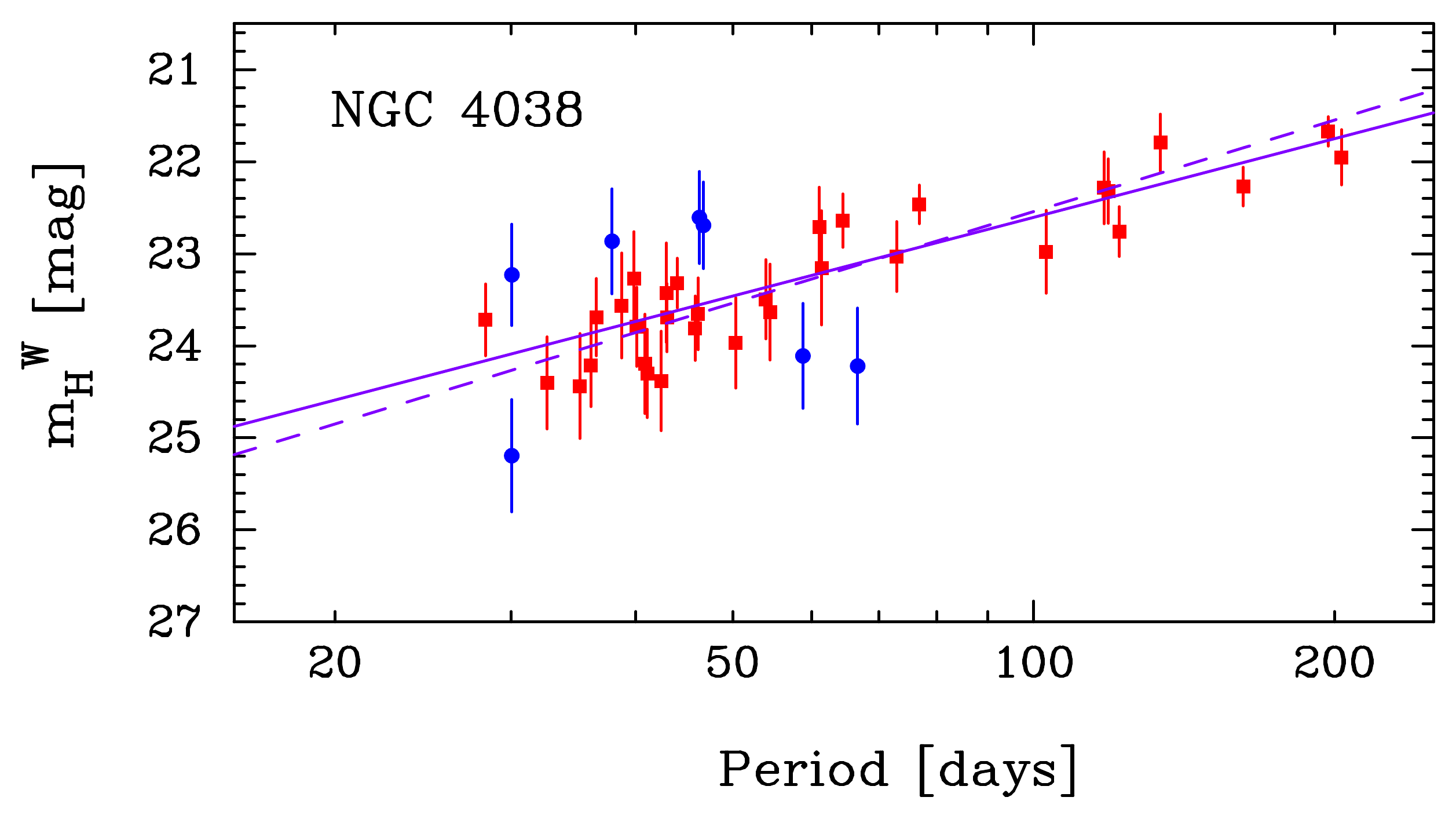} \includegraphics[width=75mm, angle=0]{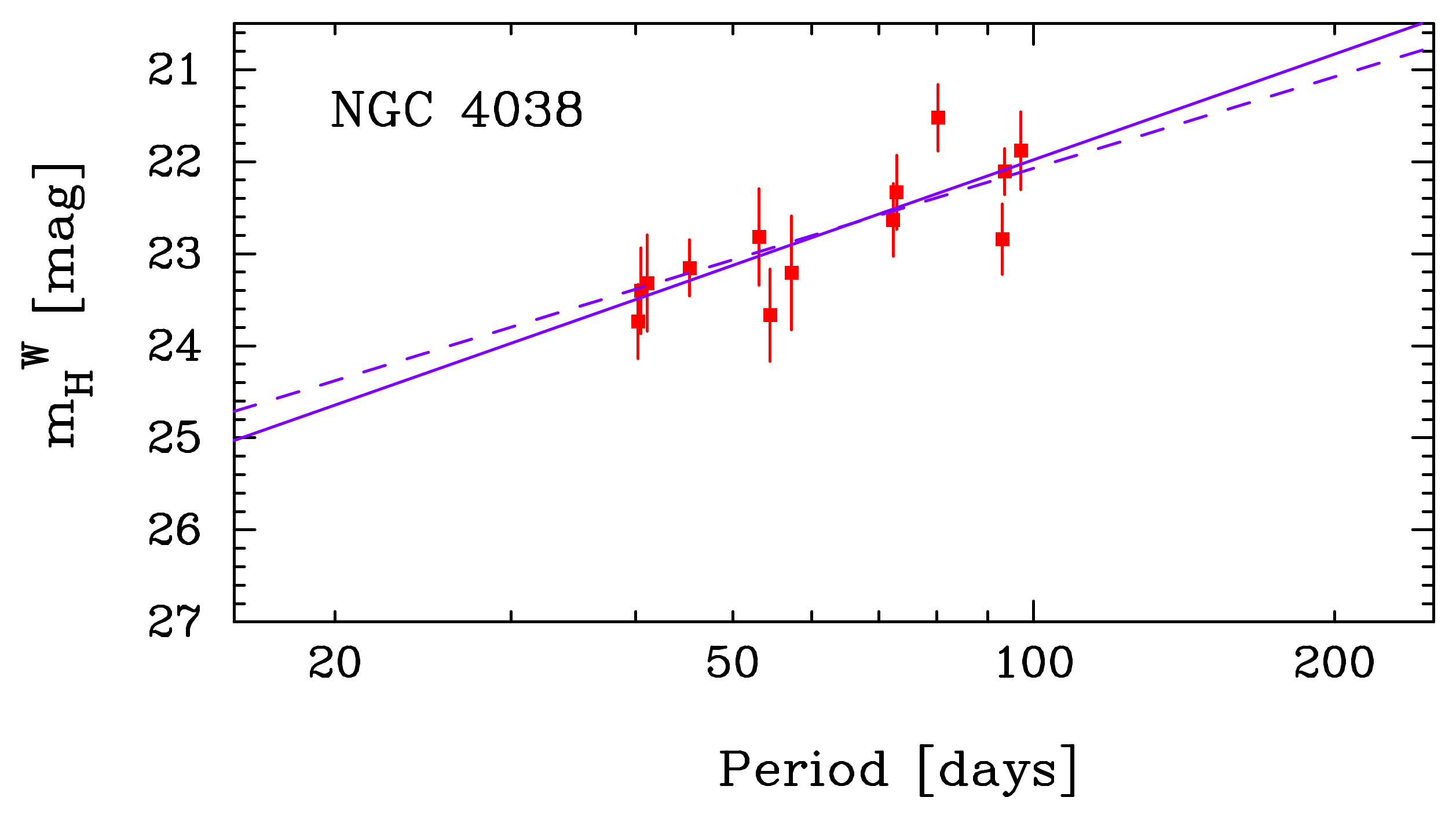}
\caption
{As Fig. \ref{fig:R4536}}

\label{fig:R4038}
\end{figure}

\begin{figure}

\includegraphics[width=150mm, angle=0]{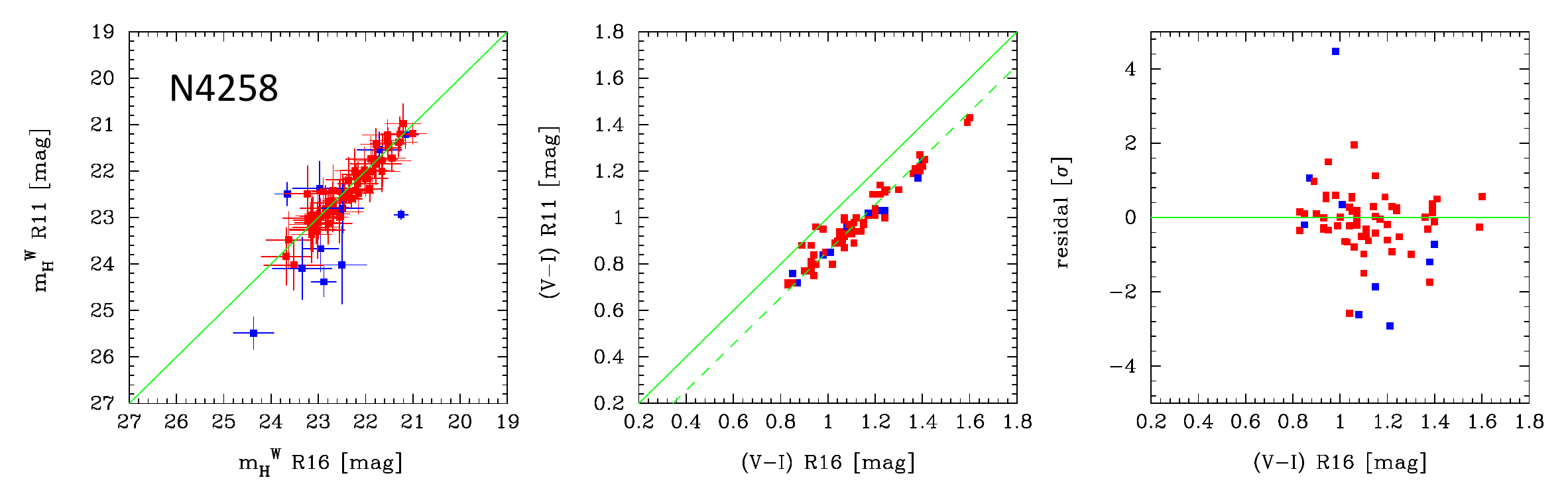}
\caption
{As Fig. \ref{fig:4536}.}
\end{figure}

\begin{figure}

\includegraphics[width=75mm, angle=0]{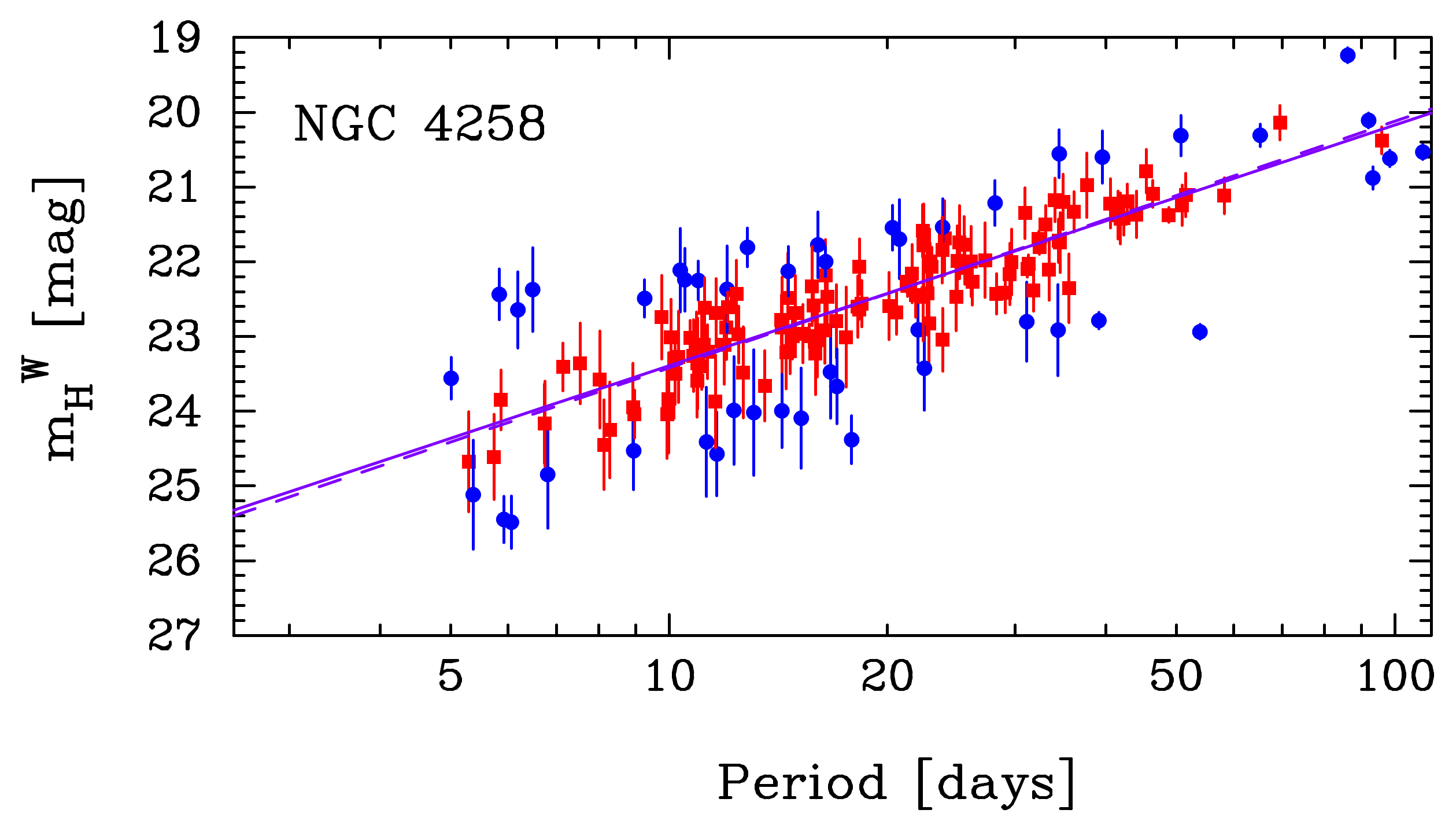} \includegraphics[width=75mm, angle=0]{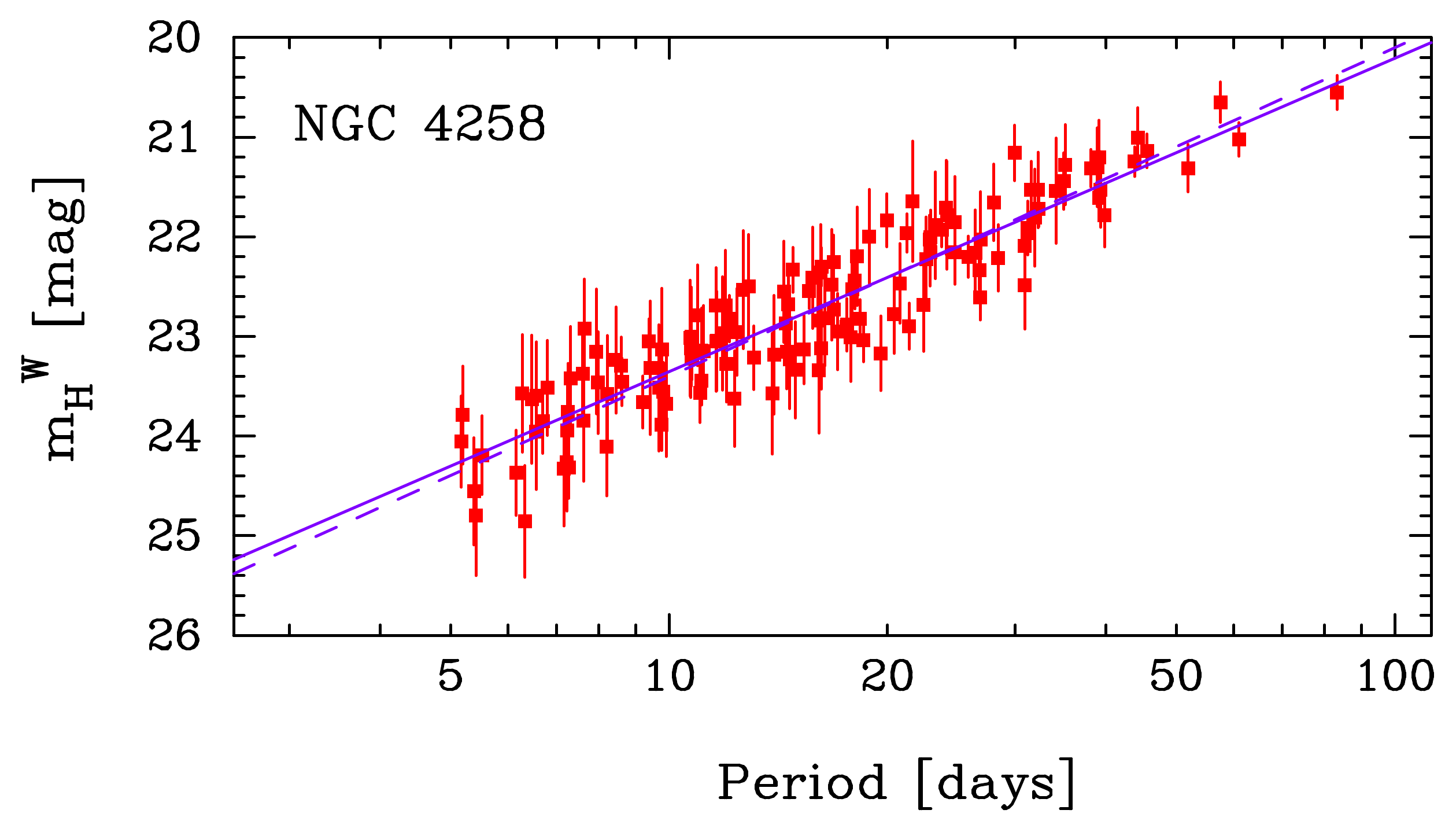}
\caption
{As Fig. \ref{fig:R4536}}

\label{fig:R4258}
\end{figure}

\newpage




\bibliographystyle{h-physrev}
\bibliography{lockdown}

\end{document}